%% file: main.tex
\documentclass[lettersize,journal]{IEEEtran}
\usepackage{cite}
\usepackage{booktabs}
\usepackage{amsmath}
\usepackage{amssymb}
\usepackage{graphicx}
\usepackage{bm}
\usepackage{multirow}
\usepackage{tikz}
\usepackage{tikz}
\newcommand*\emptycirc[1][0.6ex]{\tikz\draw (0,0) circle (#1);} 
\newcommand*\halfcirc[1][0.6ex]{%
	\begin{tikzpicture}
	\draw[fill] (0,0)-- (90:#1) arc (90:270:#1) -- cycle ;
	\draw (0,0) circle (#1);
	\end{tikzpicture}}
\newcommand*\fullcirc[1][0.6ex]{\tikz\fill (0,0) circle (#1);}

\usepackage{xspace}
\usepackage{xcolor}

\let\originaltextcolor\textcolor
\renewcommand{\textcolor}[2]{\originaltextcolor{black}{#2}} 
\usepackage{colortbl}
\usepackage{etoolbox}
\usepackage{verbatim}
\newcommand*{\minval}{0.0}

\newcommand*{\maxval}{100.0}
\newcommand{\gradient}[1]{
    \ifdimcomp{#1 pt}{<}{\minval pt}{#1}{
        \ifdimcomp{#1 pt}{>}{\maxval pt}{#1}{
            \pgfmathparse{int(100.0*((#1-\minval)/(\maxval-\minval)))}
            \xdef\tempa{\pgfmathresult}
            \cellcolor{gray!\tempa!lightgray} #1
    }}
}

\newcommand{\GPSNR}[1]{
    \ifdimcomp{#1 pt}{<}{24.0 pt}{#1}{
        \ifdimcomp{#1 pt}{>}{35.0 pt}{#1}{
            \pgfmathparse{int(100.0*((#1-24.0)/(35.0-24.0)))}
            \xdef\tempa{\pgfmathresult}
            \cellcolor{gray!\tempa!lightgray} #1
    }}
}

\newcommand{\GSSIM}[1]{
    \ifdimcomp{#1 pt}{<}{0.5 pt}{#1}{
        \ifdimcomp{#1 pt}{>}{0.76 pt}{#1}{
            \pgfmathparse{int(100.0*((#1-0.5)/(0.76-0.5)))}
            \xdef\tempa{\pgfmathresult}
            \cellcolor{gray!\tempa!lightgray} #1
    }}
}

\newcommand{\GDE}[1]{
    \ifdimcomp{#1 pt}{<}{0.66 pt}{#1}{
        \ifdimcomp{#1 pt}{>}{1.1 pt}{#1}{
            \pgfmathparse{int(100.0*((#1-0.66)/(1.1-0.66)))}
            \xdef\tempa{\pgfmathresult}
            \cellcolor{lightgray!\tempa!gray} #1
    }}
}

\newcommand{\GLPIPS}[1]{
    \ifdimcomp{#1 pt}{<}{0.17 pt}{#1}{
        \ifdimcomp{#1 pt}{>}{0.4 pt}{#1}{
            \pgfmathparse{int(100.0*((#1-0.17)/(0.4-0.17)))}
            \xdef\tempa{\pgfmathresult}
            \cellcolor{lightgray!\tempa!gray} #1
    }}
}

\newcommand{\GFID}[1]{
    \ifdimcomp{#1 pt}{<}{33 pt}{#1}{
        \ifdimcomp{#1 pt}{>}{1100 pt}{#1}{
            \pgfmathparse{int(100.0*((#1-33)/(1099-33)))}
            \xdef\tempa{\pgfmathresult}
            \cellcolor{lightgray!\tempa!gray} #1
    }}
}

\newcommand{\GTRANS}[1]{
    \ifdimcomp{#1 pt}{<}{90.3 pt}{#1}{
        \ifdimcomp{#1 pt}{>}{99.5 pt}{#1}{
            \pgfmathparse{int(100.0*((#1-90.3)/(99.5-90.3)))}
            \xdef\tempa{\pgfmathresult}
            \cellcolor{gray!\tempa!lightgray} #1
    }}
}

\newcommand{\GACC}[1]{
    \ifdimcomp{#1 pt}{<}{34.9 pt}{#1}{
        \ifdimcomp{#1 pt}{>}{100.0 pt}{#1}{
            \pgfmathparse{int(100.0*((#1-34.9)/(100.0-34.9)))}
            \xdef\tempa{\pgfmathresult}
            \cellcolor{gray!\tempa!lightgray} #1
    }}
}

\newcommand{\GAI}[1]{
    \ifdimcomp{#1 pt}{<}{4.4 pt}{#1}{
        \ifdimcomp{#1 pt}{>}{41.0 pt}{#1}{
            \pgfmathparse{int(100.0*((#1-4.4)/(41.0-4.4)))}
            \xdef\tempa{\pgfmathresult}
            \cellcolor{gray!\tempa!lightgray} #1
    }}
}

\newcommand{\GAD}[1]{
    \ifdimcomp{#1 pt}{<}{7.6 pt}{#1}{
        \ifdimcomp{#1 pt}{>}{29.5 pt}{#1}{
            \pgfmathparse{int(100.0*((#1-7.6)/(29.5-7.6)))}
            \xdef\tempa{\pgfmathresult}
            \cellcolor{lightgray!\tempa!gray} #1
    }}
}

\newcommand{\GKL}[1]{
    \ifdimcomp{#1 pt}{<}{14.8 pt}{#1}{
        \ifdimcomp{#1 pt}{>}{22.4 pt}{#1}{
            \pgfmathparse{int(100.0*((#1-14.8)/(22.4-14.8)))}
            \xdef\tempa{\pgfmathresult}
            \cellcolor{lightgray!\tempa!gray} #1
    }}
}

\usepackage{subcaption}
\usepackage{amsfonts}
\usepackage{dsfont}
\usepackage{pifont}
\newcommand{\cmark}{\ding{51}}%
\newcommand{\xmark}{\ding{55}}%

\input{tex/plot}

\usepackage{xurl}
\newcommand{\mypara}[1]{\noindent\textbf{#1.}}

\DeclareSymbolFont{cmsymbols}{OMS}{cmsy}{m}{n}
\SetSymbolFont{cmsymbols}{bold}{OMS}{cmsy}{b}{n}
\DeclareSymbolFontAlphabet{\mathcal}{cmsymbols}

\definecolor{Gray}{gray}{0.9}

\begin{document}
%
\title{Revisiting Transferable Adversarial Images: Systemization, Evaluation, and New Insights}

\author{Zhengyu Zhao$^{1}$, \quad Hanwei Zhang$^{2,3}$ \quad Renjue Li$^{4}$\\ Ronan Sicre$^5$ \quad  Laurent Amsaleg$^6$  \quad Michael Backes$^7$ \quad Qi Li$^8$ \quad Qian Wang$^9$ \quad Chao Shen$^1$\\
$^1$Xi'an Jiaotong University, China \quad $^2$Institute of Intelligent Software, China \quad $^3$Saarland University, Germany \quad\\$^4$Institute of AI for Industries/CAS, China \quad $^5$LIS - Ecole Centrale Marseille, France\\
\quad $^6$Inria/Univ Rennes/CNRS/IRISA, France \quad $^7$CISPA Helmholtz Center for Information Security, Germany \\$^8$Tsinghua University, China \quad $^9$Wuhan University, China\\
\thanks{The first three authors make equal contributions.}
\thanks{Zhengyu Zhao is the corresponding author (zhengyu.zhao@xjtu.edu.cn).}
}

\markboth{Journal of \LaTeX\ Class Files,~Vol.~14, No.~8, August~2021}%
{Shell \MakeLowercase{\textit{et al.}}: A Sample Article Using IEEEtran.cls for IEEE Journals}


\maketitle

\begin{abstract}
Transferable adversarial images raise critical security concerns for computer vision systems in real-world, black-box attack scenarios.
Although many transfer attacks have been proposed, existing research lacks a systematic and comprehensive evaluation.
In this paper, we systemize transfer attacks into five categories around the general machine learning pipeline and provide the first comprehensive evaluation, with 23 representative attacks against 11 representative defenses, including the recent, transfer-oriented defense and the real-world Google Cloud Vision.
In particular, we identify two main problems of existing evaluations: (1) for attack transferability, lack of intra-category analyses with fair hyperparameter settings, and (2) for attack stealthiness, lack of diverse measures.
Our evaluation results validate that these problems have indeed caused misleading conclusions and missing points, and addressing them leads to new, \textit{consensus-challenging} insights, such as (1) an early attack, DI, even outperforms all similar follow-up ones, (2) the state-of-the-art (white-box) defense, DiffPure, is even vulnerable to (black-box) transfer attacks, and (3) even under the same $L_p$ constraint, different attacks yield dramatically different stealthiness results regarding diverse imperceptibility metrics, finer-grained measures, and a user study.
We hope that our analyses will serve as guidance on properly evaluating transferable adversarial images and advance the design of attacks and defenses.
Code is available at \url{https://github.com/ZhengyuZhao/TransferAttackEval}.
\end{abstract}

%

\input{tex/defn}
\input{tex/Introduction}

\input{tex/CategorizeAttacks}

\input{tex/EvaluationMethodology}
\input{tex/CategorizeAnalysis}

\input{tex/TransferResults}
\input{tex/StealthResults}

\input{tex/Discussions}

\input{tex/Conclusion}



\bibliographystyle{plain}
\bibliography{ref}

\input{tex/Appendix}

\end{document}

%% file: tex/plot.tex
\usepackage{tikz}
\usetikzlibrary{arrows.meta,shapes,calc,matrix,fit,positioning,backgrounds,decorations.markings,fadings}
\usepackage{pgfplots}
\usepackage{pgfplotstable}
\pgfplotsset{compat=1.9}
\usepackage{xstring}

\usepgfplotslibrary{external}

\IfBeginWith*{\jobname}{fig/extern/}{\finalcopy}{}

\tikzstyle{every picture}+=[
	remember picture,
	every text node part/.style={align=center},
	every matrix/.append style={ampersand replacement=\&},
]
\tikzstyle{tight} = [inner sep=0pt,outer sep=0pt]
\tikzstyle{node}  = [draw,circle,tight,minimum size=12pt,anchor=center]
\tikzstyle{op}    = [draw,circle,tight]
\tikzstyle{dot}   = [fill,draw,circle,inner sep=1pt,outer sep=0]
\tikzstyle{pt}    = [fill,draw,circle,inner sep=1.5pt,outer sep=.2pt]
\tikzstyle{box}   = [draw,rectangle,inner sep=3pt]
\tikzstyle{high}  = [black!60]
\tikzstyle{group} = [high,box,opacity=.5]
\tikzstyle{dim1}  = [fill opacity=.3,text opacity=1]
\tikzstyle{dim2}  = [fill opacity=.5,text opacity=1]
\tikzstyle{dim3}  = [fill opacity=.7,text opacity=1]
\tikzstyle{rectc} = [tight,transform shape]
\tikzstyle{rect}  = [rectc,anchor=south west]

\tikzset{every mark/.append style={solid}}
\pgfplotsset{
	grid=both, width=\columnwidth, try min ticks=5,
	every axis/.append style={font=\small},
	every axis plot/.append style={thick,mark=none,mark size=1.8,tension=0.18},
	legend cell align=left, legend style={fill opacity=0.8},
	xticklabel={\pgfmathprintnumber[assume math mode=true]{\tick}},
	yticklabel={\pgfmathprintnumber[assume math mode=true]{\tick}},
	nodes near coords math/.style={
		nodes near coords={\pgfmathprintnumber[assume math mode=true]{\pgfplotspointmeta}},
	},
}

\pgfplotsset{
	dash/.style={mark=o,dashed,opacity=0.6},
	dott/.style={mark=o,dotted,opacity=0.6},
	nolim/.style={enlargelimits=false},
	plain/.style={every axis plot/.append style={},nolim,grid=none},
}

\pgfdeclarelayer{bg4}
\pgfdeclarelayer{bg3}
\pgfdeclarelayer{bg2}
\pgfdeclarelayer{bg1}
\pgfdeclarelayer{fg1}
\pgfdeclarelayer{fg2}
\pgfdeclarelayer{fg3}
\pgfdeclarelayer{fg4}
\pgfsetlayers{bg4,bg3,bg2,bg1,main,fg1,fg2,fg3,fg4}

\pgfkeys{/tikz/.cd, aspect/.store in=\aspect, aspect=1}
\pgfkeys{/tikz/.cd, depth/.store in=\depth, depth=.5}
\pgfkeys{/tikz/.cd, stepx/.store in=\step, stepx=1}

\tikzstyle{geom} = [line join=bevel,aspect=1,depth=.5,z={(\depth*\aspect,\depth)}]
\tikzstyle{wire} = [geom,draw,thick]

\def\cx[#1,#2,#3]{#1}
\def\cy[#1,#2,#3]{#2}
\def\cz[#1,#2,#3]{#3}
\def\ex[#1,#2,#3]{#1,0,0}
\def\ey[#1,#2,#3]{0,#2,0}
\def\ez[#1,#2,#3]{0,0,#3}



%% file: tex/defn.tex
\newcommand{\relu}{\operatorname{relu}}
\newcommand{\gap}{\operatorname{GAP}}
\newcommand{\up}{\operatorname{up}}

\newcommand{\cam}{\textrm{CAM}}
\newcommand{\gcam}{\textrm{Grad-CAM}}
\newcommand{\scam}{\textrm{Score-CAM}}

\newcommand{\AD}{\operatorname{AD}}
\newcommand{\AI}{\operatorname{AI}}
\newcommand{\OM}{\operatorname{OM}}
\newcommand{\LE}{\operatorname{LE}}
\newcommand{\Fo}{\operatorname{F1}}
\newcommand{\prc}{\operatorname{precision}}
\newcommand{\rec}{\operatorname{recall}}
\newcommand{\BA}{\operatorname{BoxAcc}}
\newcommand{\spg}{\operatorname{SP}}
\newcommand{\epg}{\operatorname{EP}}
\newcommand{\SM}{\operatorname{SM}}
\newcommand{\iou}{\operatorname{IoU}}

\newcommand{\alert}[1]{{\color{red}{#1}}}
\newcommand{\sm}{\scriptsize}
\newcommand{\eq}[1]{(\ref{eq:#1})}

\newcommand{\Th}[1]{\textsc{#1}}
\newcommand{\mr}[2]{\multirow{#1}{*}{#2}}
\newcommand{\mc}[2]{\multicolumn{#1}{c}{#2}}
\newcommand{\tb}[1]{\textbf{#1}}
\newcommand{\ch}{\checkmark}

\newcommand{\red}[1]{{\textcolor{red}{#1}}}
\newcommand{\blue}[1]{{\textcolor{blue}{#1}}}
\newcommand{\green}[1]{{\textcolor{green}{#1}}}
\newcommand{\gray}[1]{{\textcolor{gray}{#1}}}

\newcommand{\tran}{^\top}
\newcommand{\mtran}{^{-\top}}
\newcommand{\zcol}{\mathbf{0}}
\newcommand{\zrow}{\zcol\tran}

\newcommand{\ind}{\mathbb{1}}
\newcommand{\expect}{\mathbb{E}}
\newcommand{\nat}{\mathbb{N}}
\newcommand{\zahl}{\mathbb{Z}}
\newcommand{\real}{\mathbb{R}}
\newcommand{\proj}{\mathbb{P}}
\newcommand{\prob}{\operatorname{P}}
\newcommand{\normal}{\mathcal{N}}

\newcommand{\mif}{\textrm{if}\ }
\newcommand{\other}{\textrm{otherwise}}
\newcommand{\minimize}{\textrm{minimize}\ }
\newcommand{\maximize}{\textrm{maximize}\ }
\newcommand{\st}{\textrm{subject\ to}\ }

\newcommand{\id}{\operatorname{id}}
\newcommand{\const}{\operatorname{const}}
\newcommand{\sgn}{\operatorname{sgn}}
\newcommand{\var}{\operatorname{Var}}
\newcommand{\mean}{\operatorname{mean}}
\newcommand{\trace}{\operatorname{tr}}
\newcommand{\diag}{\operatorname{diag}}
\newcommand{\vect}{\operatorname{vec}}
\newcommand{\cov}{\operatorname{cov}}
\newcommand{\sign}{\operatorname{sign}}
\newcommand{\prj}{\operatorname{proj}}

\newcommand{\softmax}{\operatorname{softmax}}
\newcommand{\clip}{\operatorname{clip}}

\newcommand{\defn}{\mathrel{:=}}
\newcommand{\peq}{\mathrel{+\!=}}
\newcommand{\meq}{\mathrel{-\!=}}

\newcommand{\paren}[1]{\left({#1}\right)}
\newcommand{\mat}[1]{\left[{#1}\right]}
\newcommand{\set}[1]{\left\{{#1}\right\}}
\newcommand{\floor}[1]{\left\lfloor{#1}\right\rfloor}
\newcommand{\ceil}[1]{\left\lceil{#1}\right\rceil}
\newcommand{\inner}[1]{\left\langle{#1}\right\rangle}
\newcommand{\norm}[1]{\left\|{#1}\right\|}
\newcommand{\abs}[1]{\left|{#1}\right|}
\newcommand{\frob}[1]{\norm{#1}_F}
\newcommand{\card}[1]{\left|{#1}\right|\xspace}

\newcommand{\diff}{\mathrm{d}}
\newcommand{\der}[3][]{\frac{\diff^{#1}#2}{\diff#3^{#1}}}
\newcommand{\ider}[3][]{\diff^{#1}#2/\diff#3^{#1}}
\newcommand{\pder}[3][]{\frac{\partial^{#1}{#2}}{\partial{{#3}^{#1}}}}
\newcommand{\ipder}[3][]{\partial^{#1}{#2}/\partial{#3^{#1}}}
\newcommand{\dder}[3]{\frac{\partial^2{#1}}{\partial{#2}\partial{#3}}}

\newcommand{\wb}[1]{\overline{#1}}
\newcommand{\wt}[1]{\widetilde{#1}}

\def\xssp{\hspace{1pt}}
\def\ssp{\hspace{3pt}}
\def\msp{\hspace{5pt}}
\def\lsp{\hspace{12pt}}

\newcommand{\cA}{\mathcal{A}}
\newcommand{\cB}{\mathcal{B}}
\newcommand{\cC}{\mathcal{C}}
\newcommand{\cD}{\mathcal{D}}
\newcommand{\cE}{\mathcal{E}}
\newcommand{\cF}{\mathcal{F}}
\newcommand{\cG}{\mathcal{G}}
\newcommand{\cH}{\mathcal{H}}
\newcommand{\cI}{\mathcal{I}}
\newcommand{\cJ}{\mathcal{J}}
\newcommand{\cK}{\mathcal{K}}
\newcommand{\cL}{\mathcal{L}}
\newcommand{\cM}{\mathcal{M}}
\newcommand{\cN}{\mathcal{N}}
\newcommand{\cO}{\mathcal{O}}
\newcommand{\cP}{\mathcal{P}}
\newcommand{\cQ}{\mathcal{Q}}
\newcommand{\cR}{\mathcal{R}}
\newcommand{\cS}{\mathcal{S}}
\newcommand{\cT}{\mathcal{T}}
\newcommand{\cU}{\mathcal{U}}
\newcommand{\cV}{\mathcal{V}}
\newcommand{\cW}{\mathcal{W}}
\newcommand{\cX}{\mathcal{X}}
\newcommand{\cY}{\mathcal{Y}}
\newcommand{\cZ}{\mathcal{Z}}

\newcommand{\vA}{\mathbf{A}}
\newcommand{\vB}{\mathbf{B}}
\newcommand{\vC}{\mathbf{C}}
\newcommand{\vD}{\mathbf{D}}
\newcommand{\vE}{\mathbf{E}}
\newcommand{\vF}{\mathbf{F}}
\newcommand{\vG}{\mathbf{G}}
\newcommand{\vH}{\mathbf{H}}
\newcommand{\vI}{\mathbf{I}}
\newcommand{\vJ}{\mathbf{J}}
\newcommand{\vK}{\mathbf{K}}
\newcommand{\vL}{\mathbf{L}}
\newcommand{\vM}{\mathbf{M}}
\newcommand{\vN}{\mathbf{N}}
\newcommand{\vO}{\mathbf{O}}
\newcommand{\vP}{\mathbf{P}}
\newcommand{\vQ}{\mathbf{Q}}
\newcommand{\vR}{\mathbf{R}}
\newcommand{\vS}{\mathbf{S}}
\newcommand{\vT}{\mathbf{T}}
\newcommand{\vU}{\mathbf{U}}
\newcommand{\vV}{\mathbf{V}}
\newcommand{\vW}{\mathbf{W}}
\newcommand{\vX}{\mathbf{X}}
\newcommand{\vY}{\mathbf{Y}}
\newcommand{\vZ}{\mathbf{Z}}

\newcommand{\va}{\mathbf{a}}
\newcommand{\vb}{\mathbf{b}}
\newcommand{\vc}{\mathbf{c}}
\newcommand{\vd}{\mathbf{d}}
\newcommand{\ve}{\mathbf{e}}
\newcommand{\vf}{\mathbf{f}}
\newcommand{\vg}{\mathbf{g}}
\newcommand{\vh}{\mathbf{h}}
\newcommand{\vi}{\mathbf{i}}
\newcommand{\vj}{\mathbf{j}}
\newcommand{\vk}{\mathbf{k}}
\newcommand{\vl}{\mathbf{l}}
\newcommand{\vm}{\mathbf{m}}
\newcommand{\vn}{\mathbf{n}}
\newcommand{\vo}{\mathbf{o}}
\newcommand{\vp}{\mathbf{p}}
\newcommand{\vq}{\mathbf{q}}
\newcommand{\vr}{\mathbf{r}}
\newcommand{\Vs}{\mathbf{s}}
\newcommand{\vt}{\mathbf{t}}
\newcommand{\vu}{\mathbf{u}}
\newcommand{\vv}{\mathbf{v}}
\newcommand{\vw}{\mathbf{w}}
\newcommand{\vx}{\mathbf{x}}
\newcommand{\vy}{\mathbf{y}}
\newcommand{\vz}{\mathbf{z}}

\newcommand{\vone}{\mathbf{1}}
\newcommand{\vzero}{\mathbf{0}}

\newcommand{\valpha}{{\mathbf{\alpha}}}
\newcommand{\vbeta}{{\mathbf{\beta}}}
\newcommand{\vgamma}{{\mathbf{\gamma}}}
\newcommand{\vdelta}{{\mathbf{\delta}}}
\newcommand{\vepsilon}{{\mathbf{\epsilon}}}
\newcommand{\vzeta}{{\mathbf{\zeta}}}
\newcommand{\veta}{{\mathbf{\eta}}}
\newcommand{\vtheta}{{\mathbf{\theta}}}
\newcommand{\viota}{{\mathbf{\iota}}}
\newcommand{\vkappa}{{\mathbf{\kappa}}}
\newcommand{\vlambda}{{\mathbf{\lambda}}}
\newcommand{\vmu}{{\mathbf{\mu}}}
\newcommand{\vnu}{{\mathbf{\nu}}}
\newcommand{\vxi}{{\mathbf{\xi}}}
\newcommand{\vomikron}{{\mathbf{\omikron}}}
\newcommand{\vpi}{{\mathbf{\pi}}}
\newcommand{\vrho}{{\mathbf{\rho}}}
\newcommand{\vsigma}{{\mathbf{\sigma}}}
\newcommand{\vtau}{{\mathbf{\tau}}}
\newcommand{\vupsilon}{{\mathbf{\upsilon}}}
\newcommand{\vphi}{{\mathbf{\phi}}}
\newcommand{\vchi}{{\mathbf{\chi}}}
\newcommand{\vpsi}{{\mathbf{\psi}}}
\newcommand{\vomega}{{\mathbf{\omega}}}

\newcommand{\rLambda}{\mathrm{\Lambda}}
\newcommand{\rSigma}{\mathrm{\Sigma}}

\newcommand{\vLambda}{\mathbf{\rLambda}}
\newcommand{\vSigma}{\mathbf{\rSigma}}

\makeatletter
\newcommand*\bdot{\mathpalette\bdot@{.7}}
\newcommand*\bdot@[2]{\mathbin{\vcenter{\hbox{\scalebox{#2}{$\m@th#1\bullet$}}}}}
\makeatother

\makeatletter
\DeclareRobustCommand\onedot{\futurelet\@let@token\@onedot}
\def\@onedot{\ifx\@let@token.\else.\null\fi\xspace}

\def\eg{\emph{e.g}\onedot} \def\Eg{\emph{E.g}\onedot}
\def\ie{\emph{i.e}\onedot} \def\Ie{\emph{I.e}\onedot}
\def\cf{\emph{cf}\onedot} \def\Cf{\emph{Cf}\onedot}
\def\etc{\emph{etc}\onedot} \def\vs{\emph{vs}\onedot}
\def\wrt{w.r.t\onedot} \def\dof{d.o.f\onedot} \def\aka{a.k.a\onedot}
\def\etal{\emph{et al}\onedot}
\makeatother

%% file: tex/Introduction.tex
\section{Introduction}
\label{sec:intro}
\renewcommand{\thefootnote}{\arabic{footnote}}
\IEEEPARstart{D}{eep} Neural Networks (DNNs) have achieved great success in various machine learning tasks.
However, they are known to be vulnerable to adversarial examples~\cite{goodfellow2014explaining,szegedy2014intriguing}, which are intentionally perturbed model inputs to induce prediction errors.
Adversarial examples have been studied in the white-box scenario to explore the worst-case model robustness but also in the black-box scenario to understand the real-world threats to deployed models.

An intriguing property of adversarial examples that makes them threatening in the real world is their transferability.
Transfer attacks assume a realistic scenario the adversarial examples generated on a (local) surrogate model can be directly transferred to the (unknown) target model~\cite{papernot2017practical,liu2017delving}.
Such attacks require no ``bad'' query~\cite{debenedetti2023evading} interaction with the target model, and as a result, they eliminate the possibility of being easily flagged by the target model under attack~\cite{chen2020stateful}.
Due to their real-world implications, transfer attacks have attracted great attention, and an increasing number of new methods have been rapidly developed.

However, existing research still lacks a systematic evaluation of transferable adversarial images in a common, comprehensive threat model. 
Specifically, we identify two major problems of common evaluation practices as follows.

\begin{table*}[!t]

\newcommand{\tabincell}[2]{\begin{tabular}{@{}#1@{}}#2\end{tabular}}
\newcommand*{\myalign}[2]{\multicolumn{1}{#1}{#2}}
\caption{Evaluation practices in existing work compared to ours regarding attack transferability and stealthiness.}

\renewcommand{\arraystretch}{1}
      \centering
      \resizebox{1\textwidth}{!}{
        \begin{tabular}{l|l|l}
\toprule[1pt]
Evaluation&\myalign{c|}{Transferability}&\myalign{c}{Stealthiness}\\
\midrule

    \multirow{2}{*}{Existing}&\textbf{Unsystematic:} Lack of intra-category comparisons (half the time) \textcolor{red}{\xmark}&\multirow{2}{*}{Pre-fixed $L_{\infty}$ norm without comparisons \textcolor{red}{\xmark}}\\
&\textbf{Unfair:} Inconsistent hyperparameter settings for similar attacks \textcolor{red}{\xmark}&\\
\midrule
\multirow{4}{*}{Ours}&&Pre-fixed $L_{\infty}$ norm with diverse comparisons \textcolor{green}{\cmark}\\
&\textbf{Systematic:} Intra-category comparisons with detailed analyses \textcolor{green}{\cmark}&(1) Five popular imperceptibility metrics\\
&\textbf{Fair:} Consistent hyperparameter settings for similar attacks \textcolor{green}{\cmark}&(2) Two new fine-grained stealthiness measures\\
&&~~~~from the perspective of attack traceback\\
&&(3) A user study on the image quality\\

\bottomrule[1pt]
\end{tabular}
}
\label{tab:eval}
\end{table*}

\begin{table}[!t]
\caption{Overview of our evaluated transfer attacks.}
\newcommand{\tabincell}[2]{\begin{tabular}{@{}#1@{}}#2\end{tabular}}

\renewcommand{\arraystretch}{1}
      \centering
      \resizebox{1\columnwidth}{!}{
        \begin{tabular}{ccccc}
\toprule[1pt]
\tabincell{c}{Gradient\\Stabilization}&\tabincell{c}{Input\\Augmentation}&\tabincell{c}{Feature\\Disruption}&\tabincell{c}{Surrogate\\Refinement}&\tabincell{c}{Generative\\Modeling}\\
\midrule

&DI~\cite{xie2019improving} (CVPR'19)&TAP~\cite{zhou2018transferable} (ECCV'18)&SGM~\cite{wu2020skip} (ICLR'20)&GAP~\cite{poursaeed2018generative} (CVPR'18)\\
MI~\cite{dong2018boosting} (CVPR'18)&TI~\cite{dong2019evading} (CVPR'19)&AA~\cite{inkawhich2019feature} (CVPR'19)&LinBP~\cite{guo2020backpropagating} (NeurIPS'20)&CDA~\cite{naseer2019cross} (NeurIPS'19)\\
NI~\cite{lin2020nesterov} (ICLR'20)&SI~\cite{lin2020nesterov} (ICLR'20)&ILA~\cite{huang2019enhancing} (ICCV'19)&RFA~\cite{springer2021little} (NeurIPS'21)&TTP~\cite{naseer2021generating} (ICCV'21)\\
PI~\cite{wang2021boosting} (BMVC'21)&VT~\cite{wang2021enhancing} (CVPR'21)&FIA~\cite{wang2021feature} (ICCV'21)&IAA~\cite{zhu2021rethinking} (ICLR'22)&GAPF~\cite{kanth2021learning} (NeurIPS'21)\\
&Admix~\cite{wang2021admix} (ICCV'21)&NAA~\cite{zhang2022improving} (CVPR'22) &DSM~\cite{yang2022boosting} (ICTAI'23)&BIA~\cite{zhang2022beyond} (ICLR'22)\\
\bottomrule[1pt]
\end{tabular}
}
\label{tab:attack}
\end{table}

\mypara{Problem 1}
Existing evaluations on transferability are often \emph{unsystematic} and sometimes \emph{unfair}.
Specifically, they are unsystematic since there lacks intra-category attack analyses.
This is problematic particularly when both attacks are based on the same category of techniques (see our definition in Section~\ref{sec:cate-trans-attack}).
In this case, it is unclear whether a newly proposed method can be a good replacement for a previous similar method.
For example, the original work of TI~\cite{dong2019evading} does not directly compare the proposed TI to the previous method, DI~\cite{xie2019improving}, although both of them are based on input augmentations.
However, we find that actually, TI is consistently worse than DI.
Moreover, unfair hyperparameter settings are sometimes adopted.
For example, recent input augmentation-based attacks (i.e., Admix~\cite{wang2021admix}, SI~\cite{lin2020nesterov}, and VT~\cite{wang2021enhancing}) adopt multiple input copies but are directly compared to previous similar attacks that adopt only one input copy (i.e., DI~\cite{xie2019improving} and TI~\cite{dong2019evading}). 
However, we find that when adopting the same number of input copies, the previous attacks perform much better.

\mypara{Problem 2}
Existing evaluations solely compare attack strength, i.e., transferability, but largely overlook another key attack property, stealthiness.
Instead, they pre-fix the stealthiness based on simply imposing a single $L_p$ norm on the perturbations.
This is problematic since it is well known that a successful attack requires a good trade-off between attack strength and stealthiness.
More generally, (black-box) transfer attacks normally require larger perturbations than white-box attacks for success, but using a single $L_p$ norm is known to be insufficient when perturbations are not very small~\cite{sharif2018suitability,tramer2020fundamental,chen2021measuring}.
In addition, other stealthiness measures beyond the image distance have not been explored.

In this paper, we address the above two major problems by establishing new evaluation guidelines from both the perspectives of transferability and stealthiness.
To address Problem 1, we first systemize existing transfer attacks into five categories (Section~\ref{sec:cate-trans-attack}).
Based on this attack systematization, we then conduct systematic, intra-category attack comparisons (Section~\ref{sec:ana}). 
Our intra-category analyses also help determine the optimal settings of key attack hyperparameters and ensure fairness in our comprehensive attack evaluations (Section~\ref{sec:eva-res-trans}).
To address Problem 2, we compare different attacks regarding diverse stealthiness measures. 
Specifically, we adopt five different imperceptibility metrics.
Moreover, we move beyond imperceptibility and investigate other aspects of stealthiness, e.g., attack traceback with (input) image features or (output) misclassification features.
We also conducted a user study on the visual quality of adversarial images from different attacks.

Overall, we present the first large-scale evaluation of transferable adversarial images on ImageNet, covering 23 representative attacks (in 5 categories) against 11 representative defenses.
These attacks and defenses have been published in the past 5 years (i.e., 2018-2023) at top-tier machine learning and security conferences (e.g., ICLR, ICML, NeurIPS, CVPR, ICCV, ECCV, NDSS), which can represent the current state of the art in transferable adversarial examples. 
In particular, the defenses include the state-of-the-art, transferability-oriented method, PubDef~\cite{sitawarin2023defending}, and the real-world Google Cloud Vision API~\cite{gcv}.
Table~\ref{tab:eval} compares our evaluation to those in existing work in terms of attack transferability and stealthiness.
More detailed comparisons can be found in Table~\ref{tab:comp} of Appendix~\ref{app:com}.
As can be seen, existing work considers 11 attacks at most and has not conducted intra-category analyses.
In addition, only \textit{half} of transferability comparisons are considered to be fair (see detailed discussion in Section~\ref{sec:ana}).

The most relevant previous work is the recent~\cite{li2023towards}, which focuses on achieving an optimal combination of different transfer attacks through empirical comparisons.
In contrast, we focus on systemizing existing transfer attacks and address major problems in existing evaluations.
To the best of our knowledge, most existing work does not consider attack systemization, and a few recent attempts~\cite{zhang2023improving,wang2023structure} are not as complete and detailed as ours and do not provide intra-category analyses.
Another study~\cite{mao2022transfer} has provided a large-scale evaluation of attack transferability against private MLaaS platforms but only considers conventional attacks published before 2017, e.g., FGSM, CW, and DeepFool.

\begin{table}[!t]
\caption{Overview of our evaluated defenses.}
\newcommand{\tabincell}[2]{\begin{tabular}{@{}#1@{}}#2\end{tabular}}

\renewcommand{\arraystretch}{1}
      \centering
      \resizebox{1\columnwidth}{!}{
        \begin{tabular}{cccc}
\toprule[1pt]

\tabincell{c}{Input\\Pre-processing}&\tabincell{c}{Purification\\Network}&\tabincell{c}{Adversarial\\Training}&\tabincell{c}{Practical\\Systems}\\
\midrule
BDR~\cite{xu2017feature} (NDSS'18)&HGD~\cite{liao2018defense} (CVPR'18)&AT$_{\infty}$~\cite{xie2019feature} (CVPR'19)&PubDef~\cite{sitawarin2023defending} (ICLR'24)\\
PD~\cite{prakash2018deflecting} (CVPR'18)&NRP~\cite{naseer2020self} (CVPR'20)&FD$_{\infty}$~\cite{xie2019feature} (CVPR'19)&Google Cloud\\
R\&P~\cite{xie2017mitigating} (ICLR'18)&DiffPure~\cite{nie2022diffusion} (ICML'22)&AT$_{2}$~\cite{salman2020adversarially} (NeurIPS'20) &Vision API~\cite{gcv}\\

\bottomrule[1pt]
\end{tabular}
}
\label{tab:defense}
\end{table}

We draw new insights that complement or even challenge existing knowledge about transfer attacks.
Key insights into both attack transferability and stealthiness are:

\mypara{New insights into transferability}
In general, we find that the performance of different methods is highly contextual, i.e., the optimal (category of) attack/defense for one (category of) defense/attack may not be optimal for another.
In particular, we reveal that a defense may largely overfit specific attacks that generate adversarial examples similar to those used in the optimization of that defense.
For example, DiffPure~\cite{nie2022diffusion} claims the state-of-the-art white-box robustness and even beats adversarial training~\cite{madry2017towards} in their original evaluation.
On the contrary, we demonstrate that \textit{DiffPure is surprisingly vulnerable to (black-box) transfer attacks} that produce semantic perturbations (within the claimed norm bounds).
See Section~\ref{sec:eva-res-trans} for detailed results.
Our intra-category attack analyses also lead to surprising findings regarding transferability.
For example, for gradient stabilization attacks, \textit{using more iterations may even decrease the performance}.
Another example is that the earliest input augmentation-based attack, \textit{DI, actually outperforms all subsequent attacks} when they are compared fairly with the same number of input copies.

\mypara{New insights into stealthiness}
In general, we find that stealthiness may be at odds with transferability.
Specifically,~\textit{almost all transfer attacks yield worse imperceptibility than the PGD baseline}, in terms of five metrics.
This suggests that it is problematic to solely constrain all attacks with the same $L_{\infty}$ norm bound without more comprehensive comparisons.
Moreover, different attacks yield dramatically different stealthiness characteristics, in terms of imperceptibility and other aspects.
For example, the adversarial images generated from different attacks are well separable solely based on their visual features.
In addition, different attacks lead to different misclassification patterns regarding the model output distribution.
These new explorations suggest the possibility of~\textit{attack traceback} based on (input) image features or (output) misclassification features.

Overall, our extensive evaluation results validate that the existing problematic evaluations have indeed caused \emph{misleading conclusions} and \emph{missing points}, and as a result, hindered the assessment of the actual progress in this field.\footnote{Code and a curated list of transferability-related papers are available at \url{https://github.com/ZhengyuZhao/TransferAttackEval}.}
Our new insights could potentially advance the design of new transfer attacks and defenses (see details in Section~\ref{sec:eva-res-trans}). 
Based on our analyses, we recommend the following actionable practices:
\begin{itemize}

\item (Evaluation Guideline) Generate a transferability vs. iteration curve, to avoid the ``no convergence'' or ``overfitting'' of an attack.
\item (Evaluation Guideline) Adopt fair hyperparameter settings when comparing attacks from the same category.
\item (Evaluation Guideline) Measure the stealthiness from diverse aspects beyond just a single $L_p$ norm.
\item (Attack Design) Examine individual attacks in their corresponding category before integrating them but do not just use the new attacks.
\item (Defense Design) Consider diverse potential attacks when designing a defense for better generalizability.

\end{itemize}

%% file: tex/CategorizeAttacks.tex
\begin{figure}[!t]
\centering
\includegraphics[width=\columnwidth]{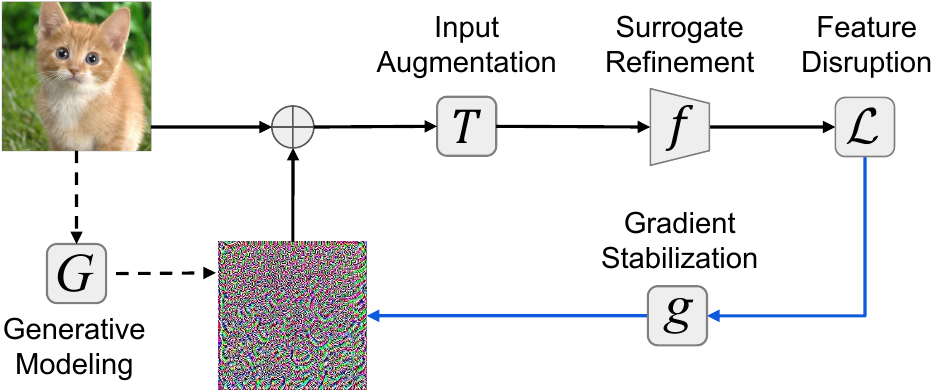}
\caption{Illustration of our attack systemization around the general machine learning pipeline, containing $T$: input, $f$: model, $\mathcal{L}$: loss, and $g$: gradient. Note that the generative modeling attack with the generator $G$ is independent.}
\label{fig:pipeline}
\end{figure}

\section{Systemization of Related Work}
\label{sec:cate-trans-attack}

In this section, we introduce our systemization of transfer attacks.
First of all, these attacks can be generally divided into two types: iterative and generative attacks.
For iterative attacks, we derive four categories: gradient stabilization, input augmentation, feature disruption, and surrogate refinement.
Each of these four categories operates on one of the four major components in the general machine learning pipeline: gradient, (image) input, (loss) output, and (surrogate) model.
The relation between these four categories and the pipeline is illustrated in Fig.~\ref{fig:pipeline}.
Due to the specific nature of generative attacks, we leave it as an independent category: generative modeling.

In general, these five categories are disjoint to each other, i.e., each attack falls into \emph{only one} category according to the technique that was claimed as the key contribution in its original work.
Different attack methods that share the same key idea fall into the same category and can be compared systematically in a single dimension.
Table~\ref{tab:key} summarizes the key ideas of the five categories.
Our attack systemization is also suggested to be reliable based on a perturbation classification task (see Section~\ref{sec:eva-res-stea}).

In the following, we present these five categories in turn.
Before that, we briefly formulate the optimization process of adversarial attacks in image classification.
Given a classifier $f(\mathbf{x}):\mathbf{x}\in\mathcal{X}\to y\in\mathcal{Y}$ that predicts a label $y$ for an original image $\mathbf{x}$, an attacker aims to create an adversarial image $\mathbf{x}'$ by perturbing $\mathbf{x}$.
For iterative attacks, which fall into our first four categories, the adversarial image is iteratively optimized based on gradients:
\begin{gather}
\label{eq:base}
\mathbf{x}'_0=\mathbf{x},~\mathbf{x}'_{i+1}=\mathbf{x}'_i+\eta\cdot\mathrm{sign}(\mathbf{g}_i),~\textrm{where}~\mathbf{g}_i=\nabla_{\mathbf{x}}\mathcal{L}.
\end{gather}
Here, $\mathbf{x}'_i$ is the updated image at the $i$-th iteration, $\mathbf{g}_i$ is the intermediate gradient vector, and $\mathcal{L}$ is the adversarial loss function.
In order to speed up the optimization and ensure the image is in the original quantized domain (e.g., with 8-bit), a sign operation is commonly added to keep only the direction of the gradients but not their magnitude~\cite{goodfellow2014explaining}.
Specifically, the gradient calculation in the baseline attack, PGD~\cite{kurakin2016adversarial,madry2017towards}, can be formulated as follows:
\begin{gather}
\label{eq:pgd}
\mathbf{g}_i=\nabla_{\mathbf{x}}\mathcal{L}_{CE}(\mathbf{x}'_i,y,f),
\end{gather}
where the loss function $\mathcal{L}_{CE}$ is the cross-entropy loss, which takes as inputs the image, label, and model.
In each of the following four attack categories, only one of the four major components (i.e., gradient update, input, loss function, and model) in Equation~\ref{eq:pgd} would be modified.

\subsection{Gradient Stabilization}
Different DNN architectures tend to yield radically different decision boundaries, yet similar test accuracy, due to their high non-linearity~\cite{liu2017delving,somepalli2022can}.
For this reason, the attack gradients calculated on a specific (source) model may cause the adversarial images to trap into local optima, resulting in low transferability to a different (target) model.
To address this issue, popular machine learning techniques that can stabilize the update directions and help the optimization escape from poor local maxima during the iterations are useful.
Existing work on transfer attacks has adopted two such techniques, momentum~\cite{polyak1964some} and Nesterov accelerated gradient (NAG)~\cite{nesterov1983method}, by adding a corresponding term into the attack optimization.

\begin{table}[!t]
\caption{Key ideas of different attack categories.}
      \centering
        \begin{tabular}{l|l}
\toprule[1pt]
Attack Categories&Key Ideas\\
\midrule

Gradient Stabilization& Stabilize the gradient update\\
Input Augmentation& Augment the input image\\
Feature Disruption& Disrupt intermediate-layer features\\
Surrogate Refinement& Refine the surrogate model\\
Generative Modeling& Learn a perturbation generator\\
\bottomrule[1pt]
\end{tabular}
\label{tab:key}
\end{table}

In this way, the attack is optimized based on gradients from not only the current iteration but also other iteration(s) determined by the additional term.
Specifically, the momentum is to accumulate previous gradients, i.e., looking back~\cite{dong2018boosting}, and the Nesterov accelerated gradient (NAG) is an improved momentum that additionally accumulates future gradients, i.e., looking ahead~\cite{lin2020nesterov}.
In particular, NAG is later examined in~\cite{wang2021boosting}, where using only the single previous gradient for looking ahead is found to work better than using all previous gradients.
In terms of the formulation, the gradient calculation in Equation~\ref{eq:pgd} becomes:
\begin{equation}
\label{eq:grad}
\begin{split}
&\textrm{MI~\cite{dong2018boosting}:}~\mathbf{g}_i=\mu\cdot\mathbf{g}_{i-1}+\nabla_{\mathbf{x}}\mathcal{L}_{CE}(\mathbf{x}'_i,y,f),\\
&\textrm{NI~\cite{lin2020nesterov}:}~\mathbf{g}_i=\mu\cdot\mathbf{g}_{i-1}+\nabla_{\mathbf{x}}\mathcal{L}_{CE}(\mathbf{x}'_i+\alpha\cdot\mathbf{g}_{i-1},y,f),\\
&\textrm{PI~\cite{wang2021boosting}:}~\mathbf{g}_i=\mu\cdot\mathbf{g}_{i-1}+\nabla_{\mathbf{x}}\mathcal{L}_{CE}(\mathbf{x}'_i+\alpha\cdot\hat{\mathbf{g}}_{i-1},y,f),\\
&\mathrm{where~~}\hat{\mathbf{g}}_{i-1}=\nabla_{\mathbf{x}}\mathcal{L}_{CE}(\mathbf{x}'_{i-1},y,f),
\end{split}
\end{equation}
where the first, momentum term in each equation is for looking back, and the second is for looking ahead.

\subsection{Input Augmentation}
Machine learning is a process where a model is optimized on a local dataset such that it can generalize to unseen test images.
Following the same principle, learning transfer attacks can be treated as a process where an adversarial image is optimized on a local surrogate model such that it can generalize/transfer to unseen target models~\cite{liang2021uncovering}. 
Due to this conceptual similarity, various techniques that are used for improving model generalizability can be exploited to improve attack transferability.
Data augmentation is one such technique that is commonly explored in existing work on transfer attacks.
To this end, the adversarial images are optimized to be invariant to a specific image transformation. 
In terms of the formulation, the gradient calculation in Equation~\ref{eq:pgd} becomes:
\begin{equation}
\label{eq:aug}
\mathbf{g}_i=\nabla_{\mathbf{x}}\mathcal{L}_{CE}(T(\mathbf{x}'_i),y,f),
\end{equation}
where the input image is first transformed by a specific image transformation $T$.

According to the idea of data augmentation, the transformation is required to be sematic-preserving.
For example, in image classification, a transformed image should still be correctly classified. 
Existing work has explored various types of image transformations, such as geometric transformations (e.g., resizing \& padding~\cite{xie2019improving} and translation~\cite{dong2019evading}), pixel value scaling~\cite{lin2020nesterov}, random noise~\cite{wang2021enhancing}, and mixing images from incorrect classes~\cite{wang2021admix}.
Recent studies also explore more complex methods that refine the above transformations~\cite{zou2020improving,yuan2022adaptive,ren2025improvingcvpr}, leverage regional (object) information~\cite{li2020regional,byun2022improving,wei2023enhancing}, or rely on frequency-domain image transformations~\cite{long2022frequency}.

\subsection{Feature Disruption}
It is understandable that adversarial attacks in image classification commonly adopt cross-entropy loss because this loss is also used for classification by default. 
However, this choice may not be optimal for transfer attacks since the output-layer, class information is often model-specific.
In contrast, features extracted from the intermediate layers are known to be more generic~\cite{yosinski2014transferable,kornblith2019similarity}.
Based on this finding, existing work on transfer attacks has proposed to replace the commonly used cross-entropy loss with some feature-based loss.
These studies aim to disrupt the feature space of an image such that the final class prediction would change.
Specifically, the distance between the adversarial image and its corresponding original image in the feature space is maximized.
In terms of the formulation, the gradient calculation in Equation~\ref{eq:pgd} becomes:
\begin{gather}
\label{eq:fea}
\mathbf{g}_i=\nabla_{\mathbf{x}}d(f_{l}(\mathbf{x}'_i),f_{l}(\mathbf{x})),
\end{gather}
where $f_{l}$ calculates the feature output at a specific intermediate layer $l$ of the classifier, and $d(\cdot,\cdot)$ calculates the feature difference between the original and adversarial images.

Early methods of feature disruption attacks simply adopt the $L_2$ distance for $d(\cdot,\cdot)$~\cite{naseer2018task,zhou2018transferable,ganeshan2019fda,huang2019enhancing,liu2019s,li2020yet}.
In this way, the resulting feature disruption is indiscriminate at any image location. 
However, these methods are sub-optimal because the model decision normally only depends on a small set of important features~\cite{zhou2016learning}.
To address this limitation, later methods~\cite{wu2020boosting,wang2021feature,zhang2022improving} instead calculate the feature difference $d(\cdot,\cdot)$ based on only important features.
In this case, the importance of different features is normally determined based on model interpretability results~\cite{selvaraju2017grad,sundararajan2017axiomatic}.
Feature disruption is also explored for the relatively challenging, targeted attacks.
In this case, rather than maximizing the distance between the original image and the adversarial image, the attack minimizes the distance between the original image and a target image~\cite{inkawhich2019feature} or a target distribution determined by a set of target images~\cite{inkawhich2020transferable,inkawhich2020perturbing}. 
Beyond feature disruption, other losses are also explored specifically for targeted attacks, such as the triplet loss~\cite{li2020towards} or logit loss~\cite{zhao2021success}.

\subsection{Surrogate Refinement}
The current CNN models are commonly learned towards high prediction accuracy but with much less attention given to transferable representations.
Several recent studies find that adversarial training could yield a model that better transfers to downstream tasks, although it inevitably trades off the accuracy of a standard model in the source domain~\cite{salman2020adversarially,deng2021adversarial,utrera2021adversarially}.
This finding encourages researchers to explore how to refine the surrogate model to improve the transferability of attacks.
In terms of the formulation, the gradient calculation in Equation~\ref{eq:pgd} becomes:
\begin{equation}
\label{eq:ref}
\mathbf{g}_i=\nabla_{\mathbf{x}}\mathcal{L}_{CE}(T(\mathbf{x}'_i),y,f'),
\end{equation}
where $f'$ demotes a refined version of the original model $f$.

The underlying assumption of surrogate refinement is a smooth surrogate model leads to transferable adversarial images.
To this end, existing studies have explored three different ways, in which a standard model is modified in terms of training procedures~\cite{springer2021little,zhang2021early,yang2022boosting} and architectural designs~\cite{guo2020backpropagating,wu2020skip,zhang2021backpropagating,zhu2021rethinking,ren2025improving}.
Specifically, in terms of training procedures, using an adversarially-trained surrogate model is helpful since robust features are known to be smoother than non-robust features~\cite{ilyas2019adversarial}.
Similarly, early stopping~\cite{zhang2021early} or training with soft annotations~\cite{yang2022boosting} are also shown to yield a smoother model. 
In terms of architectural designs, it is helpful to replace the non-linear, ReLU activation function with a smoother one~\cite{guo2020backpropagating,zhang2021backpropagating,zhu2021rethinking}.
For ResNet-like architectures, backpropagating gradients through skip connections can lead to higher transferability compared to residual modules~\cite{wu2020skip}.
This line of research also leads to rethinking the relation between the accuracy of the surrogate model and the transferability~\cite{zhang2021early,zhu2021rethinking}.


\subsection{Generative Modeling}
In addition to the above iterative attacks, existing work also relies on Generative Adversarial Networks (GANs)~\cite{goodfellow2014generative} to directly generate transferable adversarial images.
Generative modeling attacks can efficiently generate perturbations for any given image with only one forward pass.
The GAN framework consists of two networks: a generator and a discriminator.
The generator captures the data distribution, and the discriminator estimates the probability that a sample came from the training data rather than the generator.
In the typical GAN training, these two networks are simultaneously trained in a minimax two-player game.

In the context of adversarial attacks, only the generator is trained on additional (clean) data, while a pre-trained image classifier is fixed as the discriminator.
The training objective of the generator is to cause the discriminator to misclassify the generated image.
A clipping operation is normally applied after the generation to constrain the perturbation size.
Specifically, the perturbations output from the generator are initially unbounded and then clipped by a specific norm bound.
This means that once trained, the generator $G$ can be used to generate perturbations under any target norm bound $\epsilon$:
\begin{gather}
\label{eq:pgd}
\mathbf{x}'=\textrm{Clip}(G(\mathbf{x}),\epsilon).
\end{gather}

Existing studies have explored different loss functions and generator architectures. 
Specifically, the earliest generative attack~\cite{poursaeed2018generative} adopts the widely-used cross-entropy loss.
Later studies further adopt a relativistic cross-entropy loss~\cite{naseer2019cross} or intermediate-level, feature losses~\cite{kanth2021learning,zhang2022beyond}.
Class-specific~\cite{naseer2021generating} and class-conditional~\cite{yangx2022boosting} generators are designed to particularly improve targeted transferability. 

%% file: tex/EvaluationMethodology.tex
\section{Evaluation Methodology}
\label{sec:eva-met}

\subsection{Threat Model}
\label{sec:thr}
Following the common practice, we specify our threat model from three dimensions: attacker’s knowledge, attacker’s goal, and attacker’s capability~\cite{papernot2016limitations,biggio2018wild,carlini2019evaluating}.
We make sure our threat model follows the most common settings in existing work regarding five aspects, as summarized in Table~\ref{tab:circle} of Appendix~\ref{app:com}.
This threat model focuses on \textit{untargeted} attacks against \textit{public} \textit{ImageNet} models that are trained on the \textit{same} data with the convolutional neural networks (\textit{CNNs}) or Vision Transformers (\textit{ViTs}).

\mypara{Attacker's knowledge}
An attacker can have various levels of knowledge about the target model.
In the ideal, white-box case, an attacker has full control over the target model.
In the realistic, black-box case, an attacker has either query access to the target model or in our transfer setting, no access but only leverages a surrogate model.
In this work, we follow most existing work to explore the attack transferability between image classifiers that are trained on the same public dataset (here ImageNet) but with different architectures.
Cross-dataset transferability is rarely explored~\cite{naseer2019cross,zhang2022beyond} and beyond the scope of this work.

\mypara{Attacker's goal} 
In general, an attacker aims at either untargeted or targeted misclassification.
An untargeted attack aims to fool the classifier into predicting any other class than the original one, i.e., $f(\mathbf{x}')\neq{y}$.
A targeted attack aims at a specific incorrect class $t$, i.e., $f(\mathbf{x}')=t$.
Technically, an untargeted attack is equal to running a targeted attack for each possible target and taking the closest~\cite{carlini2017towards}.
Achieving the targeted goal is strictly more challenging because a specific (target) direction is required, while many (including random) directions suffice for untargeted success.
Targeted success and non-targeted success follow fundamentally different mechanisms.
Specifically, targeted success requires representation learning of the target class while untargeted success is relevant to model sensitivity to small variations~\cite{naseer2021generating,zhao2021success}.
In this work, we follow most existing work to focus on untargeted transferability while leaving the more challenging, targeted transferability for future work.

\mypara{Attacker's capability}
In addition to the adversarial effects, an attack should be constrained to stay stealthy.
Most existing work addresses this by pursuing the imperceptibility of perturbations, based on the simple, $L_p$ norms~\cite{carlini2017towards,goodfellow2014explaining,kurakin2016adversarial,szegedy2014intriguing,papernot2016limitations} or more advanced metrics~\cite{rozsa2016adversarial,Croce_2019_ICCV,luo2018towards,zhang2020smooth,xiao2018spatially,kanbak2018geometric,alaifari2018adef,wong2019wasserstein,zhao2020towards}.
There are also recent studies on ``perceptible yet stealthy'' adversarial modifications~\cite{bhattad2020Unrestricted,shamsabadi2020colorfool,zhao2020adversarial_arxiv}.
It is worth noting that the constraint on the attack budget should also be carefully set.
In particular, unreasonably limiting the iteration number of attacks is shown to be one of the major pitfalls in evaluating adversarial robustness~\cite{carlini2019evaluating,tramer2020adaptive,zhao2021success}.
In this work, we follow most existing work to constrain the perturbations to be imperceptible by using the $L_{\infty}$ norm.
We also consider more advanced perceptual metrics for measuring imperceptibility.
Beyond imperceptibility, we further look into other aspects of attack stealthiness, such as perturbation characteristics and misclassification patterns.
For the constraint on the attack budget, we ensure the attack convergence by using sufficient iterations for iterative attacks and sufficient training epochs for generative attacks.

\subsection{Experimental Setups}
\mypara{Attacks and defenses}
For each of the above five attack categories presented in Section~\ref{sec:cate-trans-attack}, we select 5 representative attacks\footnote{We only select 3 gradient stabilization attacks since they are less studied.}, resulting in a total number of 23 attacks, as summarized in Table~\ref{tab:attack}.
Following the common practice, the perturbations are constrained by the $L_{\infty}$ norm bound with $\epsilon=16/255$.
Beyond directly feeding the adversarial images into the (standard) model, we consider 9 representative defenses from 3 different categories as well as 2 practical defenses, as summarized in Table~\ref{tab:defense}.
Note that these defenses were not specifically developed against any single attack method but are generally applicable.
Detailed descriptions of these attacks and defenses are provided in Appendix~\ref{app:att_def}.

\mypara{Dataset and models} 
We follow the common threat model shown in Table~\ref{tab:circle} to use ImageNet and four CNNs with diverse architectures, i.e., InceptionV3~\cite{szegedy2016rethinking}, ResNet50~\cite{he2016deep}, DenseNet121~\cite{huang2017densely}, and VGGNet19~\cite{simonyan2015very}, as well as the Vision Transformer (ViT), i.e., ViT-B-16-224~\cite{dosovitskiy2020image}, as the target model.
We randomly select 5000 images (5 per class\footnote{Several classes contain fewer than 5 eligible images.}) from the validation set that are correctly classified by all the above four models.
All original images are resized and then cropped to the size of 299$\times$299 for Inception-V3 and 224$\times$224 for the other four models. 
Note that we do not pre-process any adversarial image since input pre-processing is discussed as a specific category of defense in our evaluation.

\subsection{Evaluation Metrics}
\mypara{Transferability}
The transferability is measured by the (untargeted) attack success rate on the target model. 
Given an attack $\mathcal{A}$ that generates an adversarial image $\mathbf{x}'_n$ for its original image with the true label $y_n$ and target classifier $f$, the transferability over $N$ test images is defined as:
\begin{gather}\label{eq:trans}
     \mathrm{Suc}(\mathcal{A})=\frac{1}{N}\sum_{n=1}^{N}\mathbf{1}\big(f(\mathbf{x}'_n)\neq y_n\big),
\end{gather}
where $\mathbf{1}(\cdot)$ is the indicator function.
Note that here the attack success rate on the (white-box) source model is not relevant to transferability.

\mypara{Stealthiness}
The imperceptibility is measured using a variety of perceptual similarity metrics: Peak Signal-to-Noise Ratio (PSNR), Structural Similarity Index Measure (SSIM)~\cite{wang2004image}, $\Delta E$~\cite{luo2001development,zhao2020towards}, Learned Perceptual Image Patch Similarity (LPIPS)~\cite{zhang2018unreasonable}, and Frechet Inception Distance (FID)~\cite{szegedy2016rethinking}.
Detailed formulations of these metrics can be found in Appendix~\ref{app:metric}.
Beyond the above five imperceptibility measures, we also look into other, finer-grained stealthiness characteristics regarding the perturbation and misclassification patterns.
Detailed descriptions of these characteristics can be found in Section~\ref{sec:eva-res-stea}.

%% file: tex/CategorizeAnalysis.tex
\section{Intra-Category Analyses}
\label{sec:ana}

In this section, we conduct systematic, one-to-one analyses of similar attacks in each of the five categories.
In this way, similar attacks can be compared along a single dimension with fair settings of key hyperparameters.
In particular, each attack is implemented with only its core technique, although it may integrate others by default when it was originally proposed.


\begin{figure}[!t]
\centering
\includegraphics[width=\columnwidth]{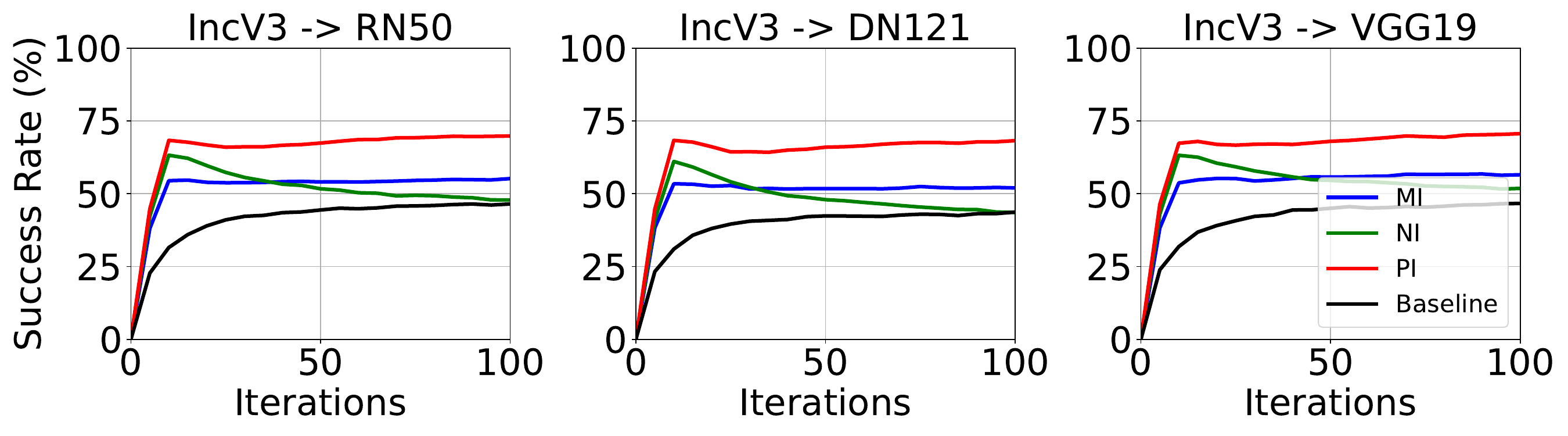}
\includegraphics[width=\columnwidth]{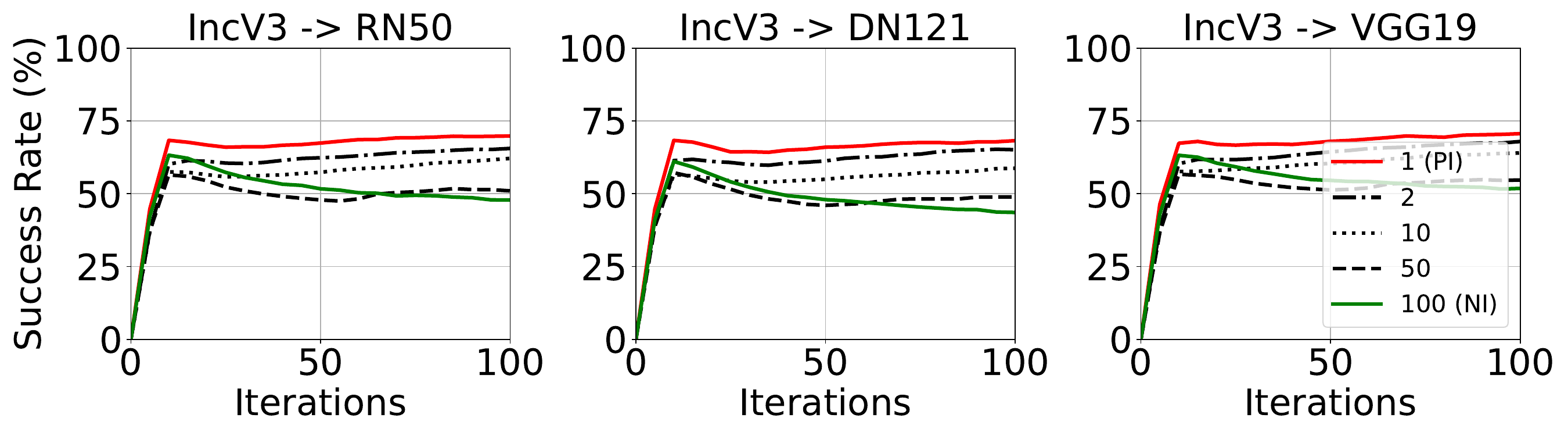}
\caption{\textbf{Results for gradient stabilization attacks.} \textbf{Top:} Transferability vs. iteration curve. Additional results on another three surrogates with consistent patterns are in Fig.~\ref{fig:gradient_app} of Appendix~\ref{app:add_res}.
\textbf{Bottom:} Impact of the number of looking-ahead iterations.} 
\label{fig:gradient}
\end{figure}

\subsection{Analysis of Gradient Stabilization Attacks}
\label{sec:attack-gs}
Fig.~\ref{fig:gradient} Top shows the transferability of the three gradient stabilization attacks under various iterations.
As can be seen, all three attacks converge very fast, within 10 iterations, since they accumulate gradients with the momentum.
However, using more iterations does not necessarily improve, but may even cause the ``overfitting'' phenomenon.
\textcolor{blue}{However, this ``overfitting'' phenomenon was previously overlooked since existing studies directly adopt only a few iterations (for all different categories of attacks). Such an ``overfitting'' phenomenon is also found to be contrastive to ``no convergence'' of the other categories of attacks (see detailed discussion in the following subsections).} 

Since the above ``overfitting'' phenomenon is newly found, we further explore whether it also applies to the more challenging, targeted attacks. 
We find that although the results of MI and NI are always lower than 3\%, they still suffer from the same ``overfitting'' problem.
In contrast, PI performs better, and its targeted transferability finally reaches 9.4\% on average.
This difference between PI for non-targeted and targeted transferability might be because targeted transferability especially requires an accurate direction~\cite{zhao2021success}, and PI achieves this through additional looking ahead~\cite{wang2021boosting}.

When comparing different attacks, we can observe that PI and NI perform better than MI due to looking ahead, but NI is better only at the beginning.
The only difference between PI and NI is the maximum number of previous iterations used to look ahead.
To figure out the impact of the number of look-ahead iterations, we repeat the experiments with values from 1 (i.e., PI) to 100 (i.e., NI).
Fig.~\ref{fig:gradient} Bottom shows that the performance drops more as more previous iterations are incorporated to look ahead, and the optimal performance is actually achieved by PI.

\begin{figure}[!t]
\centering
\includegraphics[width=\columnwidth]{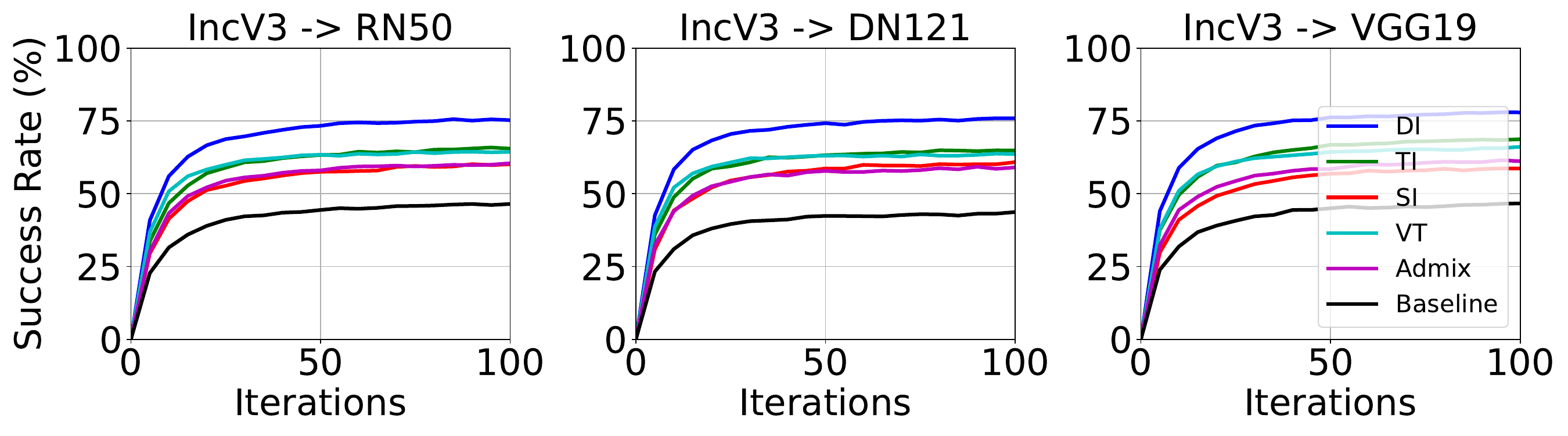}
\includegraphics[width=\columnwidth]{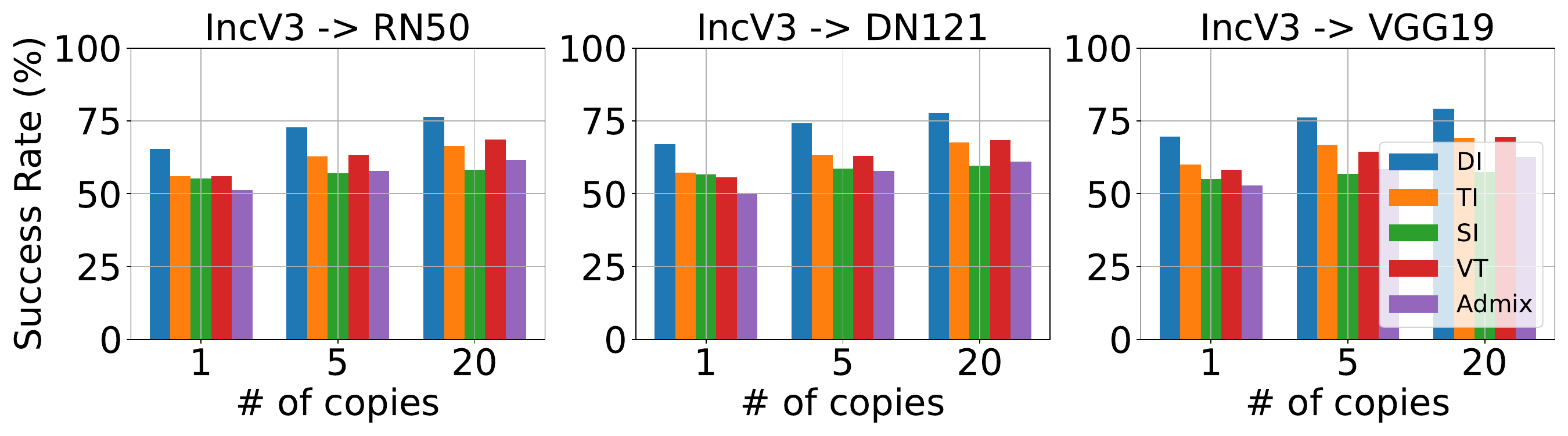}
\caption{\textbf{Results for input augmentation attacks.} \textbf{Top:} Transferability vs. iteration curve. All attacks use five input copies. Additional results on another three surrogates with consistent patterns are in Fig.~\ref{fig:copy_5_app} of Appendix~\ref{app:add_res}. \textbf{Bottom:} Impact of the number of input copies.}
\label{fig:copy_5}
\end{figure}

\subsection{Analysis of Input Augmentation Attacks}
\label{sec:attack-ia}
Fig.~\ref{fig:copy_5} Top shows the transferability of the five input augmentation attacks under various iterations.
Previous evaluations of input augmentation attacks often conduct unfair comparisons where different attacks may leverage a different number of random input copies~\cite{wang2021admix}.
In contrast, our comparisons are under the same number of random input copies. 
We find that, surprisingly, the earliest attack methods, DI and TI, always perform the best.
However, the latest method, Admix, achieves very low transferability. 

We explain the above new finding by comparing the input diversity caused by different attacks since higher input diversity is commonly believed to yield higher transferability~\cite{xie2019improving,long2022frequency}.
Specifically, we quantify the input diversity of the five attacks by the change of the logit value of the top-1 model prediction caused by their specific augmentations.
We calculate the input diversity over 5000 images with 10 repeated runs. 
We get 0.23 for DI, 0.0017 for TI, 0.00039 for SI, 0.000091 for VT, and -0.035 for Admix.
The highest input diversity of DI explains its highest transferability and the lowest input diversity of Admix explains its lowest transferability.
In particular, Admix even decreases the original logit value, denoting that it is not a suitable choice for the required label-preserving augmentation.

We further explore the impact of the number of random input copies.
Fig.~\ref{fig:copy_5} Bottom shows that the transferability of all attacks is improved with more copies.
\textcolor{blue}{Therefore, we would get misleading conclusions if we follow existing studies (e.g., Admix, SI, and VT) to (unfairly) set different numbers of copies for different attacks.
For example, for IncV3 $\rightarrow$ RN50, VT with 20 copies is shown to be better than DI with only one copy (68.62\% vs. 65.38\%), but when they are compared fairly with 20 copies, VT is far behind (68.62\% vs. 76.46\%).}

Specifically, DI and TI consistently perform the best in all settings.
This finding suggests the global superiority of spatial transformations (i.e., resizing\&padding in DI and translation in TI) over other types of image transformations, such as pixel scaling in SI, additive noise in VT, and image composition in Admix.
Note that more copies generally consume more computational resources.

\subsection{Analysis of Feature Disruption Attacks}
\begin{figure}[!t]
\centering
\includegraphics[width=\columnwidth]{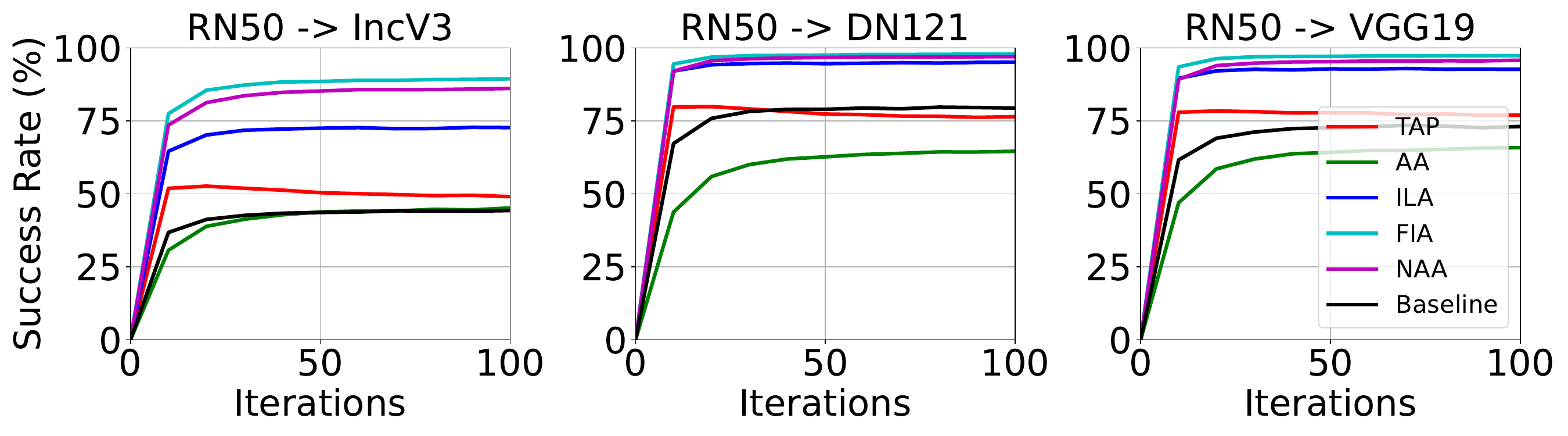}
\includegraphics[width=\columnwidth]{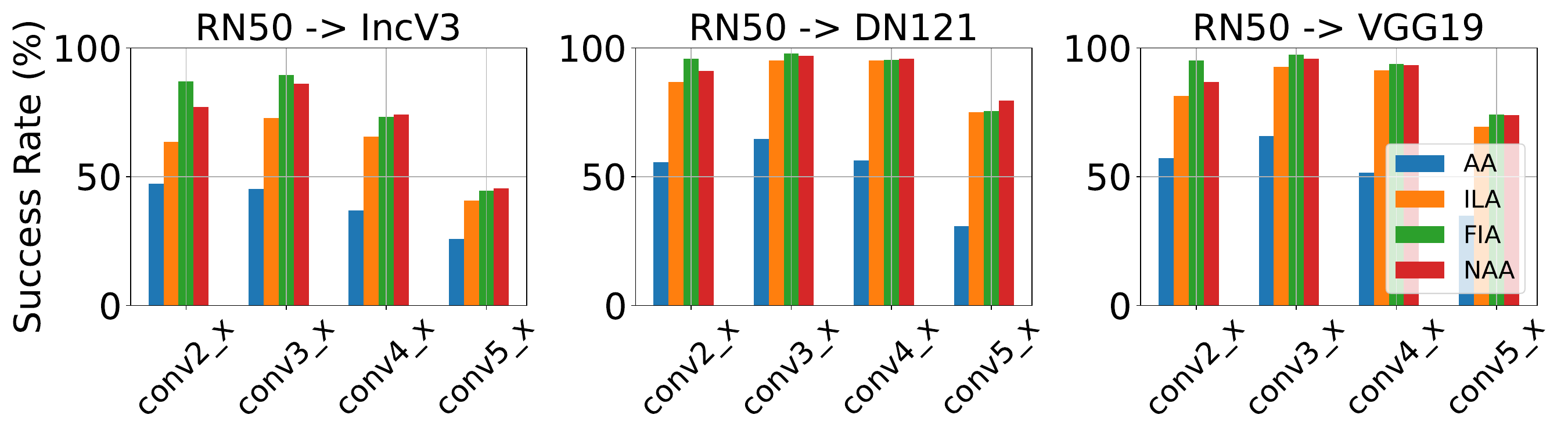}
\caption{\textbf{Results for feature disruption attacks.} \textbf{Top:} Transferability vs. iteration curve. TAP uses all layers and the others use ``conv3\_x''. \textbf{Bottom:} Impact of the layers.}
\label{fig:intermediate_resnet50}
\end{figure}

Fig.~\ref{fig:intermediate_resnet50} Top shows the transferability of the five feature disruption attacks under various iterations.
We see that FIA and NAA, which exploit feature importance, perform the best.
In addition, ILA and TAP perform well, by incorporating the CE loss into their feature-level optimizations.
Finally, AA performs even worse than the PGD baseline since it only uses a plain feature loss but ignores the CE loss.

Since all these attacks, except TAP, only disrupt features in a specific layer, we explore the impact of layer choice.
As can be seen from Fig.~\ref{fig:intermediate_resnet50} Bottom, the last layer (``conv5\_x'') performs much worse than the early layers.
This might come from the last layer being too complex and model-specific, which results in poor generalizability to unseen models.
Moreover, the mid layer (``conv3\_x'') always achieves the best performance since it can learn more semantic features than earlier layers, which are known to capture low-level visual attributes, e.g., colors and textures~\cite{zeiler2014visualizing}.
\textcolor{blue}{This is consistent with existing conclusions on layer selection.}

\subsection{Analysis of Surrogate Refinement Attacks}
\begin{figure}[!t]
\centering
\includegraphics[width=\columnwidth]{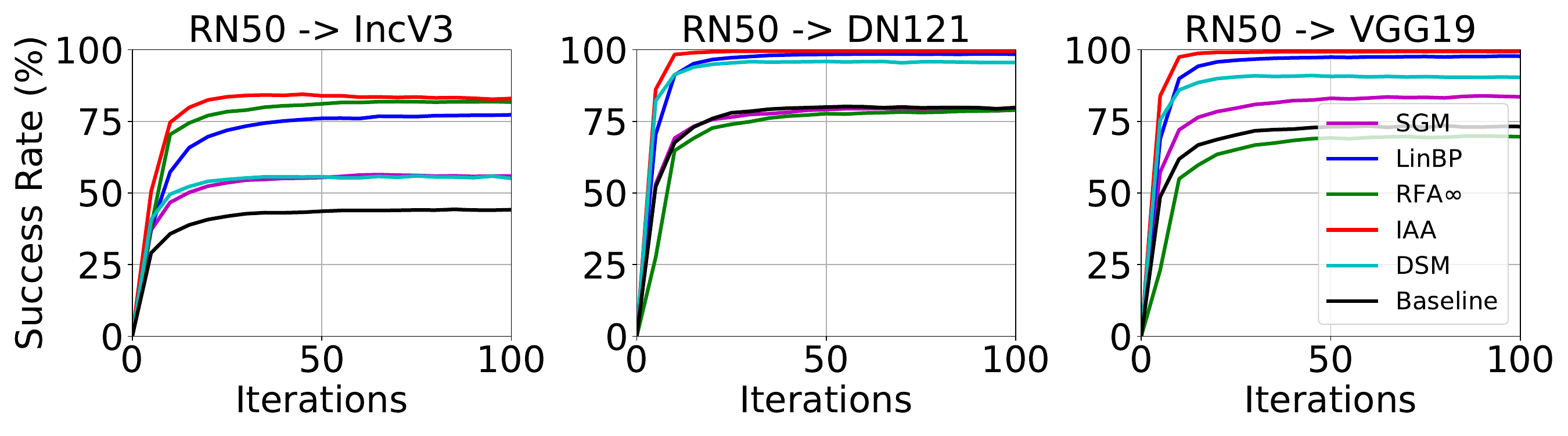}
\caption{\textbf{Results for surrogate refinement attacks (1/2):} Transferability vs. iteration curve.
}
\label{fig:surrogateR_resnet50}
\end{figure}

Fig.~\ref{fig:surrogateR_resnet50} shows the transferability of the five surrogate refinement attacks under various iterations.
Here ResNet50 is used as the surrogate model for a fair comparison since SGM~\cite{wu2020skip} can only be applied to architectures with skip connections.
We see that IAA achieves the best results mainly because it does not simply integrate specific components (i.e., skip connections and continuous activation functions) but optimizes their hyperparameters towards high transferability.
In addition, RFA$_{\infty}$ achieves lower performance (sometimes even lower than the baseline attack) since it solely uses a robust surrogate model, which is known to have different representation characteristics from the (target) standard model~\cite{ilyas2019adversarial,santurkar2019computer,salman2020adversarially}.

Model refinement opens a way to understand the impact of specific properties of the surrogate model on transferability.
Thus we look into model properties in terms of two common metrics: accuracy and interpretability, as well as another specific measure, i.e., the model similarity between the surrogate and the target models.
\textcolor{blue}{Existing work has not analyzed such impact but solely focuses on improving transferability.}
Specifically, interpretability is measured by Average Increase (AI) and Average Drop (AD)~\cite{chattopadhay2018grad} based on GradCAM~\cite{selvaraju2017grad}.
Detailed definitions of these interpretability measures are provided in Appendix~\ref{app:inter}.
For model similarity, we calculate the Kullback-Leibler (KL) divergence of output features following~\cite{inkawhich2020transferable}.
For LinBP and SGM, we refine the surrogate model not only during backpropagation as in their original implementations but also during the forward pass.
This makes sure that the refinement can also potentially influence the model properties besides transferability.

\begin{table}[!t]
\caption{\textbf{Results for surrogate refinement attacks (2/2):} Impact of the surrogate property.
DN121 is the target model.}
\newcommand{\tabincell}[2]{\begin{tabular}{@{}#1@{}}#2\end{tabular}}

\renewcommand{\arraystretch}{1}
      \centering
        \begin{tabular}{l|c|cccc}
\toprule[1pt]
    Attacks   &  Transfer$\uparrow$ &Acc$\uparrow$  &AI$\uparrow$ &AD$\downarrow$      &KL$\downarrow$\\
\midrule

Original&79.6 &100.0 &37.4 &8.4 &14.8 \\
\midrule
SGM      &\GTRANS{90.3}&\GACC{100.0}&\GAI{37.4}&\GAD{8.4 } &\GKL{14.8} \\
LinBP    &\GTRANS{98.1}&\GACC{100.0}&\GAI{37.4}&\GAD{8.4 } &\GKL{14.8} \\
RFA      &\GTRANS{99.1}&\GACC{96.4 }&\GAI{33.8}&\GAD{14.2} &\GKL{16.0} \\
IAA      &\GTRANS{99.5}&\GACC{34.9 }&\GAI{4.4 }&\GAD{29.5} &\GKL{22.4} \\
DSM      &\GTRANS{95.7}&\GACC{97.6 }&\GAI{41.0}&\GAD{7.6 } &\GKL{15.9} \\
\bottomrule[1pt]
\end{tabular}
\label{tab:surrogateR_resnet50}
\end{table}

As can be seen from Table~\ref{tab:surrogateR_resnet50}, there is a clear negative correlation between transferability and other model properties.
In particular, IAA ranks the highest in terms of transferability but the lowest in terms of all the other four measures of model properties.
This is aligned with the previous finding in transfer learning that a pre-trained model that performs well in the target domain does not necessarily perform well in the source domain~\cite{salman2020adversarially}.
In addition, the refinement in LinBP and SGM has little impact on all model properties.\footnote{There are very small differences after the second decimal place.}
This indicates that modifying only the activation function or the hyperparameters of the residual block does not influence the model properties.
Specifically, the GradCAM calculations in AI and AD only involve the gradients with respect to the last layer and the ReLU function is applied by default.
In addition, for KL, since both the surrogate model and the target model achieve an accuracy of 100\%, the above modifications yield small changes in the logit values for the (true) prediction.

\subsection{Analysis of Generative Modeling Attacks}
\begin{figure}[!t]
\centering
\includegraphics[width=\columnwidth]{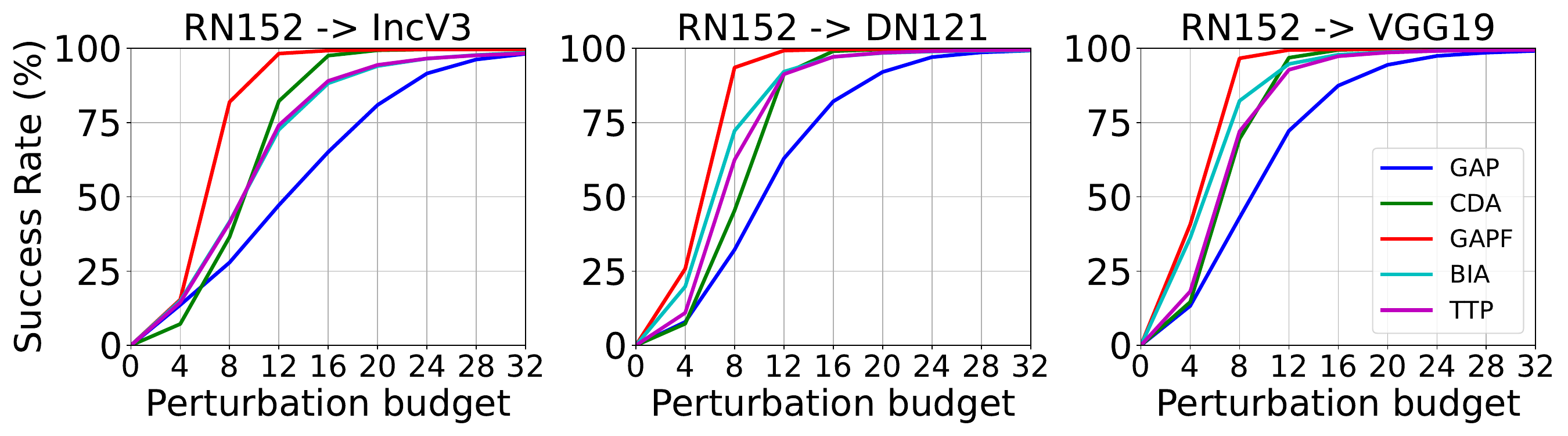}
\includegraphics[width=\columnwidth]{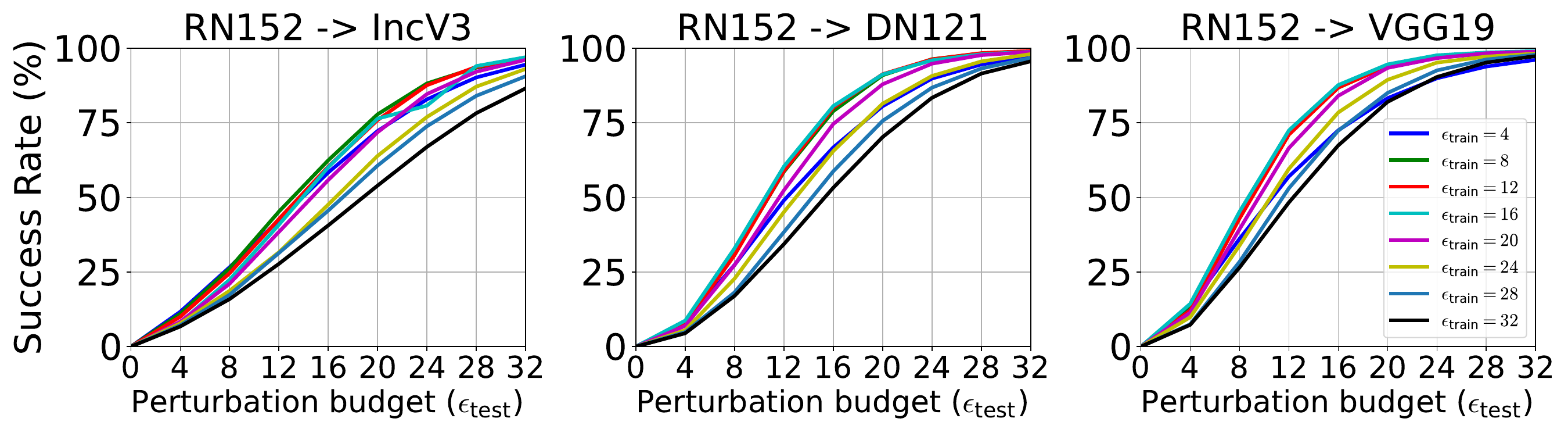}
\caption{\textbf{Results for generative modeling attacks.} \textbf{Top:} Transferability vs. perturbation budget curve. \textbf{Bottom:} Impact of the training perturbation bound, for GAP attack.}
\label{fig:gen_test}
\end{figure}

As discussed above, generative modeling attacks can be used to generate perturbations under various constraints by simply specifying the corresponding norm bound in the clipping operation.
Fig.~\ref{fig:gen_test} Top shows the transferability of the five generative modeling attacks under various perturbation bounds.
For the targeted attack TTP, we calculate its untargeted transferability over 5000 targeted adversarial images that are generated following the 10-Targets setting~\cite{naseer2021generating}.

As can be seen, the transferability generally increases as the perturbation constraint is relaxed.
Consistent with our discussion about feature disruption attacks, feature-level losses (i.e., GAPF and BIA) outperform output-level losses (i.e., GAP and CDA).
More specifically, GAPF outperforms BIA probably because GAPF leverages the original, multi-channel feature maps but BIA compresses the feature maps by cross-channel pooling.
CDA outperforms GAP since the Relative CE loss in CDA is known to be better than the CE loss in GAP~\cite{naseer2019cross}.
However, although TTP also adopts a feature-level loss, it falls far behind GAPF and BIA, especially for a small perturbation size. 
This might be because TTP, as a targeted attack, relies on embedding semantic features (of the target) in the perturbations, which are shown to be difficult when the perturbations are constrained to be small~\cite{naseer2021generating,zhao2021success}.

\input{tex/colorTableFull}

In addition to the generation/testing stage, the clipping operation is also applied when training the generator.
Existing work commonly adopts the perturbation bound $\epsilon_{\textrm{train}}=10$ without exploring the impact of this training perturbation bound on the transferability.
To fill this gap, we train the GAP generator with various $\epsilon_{\textrm{train}}$ and test these generators across different perturbation bounds $\epsilon_{\textrm{test}}$.
Fig.~\ref{fig:gen_test} Bottom shows that in general, adjusting $\epsilon_{\textrm{train}}$ has a substantial impact on the transferability.
For example, when the $\epsilon_{\textrm{test}}$ is set to 16, varying the $\epsilon_{\textrm{train}}$ from 4 to 32 yields a difference in transferability by about 25\%.
In addition, using a moderate $\epsilon_{\textrm{train}}$ (i.e. 8-16) consistently leads to the best results for all possible $\epsilon_{\textrm{test}}$.
This suggests that it makes sense to set $\epsilon_{\textrm{train}}=10$ as in existing work.
However, this finding is somewhat unexpected since choosing a matched parameter (here, $\epsilon_{\textrm{train}}=\epsilon_{\textrm{test}}$) performs the best in normal machine learning.

\textcolor{blue}{Additional experimental results in Appendix~\ref{app:add_res} confirm that our main intra-category insights generalize to more advanced models and practical scenarios.
For example, as shown in Fig.~\ref{fig:GS_revision}, for gradient stabilization attacks, ``using more iterations may even decrease the performance'' still holds ViT and PubDef.
As shown in Fig.~\ref{fig:inte}, the relative performance of attacks within a category remains largely unchanged after integrating another category of attack, which is practically implemented for further attack improvement.}

%% file: tex/colorTableFull.tex
\begin{table*}[!t]
\caption{Attack transferability in terms of success rate (\%). ResNet-50 is used as the surrogate. Additional results under $L_{\infty}=8/255$ in Table~\ref{tab:trans_linf8} of Appendix~\ref{app:add_res} show consistent patterns. \textcolor{blue}{Note that GCV is not directly comparable to the others because it adopts a different label set from the standard ImageNet-1K. A darker cell means a better attack/worse defense.}}
\newcommand{\tabincell}[2]{\begin{tabular}{@{}#1@{}}#2\end{tabular}}

      \centering
      \resizebox{\textwidth}{!}{
        \begin{tabular}{l|cccc|ccccccccccc}
\toprule[1pt]
\multirow{2}{*}{Attacks}&\multicolumn{4}{c|}{Without Defenses}&\multicolumn{3}{c}{Input Pre-processing}&\multicolumn{3}{c}{Purification Network}&\multicolumn{3}{c}{Adversarial Training}&\multicolumn{2}{c}{Real-World Model}\\
&DN121&VGG19&IncV3&ViT&BDR&PD&R\&P&HGD&NRP&DiffPure&AT$_{\infty}$&FD$_{\infty}$&AT$_{2}$&PubDef&GCV\\
\midrule
Clean Acc&100.0&100.0&100.0&94.7&89.1&97.3&94.1&98.0&90.2&91.7&77.8&81.4&77.0&\textcolor{blue}{95.4}&\textcolor{blue}{100}\\
PGD&\gradient{79.6}&\gradient{72.7}&\gradient{43.6}&\gradient{21.2}&\gradient{100.0}&\gradient{100.0}&\gradient{98.2}&\gradient{84.5}&\gradient{78.9}&\gradient{13.1}&\gradient{22.1}&\gradient{18.6}&\gradient{61.2}&\textcolor{blue}{\gradient{48.7}}&\textcolor{blue}{\gradient{55.0}}\\
\hline
MI~\cite{dong2018boosting}&\gradient{85.7}&\gradient{78.1}&\gradient{55.8}&\gradient{31.4}&\gradient{100.0}&\gradient{100.0}&\gradient{98.6}&\gradient{87.5}&\gradient{49.4}&\gradient{21.4}&\gradient{22.6}&\gradient{18.9}&\gradient{64.9}&\textcolor{blue}{\gradient{60.8}}&\textcolor{blue}{\gradient{61.0}}\\
NI~\cite{lin2020nesterov} &\gradient{87.2}&\gradient{82.7}&\gradient{60.4}&\gradient{31.1}&\gradient{100.0}&\gradient{100.0}&\gradient{99.2}&\gradient{88.2}&\underline{\gradient{83.3}}&\gradient{20.0}&\gradient{22.4}&\gradient{19.1}&\gradient{64.3}&\textcolor{blue}{\gradient{64.9}}&\textcolor{blue}{\gradient{63.0}}\\
PI~\cite{wang2021boosting}&\gradient{92.1}&\gradient{87.7}&\gradient{66.0}&\gradient{33.3}&\gradient{100.0}&\gradient{100.0}&\gradient{99.7}&\gradient{81.3}&\textbf{\gradient{93.3}}&\gradient{20.4}&\gradient{22.8}&\gradient{19.2}&\gradient{64.4}&\textcolor{blue}{\gradient{73.1}}&\textcolor{blue}{\gradient{59.0}}\\

\hline
DI~\cite{xie2019improving} &\gradient{99.0}&\gradient{99.1}&\gradient{69.8}&\gradient{36.9}&\gradient{100.0}&\gradient{100.0}&\gradient{100.0}&\gradient{98.9}&\gradient{81.4}&\gradient{16.1}&\gradient{22.9}&\gradient{19.5}&\gradient{62.3}&\textcolor{blue}{\gradient{85.5}}&\textcolor{blue}{\gradient{65.0}}\\
TI~\cite{dong2019evading}  &\gradient{96.9}&\gradient{96.1}&\gradient{63.0}&\gradient{34.5}&\gradient{100.0}&\gradient{100.0}&\gradient{100.0}&\gradient{97.9}&\gradient{78.4}&\gradient{16.1}&\gradient{22.9}&\gradient{19.1}&\gradient{62.3}&\textcolor{blue}{\gradient{78.4}}&\textcolor{blue}{\gradient{62.0}}\\
SI~\cite{lin2020nesterov}  &\gradient{93.8}&\gradient{85.6}&\gradient{61.8}&\gradient{29.5}&\gradient{100.0}&\gradient{100.0}&\gradient{99.2}&\gradient{95.2}&\gradient{72.6}&\gradient{14.6}&\gradient{22.6}&\gradient{19.2}&\gradient{62.6}&\textcolor{blue}{\gradient{70.9}}&\textcolor{blue}{\gradient{66.0}}\\
VT~\cite{wang2021enhancing}&\gradient{95.4}&\gradient{92.5}&\gradient{67.1}&\gradient{35.9}&\gradient{100.0}&\gradient{100.0}&\gradient{99.8}&\gradient{97.8}&\gradient{80.5}&\gradient{19.1}&\gradient{22.8}&\gradient{19.1}&\gradient{63.3}&\textcolor{blue}{\gradient{79.1}}&\textcolor{blue}{\gradient{61.0}}\\
Admix~\cite{wang2021admix} &\gradient{86.7}&\gradient{83.5}&\gradient{53.4}&\gradient{26.5}&\gradient{100.0}&\gradient{95.4}&\gradient{95.0}&\gradient{89.9}&\gradient{80.0}&\gradient{14.1}&\gradient{22.7}&\gradient{19.0}&\gradient{62.1}&\textcolor{blue}{\gradient{63.7}}&\textcolor{blue}{\gradient{67.0}}\\
\hline
TAP~\cite{zhou2018transferable} &\gradient{77.0}&\gradient{77.6}&\gradient{50.3}&\gradient{28.0}&\gradient{100.0}&\gradient{100.0}&\gradient{95.6}&\gradient{80.5}&\gradient{40.9}&\gradient{15.8}&\gradient{22.6}&\gradient{19.4}&\gradient{64.4}&\textcolor{blue}{\gradient{56.2}}&\textcolor{blue}{\gradient{67.0}}\\
AA~\cite{inkawhich2019feature}  &\gradient{61.8}&\gradient{64.1}&\gradient{43.5}&\gradient{26.1}&\gradient{91.0}&\gradient{88.4}&\gradient{81.4}&\gradient{67.8}&\gradient{14.7}&\gradient{20.6}&\gradient{24.9}&\gradient{21.4}&\gradient{65.1}&\textcolor{blue}{\gradient{47.0}}&\textcolor{blue}{\gradient{67.0}}\\
ILA~\cite{huang2019enhancing}   &\gradient{94.5}&\gradient{92.6}&\gradient{72.3}&\gradient{38.3}&\gradient{100.0}&\gradient{100.0}&\gradient{99.1}&\gradient{94.8}&\underline{\gradient{83.5}}&\gradient{18.2}&\gradient{22.6}&\gradient{19.1}&\gradient{63.9}&\textcolor{blue}{\gradient{86.0}}&\textcolor{blue}{\gradient{62.0}}\\
FIA~\cite{wang2021feature}      &\gradient{97.5}&\gradient{97.1}&\gradient{88.4}&\gradient{61.8}&\gradient{100.0}&\gradient{99.9}&\gradient{99.6}&\gradient{98.1}&\gradient{71.6}&\gradient{28.5}&\gradient{25.1}&\gradient{21.3}&\gradient{68.2}&\textcolor{blue}{\gradient{92.6}}&\textcolor{blue}{\gradient{74.0}}\\
NAA~\cite{zhang2022improving}   &\gradient{96.7}&\gradient{95.3}&\gradient{85.0}&\gradient{55.0}&\gradient{99.9}&\gradient{99.9}&\gradient{99.1}&\gradient{97.2}&\gradient{76.8}&\gradient{29.2}&\gradient{24.8}&\gradient{21.3}&\gradient{66.2}&\textcolor{blue}{\gradient{89.2}}&\textcolor{blue}{\gradient{73.0}}\\
\hline
SGM~\cite{wu2020skip}               &\gradient{79.1}&\gradient{83.0}&\gradient{55.4 }&\gradient{31.8  }&\gradient{90.4}&\gradient{100.0}&\gradient{97.6}&\gradient{94.5}&\gradient{83.1}&\gradient{14.9}&\gradient{22.7}&\gradient{18.9}&\gradient{62.2}&\textcolor{blue}{\gradient{65.8}}&\textcolor{blue}{\gradient{60.0}}\\
LinBP~\cite{guo2020backpropagating} &\gradient{98.1}&\gradient{97.4}&\gradient{82.8}&\gradient{40.0}&\gradient{96.6}&\gradient{100.0}&\gradient{99.2}&\gradient{98.3}&\gradient{66.8}&\gradient{17.7}&\gradient{23.1}&\gradient{19.4}&\gradient{63.8}&\textcolor{blue}{\gradient{89.4}}&\textcolor{blue}{\gradient{68.0}}\\
RFA$_2$~\cite{springer2021little}       &\gradient{99.1}&\gradient{96.7}&\gradient{89.2}&\gradient{63.1}&\gradient{98.9}&\gradient{99.5}&\gradient{97.9}&\gradient{99.4}&\gradient{34.8}&\underline{\gradient{43.8}}&\gradient{24.8}&\gradient{21.2}&\gradient{67.6}&\textcolor{blue}{\gradient{93.9}}&\textcolor{blue}{\gradient{66.0}}\\
RFA$_{\infty}$~\cite{springer2021little}     &\gradient{77.7}&\gradient{68.7}&\gradient{81.1}&\underline{\gradient{73.3}}&\gradient{81.7}&\gradient{80.4}&\gradient{77.5}&\gradient{72.8}&\gradient{21.0}&\textbf{\gradient{69.4}}&\textbf{\gradient{62.0}}&\textbf{\gradient{53.4}}&\textbf{\gradient{87.3}}&\textcolor{blue}{\gradient{62.7}}&\textcolor{blue}{\gradient{77.0}}\\
IAA~\cite{zhu2021rethinking}        &\underline{\gradient{99.5}}&\underline{\gradient{99.2}}&\gradient{83.9}&\gradient{50.2}&\gradient{98.9}&\gradient{100.0}&\gradient{99.9}&\textbf{\gradient{99.9}}&\gradient{65.4}&\gradient{20.6}&\gradient{23.0}&\gradient{19.5}&\gradient{65.5}&\textcolor{blue}{\underline{\gradient{95.7}}}&\textcolor{blue}{\gradient{79.0}}\\
DSM~\cite{yang2022boosting}         &\gradient{95.7}&\gradient{90.8}&\gradient{57.3}&\gradient{27.3}&\gradient{97.4}&\gradient{99.1}&\gradient{92.6}&\gradient{98.1}&\gradient{38.2}&\gradient{14.9}&\gradient{22.7}&\gradient{19.1}&\gradient{62.3}&\textcolor{blue}{\gradient{74.6}}&\textcolor{blue}{\gradient{64.0}}\\
\hline
GAP~\cite{poursaeed2018generative} &\gradient{82.1}&\gradient{87.4}&\gradient{65.1}&\gradient{34.8}&\gradient{85.1}&\gradient{87.0}&\gradient{83.5}&\gradient{93.9}&\gradient{2.0}&\gradient{16.8}&\gradient{21.7}&\gradient{18.2}&\gradient{64.2}&\textcolor{blue}{\gradient{66.3}}&\textcolor{blue}{\gradient{80.0}}\\
CDA~\cite{naseer2019cross}         &\underline{\gradient{99.2}}&\underline{\gradient{99.2}}&\underline{\gradient{97.8}}&\textbf{\gradient{82.1}}&\gradient{97.6}&\gradient{100.0}&\gradient{98.3}&\underline{\gradient{99.9}}&\gradient{3.9}&\underline{\gradient{62.0}}&\underline{\gradient{26.0}}&\underline{\gradient{22.6}}&\underline{\gradient{73.2}}&\textcolor{blue}{\underline{\gradient{98.1}}}&\textcolor{blue}{\underline{\gradient{86.0}}}\\
GAPF~\cite{kanth2021learning}      &\textbf{\gradient{99.6}}&\textbf{\gradient{99.6}}&\textbf{\gradient{99.2}}&\underline{\gradient{72.5}}&\gradient{99.5}&\gradient{99.6}&\gradient{99.8}&\underline{\gradient{99.8}}&\gradient{14.3}&\gradient{28.9}&\gradient{23.4}&\gradient{20.2}&\gradient{67.7}&\textcolor{blue}{\textbf{\gradient{99.3}}}&\textcolor{blue}{\textbf{\gradient{94.0}}}\\
BIA~\cite{zhang2022beyond}         &\gradient{97.1}&\gradient{97.8}&\gradient{88.2}&\gradient{50.3}&\gradient{96.9}&\gradient{95.4}&\gradient{96.6}&\gradient{98.7}&\gradient{2.6}&\gradient{20.0}&\underline{\gradient{25.3}}&\underline{\gradient{22.2}}&\gradient{66.7}&\textcolor{blue}{\gradient{94.9}}&\textcolor{blue}{\underline{\gradient{88.0}}}\\
TTP~\cite{naseer2021generating}    &\gradient{97.1}&\gradient{97.3}&\gradient{89.0}&\gradient{64.3}&\gradient{97.6}&\gradient{98.2}&\gradient{97.9}&\gradient{95.4}&\gradient{36.2}&\gradient{37.4}&\underline{\gradient{25.3}}&\gradient{21.3}&\underline{\gradient{68.7}}&\textcolor{blue}{\gradient{90.0}}&\textcolor{blue}{\gradient{67.0}}\\
\bottomrule[1pt]
\end{tabular}
}
\label{tab:transfer}
\end{table*}

%% file: tex/TransferResults.tex
\section{Comprehensive Evaluation}
\label{sec:eva-res-trans}
Through the above systematic intra-category analyses, we can determine the optimal choices of the key attack hyperparameters to achieve a fair large-scale evaluation of attack transferability and stealthiness.
\textcolor{blue}{\textit{Specifically, the key attack hyperparameters refer to the number of iterations for all except generative model attacks, the number of input copies for input augmentation attacks, the choice of layers for feature disruption attacks, and the training perturbation bound $\epsilon_{train}$ for generative model attacks.}}

In general, different attacks under the same category have the same optimal hyperparameters.
Specifically, for gradient stabilization attacks, we set the iteration number to 10 to avoid the ``overfitting'' phenomenon, and for the other three categories of iterative attacks, we set the iteration number to 50 to ensure attack convergence.
For input augmentation attacks, we set the number of copies to 5 for a good trade-off between attack strength and efficiency. 
For feature disruption attacks, we use the mid layer (``conv3\_x''), which always yields the best performance.
For generative modeling attacks, we set $\epsilon_{\textrm{train}}=10$, which is commonly adopted in existing work and validated to be optimal by our intra-category analyses.
For other attack hyperparameters not discussed in our intra-category analyses and defense hyperparameters, we follow the common settings as summarized in Table~\ref{tab:hyper} of Appendix~\ref{app:hyper}.

We make sure our findings are stable by repeating our experiments 5 times on 1000 images, each of which is selected from each of the 1000 ImageNet classes.
The results confirm that 16 attacks always yield identical numbers because they are deterministic by design. 
For the other 7 attacks, i.e., the 5 input augmentation attacks, AA, and FIA, the standard deviations are always very small, i.e., standard deviations within [0.001, 0.013] for attack success rates within [0.464, 0.993].
Note that in contrast, there is no statistical analysis in any original work of the 23 attacks.

\subsection{New Insights into Transferability}
Table~\ref{tab:transfer} summarizes the transferability results in both the standard and defense settings.
These results either lead to new findings or complement existing findings that were only validated in a limited set of attacks and defenses.

\noindent\textbf{General conclusion:} \textit{The transferability of an attack against a defense is highly contextual.}
An attack/defense that is the best against one specific defense/attack may perform very badly against another.
For example, the transferability of CDA to 9 different defenses varies largely, from 3.9\% to 100.0\%.
Similarly, a defense that achieves the highest robustness to one specific attack may be very vulnerable to another attack.
For example, the robustness of NRP to 24 different attacks varies largely, from 2.0\% to 93.3\%.

\begin{figure}[!t]
\centering
\begin{subfigure}[b]{0.2\columnwidth}
\includegraphics[width=\columnwidth]{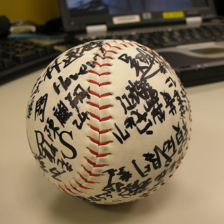}
\end{subfigure}
\hspace{-0.15cm}
\begin{subfigure}[b]{0.2\columnwidth}
\includegraphics[width=\columnwidth]{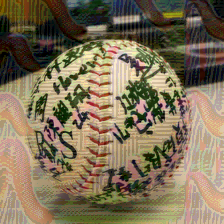}
\end{subfigure}
\hspace{-0.15cm}
\begin{subfigure}[b]{0.2\columnwidth}
\includegraphics[width=\columnwidth]{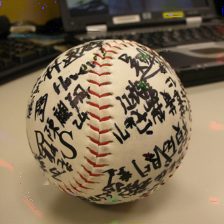}
\end{subfigure}
\hspace{-0.15cm}
\begin{subfigure}[b]{0.2\columnwidth}
\includegraphics[width=\columnwidth]{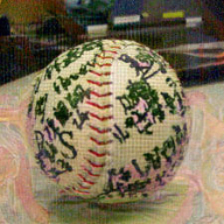}
\end{subfigure}
\vspace{0.08cm}
\hspace{-0.12cm}

\begin{subfigure}[b]{0.2\columnwidth}
\includegraphics[width=\columnwidth]{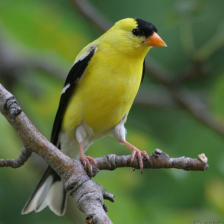}
\caption*{Original}
\end{subfigure}
\hspace{-0.15cm}
\begin{subfigure}[b]{0.2\columnwidth}
\includegraphics[width=\columnwidth]{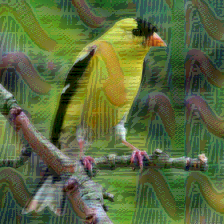}
\caption*{CDA}
\end{subfigure}
\hspace{-0.15cm}
\begin{subfigure}[b]{0.2\columnwidth}
\includegraphics[width=\columnwidth]{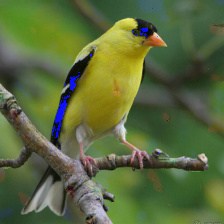}
\caption*{with NRP}
\end{subfigure}
\hspace{-0.15cm}
\begin{subfigure}[b]{0.2\columnwidth}
\includegraphics[width=\columnwidth]{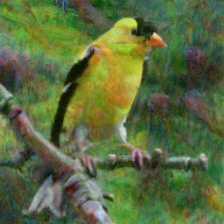}
\caption*{with DiffPure}
\end{subfigure}

\caption{Purifying CDA adversarial images. NRP works well but DiffPure does not due to the defense overfitting.}
\label{fig:pure}
\end{figure}

\begin{figure*}[!t]
\newcommand{\sizeS}{0.08}
\newcommand{\fig}[2][1]{\includegraphics[width=#1\linewidth]{figures/visualize_examples/#2}}
\tiny
\centering
\setlength{\tabcolsep}{1pt}
\begin{tabular}{cccccccccccc}
\fig[\sizeS]{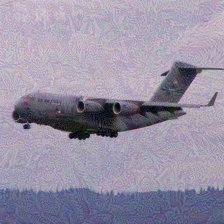} &
\fig[\sizeS]{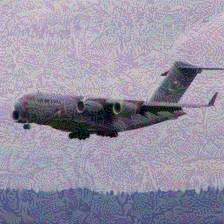} &
\fig[\sizeS]{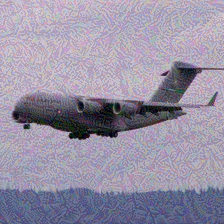} &
\fig[\sizeS]{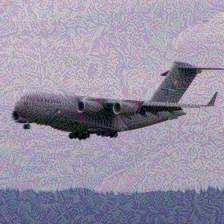} &
\fig[\sizeS]{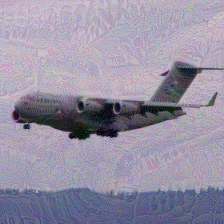} &
\fig[\sizeS]{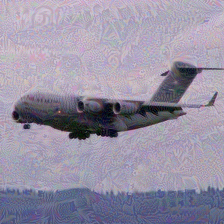} &
\fig[\sizeS]{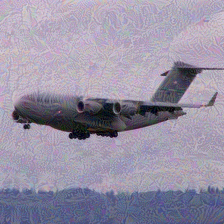} &
\fig[\sizeS]{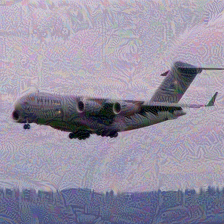} &
\fig[\sizeS]{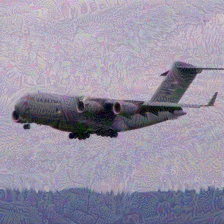} &
\fig[\sizeS]{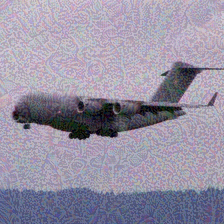} &
\fig[\sizeS]{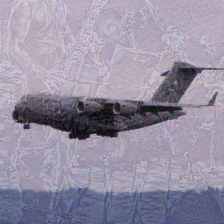} &
\fig[\sizeS]{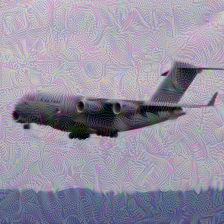} \\
\fig[\sizeS]{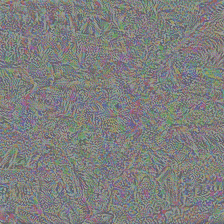} &
\fig[\sizeS]{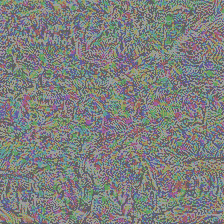} &
\fig[\sizeS]{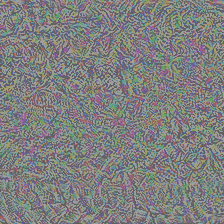} &
\fig[\sizeS]{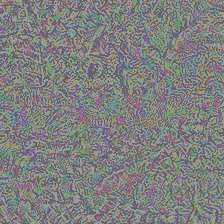} &
\fig[\sizeS]{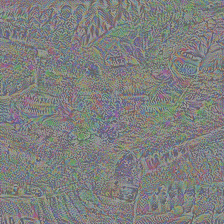} &
\fig[\sizeS]{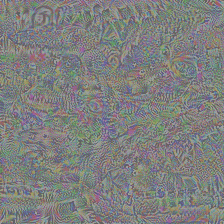} &
\fig[\sizeS]{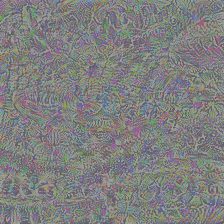} &
\fig[\sizeS]{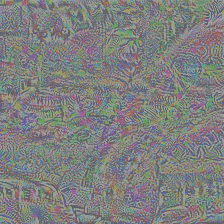} &
\fig[\sizeS]{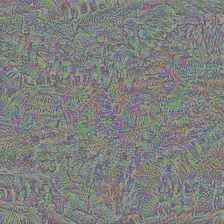} &
\fig[\sizeS]{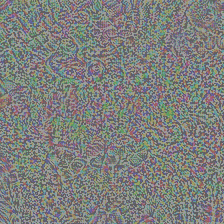} &
\fig[\sizeS]{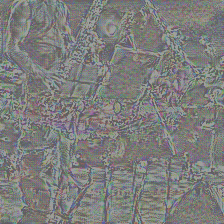} &
\fig[\sizeS]{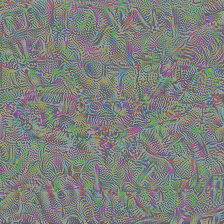} \\
\vspace{0.2cm}
\small{PGD} & \small{MI} & \small{NI} & \small{PI} & \small{DI} & \small{TI} & \small{SI} & \small{VT} & \small{Admix} & \small{TAP} & \small{AA} & \small{ILA} \\
\fig[\sizeS]{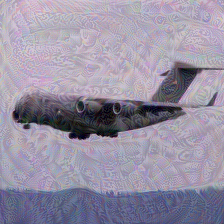} &
\fig[\sizeS]{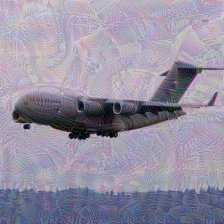} &
\fig[\sizeS]{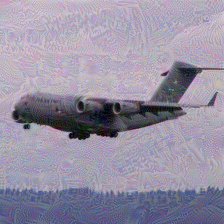} &
\fig[\sizeS]{data_resnet50_LinBP_ILSVRC2012_val_00047539_adv.png} &
\fig[\sizeS]{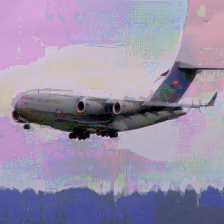} &
\fig[\sizeS]{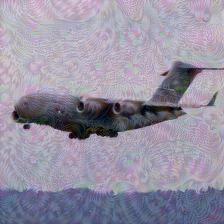} &
\fig[\sizeS]{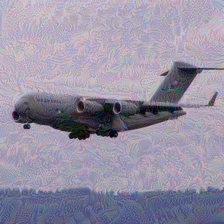} &
\fig[\sizeS]{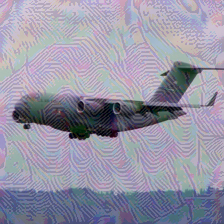} &
\fig[\sizeS]{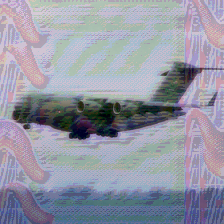} &
\fig[\sizeS]{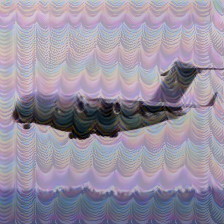} &
\fig[\sizeS]{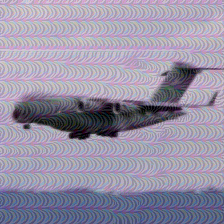} &
\fig[\sizeS]{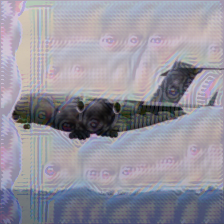}\\
\fig[\sizeS]{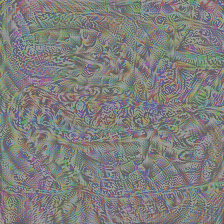} &
\fig[\sizeS]{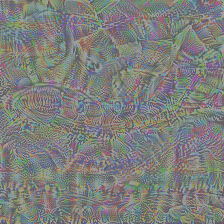} &
\fig[\sizeS]{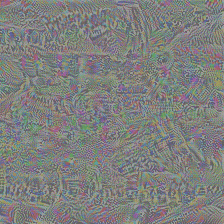} &
\fig[\sizeS]{data_resnet50_LinBP_ILSVRC2012_val_00047539_pt.png} &
\fig[\sizeS]{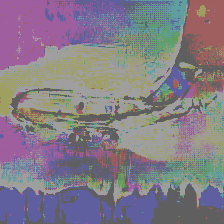} &
\fig[\sizeS]{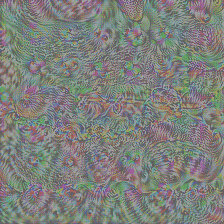} &
\fig[\sizeS]{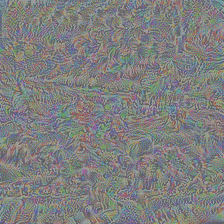} &
\fig[\sizeS]{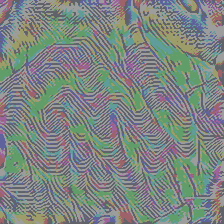} &
\fig[\sizeS]{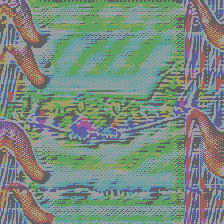} &
\fig[\sizeS]{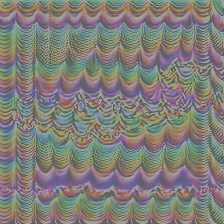} &
\fig[\sizeS]{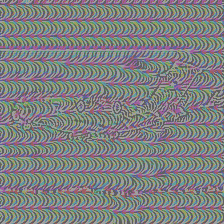} &
\fig[\sizeS]{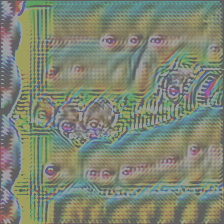} \\
\small{FIA} & \small{NAA} & \small{SGM} & \small{LinBP} & \small{RFA} & \small{IAA} & \small{DSM} & \small{GAP} & \small{CDA} & \small{GAPF} & \small{BIA} & \small{TTP}

\end{tabular}
\caption{Visualizations of adversarial examples generated by the 24 different attacks with their corresponding perturbations. Additional examples can be found in our GitHub repository.}
\label{fig:vis}
\end{figure*}

In particular, we reveal that this contextual performance is related to the overfitting of a defense to the type of perturbations used in its optimization.
Specifically, the purification network-based and adversarial training-based defenses rely on training defensive models with additional adversarial data.
In this way, the resulting defensive model only ensures high robustness against the perturbations that have been used for generating the adversarial data.
We take NRP vs. DiffPure as an example to explain this finding.
Specifically, NRP is trained with semantic perturbations~\cite{naseer2020self} and so it is very effective in mitigating low-frequency, semantic perturbations, e.g., those generated by generative modeling attacks (see perturbation visualizations in Fig.~\ref{fig:vis}).
However, it performs much worse against high-frequency perturbations, e.g., those generated by gradient stabilization and input augmentation attacks.
In contrast, DiffPure relies on the diffusion process with (high-frequency) Gaussian noise.
Therefore, it is not capable of purifying low-frequency, semantic perturbations.
Fig.~\ref{fig:pure} shows that NRP can purify the CDA adversarial images but DiffPure cannot do it well.

The substantial vulnerability of DiffPure to black-box (transferable) attacks is indeed very surprising since DiffPure was initially claimed to provide state-of-the-art robustness to white-box (adaptive) attacks, even surpassing adversarial training~\cite{nie2022diffusion}. 
Our finding suggests that \textit{DiffPure provides a false sense of security and requires a rightful evaluation following recommendations from}~\cite{athalye2018obfuscated,carlini2019evaluating,tramer2020adaptive}.
In addition, adversarial training is not effective against RFA since RFA has learned how to distort robust features by adopting an adversarially trained surrogate model.
Note that the norm bound used in RFA is $2\times$ larger than that in AT$_{\infty}$ and FD$_{\infty}$.

\mypara{Specific observations}
\textit{Adversarial training (AT) is effective but input pre-processing is not.} Both the two $L_{\infty}$ AT defenses, AT$_{\infty}$ and FD$_{\infty}$, are consistently effective against all attacks except RFA$_{\infty}$, which uses an AT surrogate model for generating robust perturbations.
The performance of AT$_{2}$ is also not sensitive to the attack method but sub-optimal due to the use of a different, $L_2$ norm.
In contrast, input pre-processing defenses are generally not useful, although they are known to be effective against (non-adaptive) white-box attacks where the perturbations are much smaller~\cite{xu2017feature,xie2017mitigating}.
However, adversarial training is known to be computationally expensive~\cite{shafahi2019adversarial,wong2020fast}.

\textit{Changing the model architecture is more useful than applying certain defenses} (although may not be practically desired). Previous work has noticed that transferring to Inception-v3 is especially hard~\cite{inkawhich2020transferable,inkawhich2020perturbing,zhao2021success}.
This might be because the Inception architecture contains relatively complex components, e.g., multiple-size convolution and two auxiliary classifiers.
Our results extend this finding from $\le3$ to 23 attacks. 
Furthermore, transferring from CNNs to ViT is also known to be hard~\cite{wei2022towards,wang2022generating,zhang2023transferable}.
Based on our results, we can observe that adopting ViT as the target model ranks about 4th on average among the 9 defenses. 

\textit{New model design largely boosts transferability.}
The generative modeling attacks and some of the surrogate refinement attacks are based on designing new surrogate or generative models, while the other attacks adopt off-the-shelf surrogate models.
In almost all cases, the former attacks outperform the latter.
However, they inevitably consume additional computational and data resources.

\mypara{Real-world attacks}
\textcolor{blue}{Beyond the single CNN or ViT models, we further evaluate attacks against two real-world models: a specifically designed defensive model against transfer-based attacks, PubDef~\cite{sitawarin2023defending}, and a production model, Google Cloud Vision (GCV)~\cite{gcv}, of which the architecture and training data are completely unknown to the attacker. 
Specifically, PubDef trains a defense model with adversarial examples generated on a heuristically selected set of 24 public pre-trained models, including CNNs, ViTs, Swin, BeiT, ConvMixer, zero-shot CLIP, etc.
The results on PubDef follow a similar pattern with those on other models.\footnote{For PubDef, the absolute numbers are not directly comparable to those in their original paper because we attack with $L_{\infty}=16$ rather than $4$.}
For example, generative attacks are still the overall best, and the earliest method, DI, is still the best input augmentation attack.}

\textcolor{blue}{For GCV, since it adopts a different label set from the standard ImageNet-1K, we treat an image to be successfully attacked when it yields a different top-1 prediction from the original image.
In this case, the clean accuracy of GCV is 100\%.
Considering practical budget constraints, we test with 100 randomly selected adversarial images per attack.
The results on GCV still confirm the best performance of generative attacks.
However, compared to other models (that are trained on ImageNet), GCV yields relatively stable robustness against other attacks.}

%% file: tex/StealthResults.tex
\subsection{New Insights into Stealthiness}
\label{sec:eva-res-stea}





Here we compare different attacks on their imperceptibility and also other finer-grained stealthiness characteristics.
In particular, we conduct a user study to validate our findings.
\noindent\textbf{General conclusion:} \textit{Stealthiness may be at odds with transferability.}
Fig.~\ref{fig:vis} visualizes the adversarial images with their corresponding perturbations for different attacks.
Solely looking at these examples, we notice that different attacks yield dramatically different perturbations, although they are constrained by the same $L_{\infty}$ norm bound with $\epsilon=16$.
In the following, we will demonstrate in more detail the distinct stealthiness characteristics of different attacks.

\begin{table}[!t]
\caption{Imperceptibility regarding five metrics. Additional results under $L_{\infty}=8/255$ with consistent patterns are in Table~\ref{tab:imper_linf8} of Appendix~\ref{app:add_res}.}
\newcommand{\tabincell}[2]{\begin{tabular}{@{}#1@{}}#2\end{tabular}}

\renewcommand{\arraystretch}{1}
      \centering
      \resizebox{\columnwidth}{!}{
        \begin{tabular}{l|cccccc}
\toprule[1pt]
Attacks&PSNR$\uparrow$&SSIM$\uparrow$&$\Delta E$$\downarrow$&LPIPS$\downarrow$&FID$\downarrow$\\
\midrule
PGD&\textbf{\GPSNR{28.112}}&\underline{\GSSIM{0.714}}&\textbf{\GDE{0.662}}&\textbf{\GLPIPS{0.175}}&\textbf{\GFID{33.823}}\\
\hline
MI~\cite{dong2018boosting}&\GPSNR{24.853}&\GSSIM{0.576}&\GDE{0.901}&\GLPIPS{0.290}&\GFID{47.570}\\
NI~\cite{lin2020nesterov} &\GPSNR{26.030}&\GSSIM{0.619}&\GDE{0.742}&\GLPIPS{0.300}&\GFID{47.637}\\
PI~\cite{wang2021boosting}&\GPSNR{26.015}&\GSSIM{0.619}&\GDE{0.741}&\GLPIPS{0.303}&\GFID{50.051}\\
\hline
DI~\cite{xie2019improving} &\underline{\GPSNR{27.987}}&\GSSIM{0.713}&\underline{\GDE{0.666}}&\underline{\GLPIPS{0.176}}&\GFID{66.075}\\
TI~\cite{dong2019evading} &\underline{\GPSNR{27.903}}&\GSSIM{0.712}&\GDE{0.671}&\GLPIPS{0.178}&\GFID{54.687}\\
SI~\cite{lin2020nesterov} &\GPSNR{27.712}&\GSSIM{0.704}&\GDE{0.668}&\GLPIPS{0.187}&\GFID{51.098}\\
VT~\cite{wang2021enhancing}&\GPSNR{27.264}&\GSSIM{0.693}&\GDE{0.710}&\GLPIPS{0.200}&\GFID{51.858}\\
Admix~\cite{wang2021admix} &\GPSNR{27.817}&\GSSIM{0.703}&\GDE{0.670}&\GLPIPS{0.187}&\GFID{43.052}\\
\hline
TAP~\cite{zhou2018transferable} &\GPSNR{25.710}&\GSSIM{0.560}&\GDE{0.743}&\GLPIPS{0.314}&\GFID{84.705}\\
AA~\cite{inkawhich2019feature}&\GPSNR{26.613}&\GSSIM{0.665}&\GDE{0.793}&\GLPIPS{0.246}&\GFID{71.301}\\
ILA~\cite{huang2019enhancing}&\GPSNR{26.222}&\GSSIM{0.625}&\GDE{0.665}&\GLPIPS{0.223}&\GFID{119.047}\\
FIA~\cite{wang2021feature} &\GPSNR{26.533}&\GSSIM{0.647}&\GDE{0.755}&\GLPIPS{0.225}&\GFID{139.522}\\
NAA~\cite{zhang2022improving}&\GPSNR{26.634}&\GSSIM{0.653}&\GDE{0.757}&\GLPIPS{0.207}&\GFID{109.383}\\
\hline
SGM~\cite{wu2020skip} &\GPSNR{27.586}&\GSSIM{0.704}&\underline{\GDE{0.663}}&\GLPIPS{0.180}&\underline{\GFID{40.637}}\\
LinBP~\cite{guo2020backpropagating}&\GPSNR{26.884}&\GSSIM{0.680}&\GDE{0.708}&\GLPIPS{0.196}&\GFID{87.930}\\
RFA$_{2}$~\cite{springer2021little} &\GPSNR{27.113}&\underline{\GSSIM{0.720}}&\GDE{0.701}&\GLPIPS{0.185}&\GFID{70.065}\\
RFA$_{\infty}$~\cite{springer2021little} &\GPSNR{24.542}&\textbf{\GSSIM{0.751}}&\GDE{0.906}&\GLPIPS{0.226}&\GFID{60.349}\\
IAA~\cite{zhu2021rethinking} &\GPSNR{26.480}&\GSSIM{0.671}&\GDE{0.773}&\GLPIPS{0.205}&\GFID{145.231}\\
DSM~\cite{yang2022boosting}&\GPSNR{27.947}&\GSSIM{0.712}&\GDE{0.667}&\underline{\GLPIPS{0.177}}&\underline{\GFID{45.909}}\\
\hline
GAP~\cite{poursaeed2018generative}&\GPSNR{25.184}&\GSSIM{0.611}&\GDE{0.847}&\GLPIPS{0.285}&\GFID{145.964}\\
CDA~\cite{naseer2019cross}&\GPSNR{24.200}&\GSSIM{0.605}&\GDE{1.022}&\GLPIPS{0.337}&\GFID{1098.624}\\
GAPF~\cite{kanth2021learning} &\GPSNR{25.423}&\GSSIM{0.641}&\GDE{0.846}&\GLPIPS{0.250}&\GFID{215.241}\\
BIA~\cite{zhang2022beyond}  &\GPSNR{24.590}&\GSSIM{0.537}&\GDE{0.883}&\GLPIPS{0.400}&\GFID{230.470}\\
TTP~\cite{naseer2021generating} &\GPSNR{25.813}&\GSSIM{0.678}&\GDE{0.764}&\GLPIPS{0.270}&\GFID{209.802}\\

\bottomrule[1pt]
\end{tabular}
}
\label{tab:imper}
\end{table}

\mypara{Imperceptibility}
Table~\ref{tab:imper} reports the imperceptibility of different attacks regarding five diverse perceptual similarity metrics.
First of all, almost all the 23 transfer attacks are more perceptible than the PGD baseline although their perturbations are constrained in the same way.
This means that the current transfer attacks indeed sacrifice imperceptibility for improving transferability.
\textcolor{blue}{One exception is the SSIM performance of RFA, which is even better than the PGD baseline. This might be because RFA mainly introduces smooth perturbations but does not break image structures too much, as confirmed by the visualizations in Fig.~\ref{fig:vis}. We further calculate the luminance, contrast, and structure components of SSIM for those adversarial images, and the results confirm that SSIM yields a structure score of 0.916 (and contrast score of 0.893), which is 8.2\% (13.6\%) higher than the second best.}

Moreover, among the 23 transfer attacks, imperceptibility varies greatly and negatively correlates with the transferability reported in Table~\ref{tab:transfer}. 
In particular, certain generative modeling attacks are the most transferable but also the most perceptible.
In addition, generative modeling attacks yield especially large FID scores.
This can be explained by the dramatic distribution shift between their adversarial and original images due to their image-agnostic perturbations~\cite{naseer2021generating,kanth2021learning}.
More generally, the poor imperceptibility performance suggests that using only the $L_{\infty}$ constraint is not sufficient.
This conclusion is consistent with that for white-box attacks~\cite{sharif2018suitability,zhao2020towards} and calls for a more advanced design of attack constraints~\cite{laidlaw2020perceptual}. 

\begin{figure}[!t]
\centering
\includegraphics[width=\columnwidth]{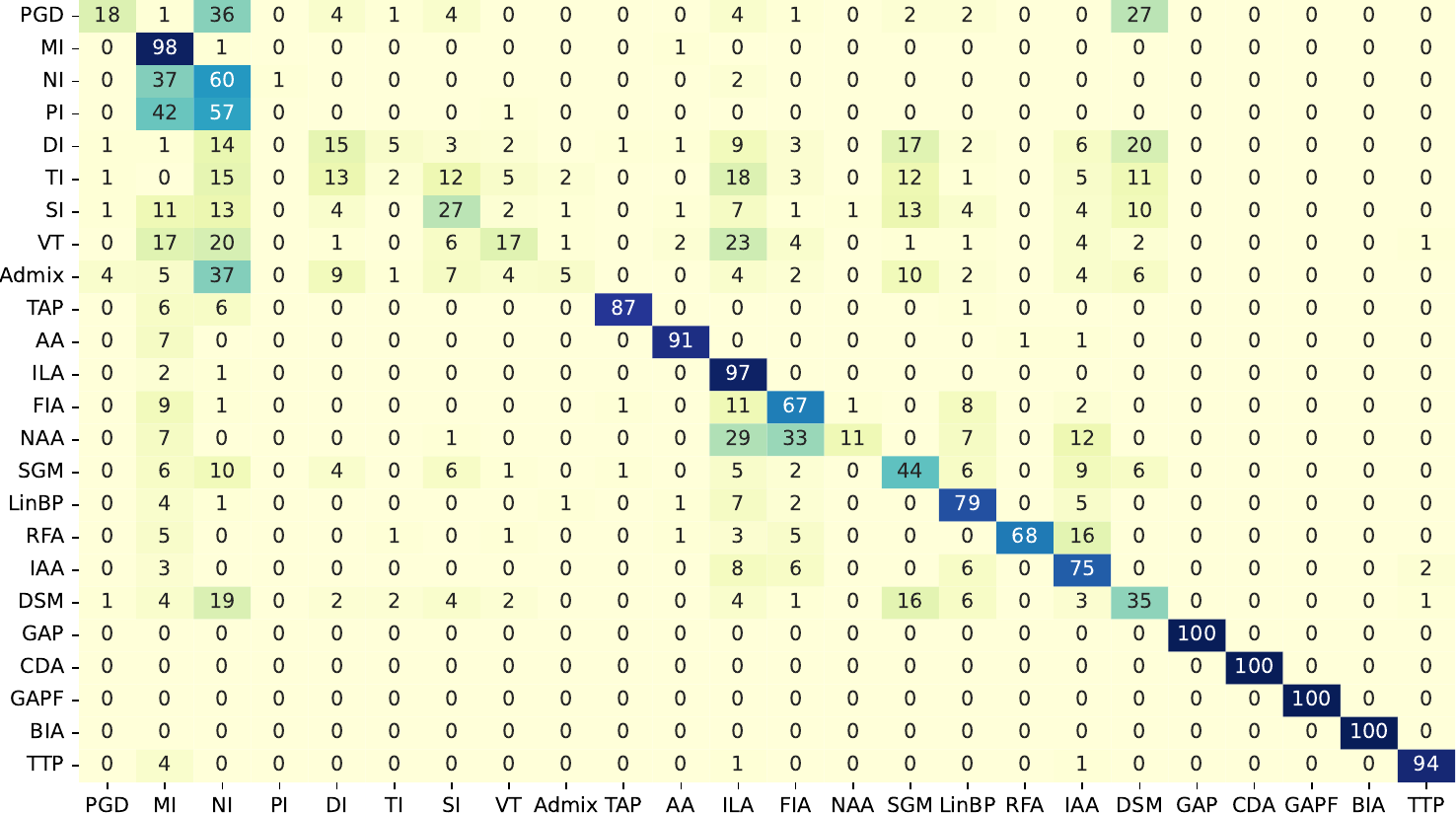}
\caption{Attack traceback with image features.}
\label{fig:confusion}
\end{figure}

\begin{figure*}[!t]
\centering
\includegraphics[width=0.52\textwidth,height=3cm]{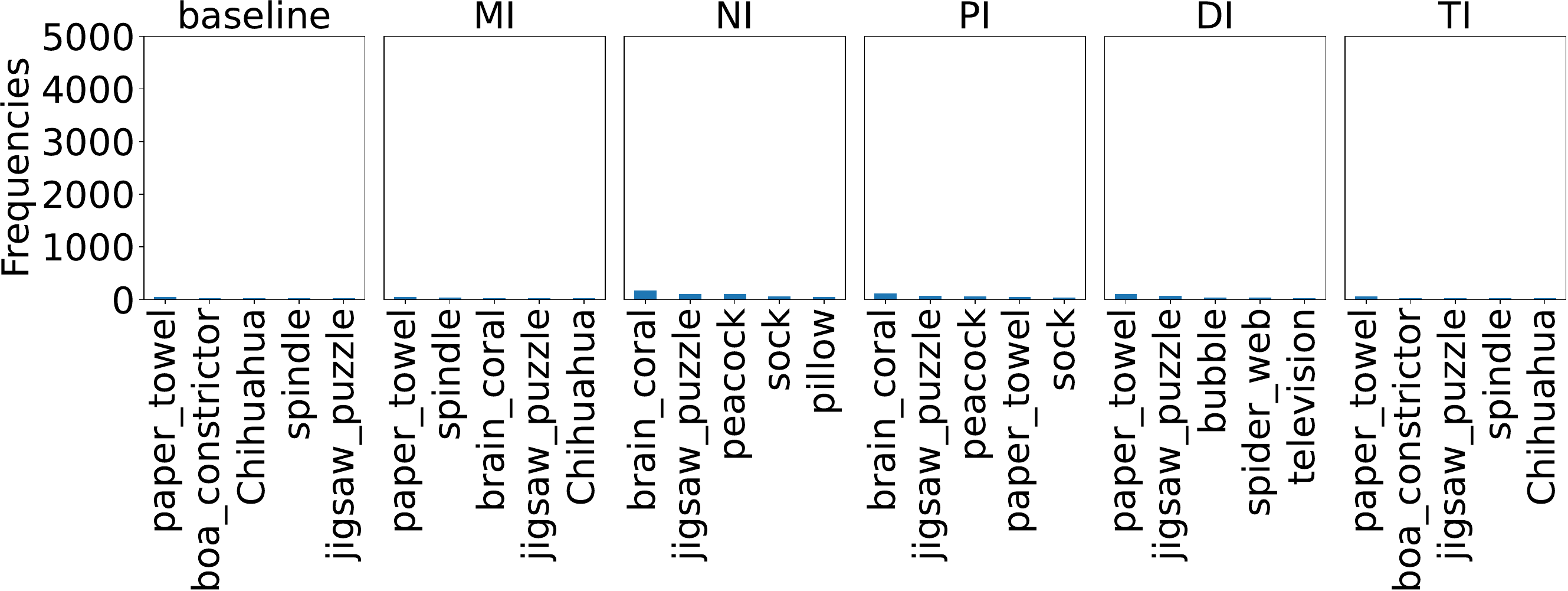}
\hspace{-0.15cm}
\includegraphics[width=0.47\textwidth,height=3cm]{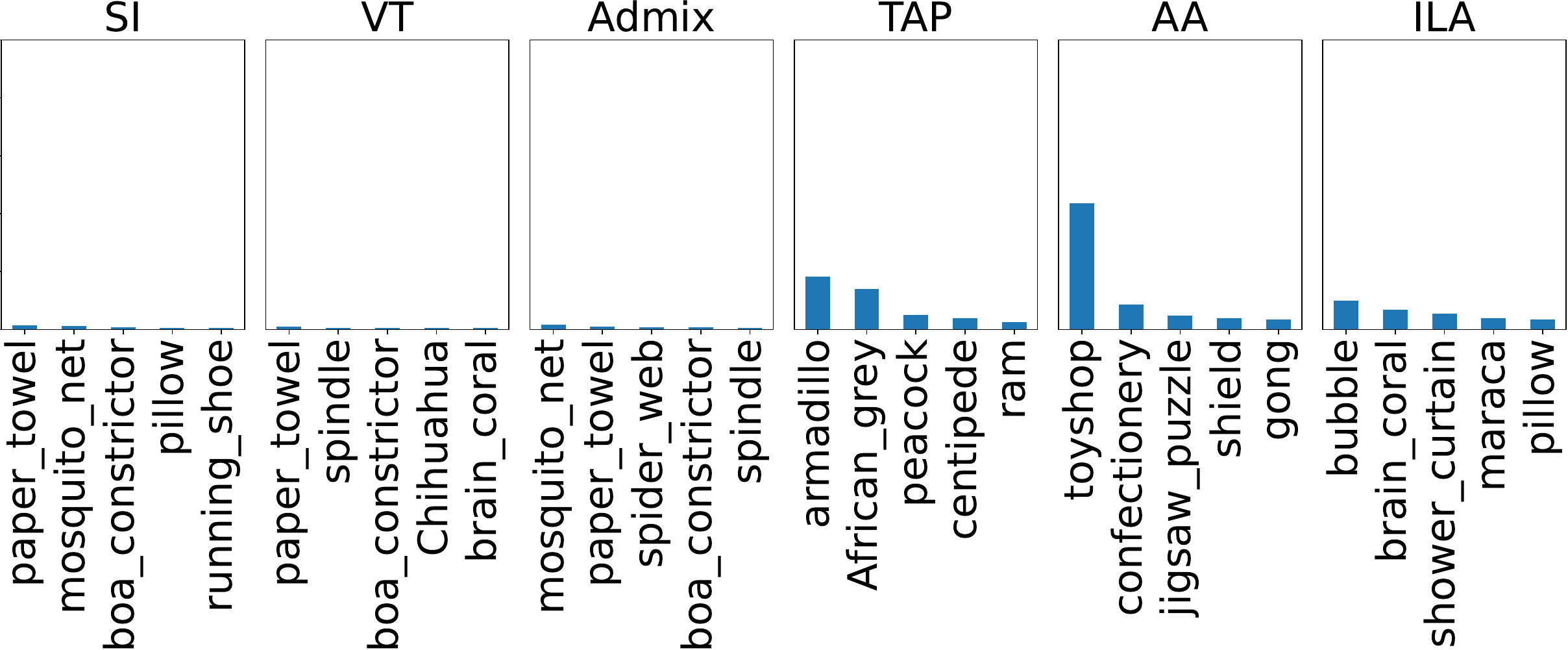}
\vspace{0.15cm}
\hspace{-0.15cm}

\includegraphics[width=0.52\textwidth,height=3.5cm]{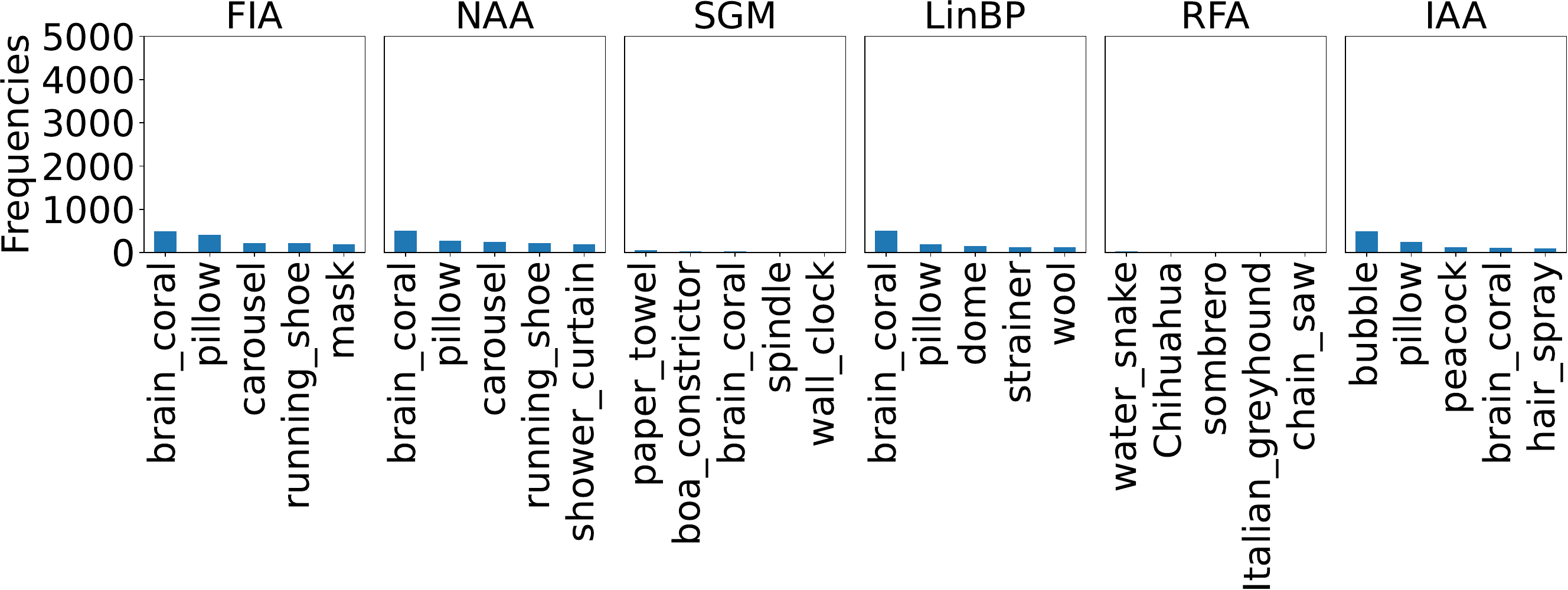}
\hspace{-0.15cm}
\includegraphics[width=0.47\textwidth,height=3.47cm]{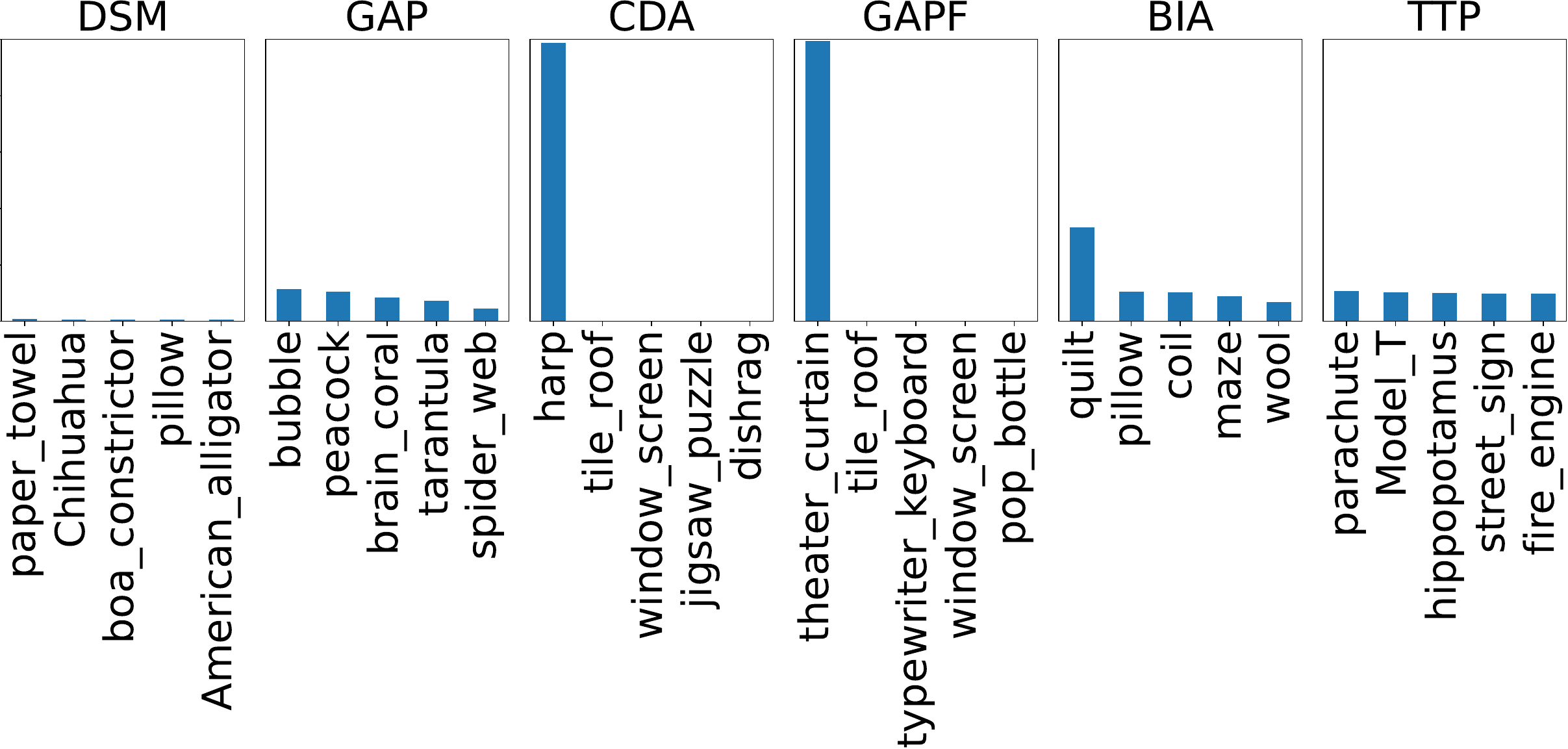}
\caption{Distributions of five most frequently predicted classes with their names.}
\label{fig:dis_name_full}

\end{figure*}

The above results demonstrate that different attacks indeed yield dramatically different imperceptibility.
In the following, we further compare their stealthiness at finer-grained levels.
To this end, we conduct ``\textit{attack traceback}'', which aims to trace back the specific attack based on certain information.
In general, an attack is less stealthy when it is easier to trace back.
The ``attack traceback'' is basically a classification problem, and in our case, with 24 classes.
Specifically, we randomly select 500 original images and then perturb them using the 24 different attacks, resulting in 12000 adversarial images in total.
For each attack, 400 adversarial images are used for training and the rest 100 for testing.
we conduct the attack classification with two types of features: (input) image features and (output) misclassification features.

\mypara{Attack traceback with image features} 
We use ResNet-18 as the backbone for the 24-class classifier with image features.
As can be seen from Fig.~\ref{fig:confusion}, different attacks can be well differentiated solely based on image features, with an overall accuracy of 60.43\%, much higher than a random guess, i.e., 1/24.
Moreover, the category-wise accuracy is even higher, i.e., 75.61\%.
This suggests that our attack systemization is reliable since it captures the signature of each specific category.

More specifically, we note low accuracy for PI and input augmentation attacks but perfect accuracy for generative modeling attacks.
For PI, the low accuracy can be explained by its technical similarity to NI/MI and their resulting similar perturbations, as shown in Fig.~\ref{fig:vis}.
For generative modeling attacks, the perfect accuracy might be explained by their special perturbations.
For input augmentation attacks, the perturbations are somewhat a mixture of low- and high-frequency components.
So they are harder to distinguish from multiple other attacks.
In general, it should be noted that perturbation patterns are hard to characterize, compared to objects that are semantically meaningful to humans.

\mypara{Attack traceback with misclassification features} 
We train an SVM classifier with an RBF kernel and adopt the top-$N$ output class IDs as the misclassification features.
Table~\ref{tab:mis} shows the classification accuracy as a function of $N$.
As can be seen, too small $N$ is insufficient to represent the attack and too large $N$ may introduce noisy information. 
In particular, a moderate choice of $N=5$ yields the optimal accuracy, i.e., 18.5\%, which is also much higher than a random guess, i.e., 1/24.
However, it is still far behind the image features, which have much higher dimensions, i.e., $224\times224=50176$.

We further examine the class distributions of the output predictions.
To this end, for each attack, we calculate the frequencies of different mispredicted classes over 5000 adversarial images.
As can be seen from Fig.~\ref{fig:dis_name_full}, different attack categories lead to different class distributions.
Specifically, gradient stabilization and input augmentation attacks lead to uniform class distributions.
In contrast, the other attacks yield more concentrated distributions.
Moreover, the most frequent class usually corresponds to semantics that contain rich textures, e.g., ``brain coral'' and ``jigsaw puzzle''.
This somewhat confirms the high-frequency nature of perturbations.
Specifically, a few classes dominate the prediction of generative modeling attacks.
For example, almost all CDA adversarial images are predicted into ``harp''.
These findings suggest that generative modeling attacks are very easy to trace back, even with only the Top-1 prediction.

\begin{table}[!t]
    \caption{Attack traceback with misclassification features.}
    \centering
    \begin{tabular}{l|ccccccc}
    \toprule[1pt]
        $N$ & 1 & 3 & 5 & 10 & 100 & 1000\\
        \midrule
        Acc (\%) & 14.6 & 17.9 & 18.5 & 18.1 & 17.4 & 17.0\\
    \bottomrule[1pt]
    \end{tabular}
    \label{tab:mis}
\end{table}

\begin{figure*}[!t]
\centering
\includegraphics[width=0.92\columnwidth]{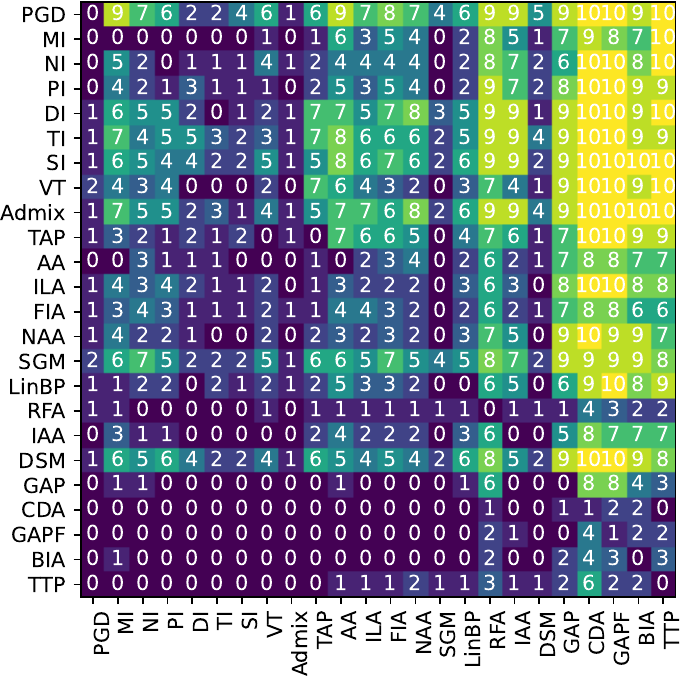}
\includegraphics[width=0.92\columnwidth]{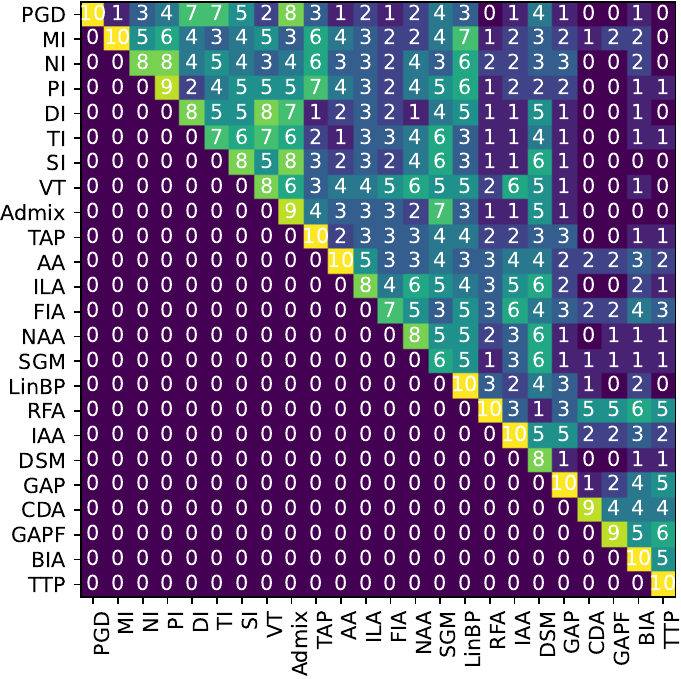}

\caption{User study results that further support our findings on stealthiness. \textbf{Left:} Visual quality comparison between any two attacks. Each cell at (Row, Column) means the frequencies of \textit{the attack in that Row with higher visual quality than the attack in that Column}. Note that all diagonal values are supposed to be zeros since the same attack yields the same quality. \textbf{Right:} Similarity between any two attacks.}
\label{fig:human}
\end{figure*}

\mypara{User study}
We conducted user studies to validate our findings on stealthiness with 3000 pairwise attack comparisons judged by 10 graduate students with sufficient knowledge about adversarial examples. 
We pay $\$$5/hour per participant, which is higher than the minimum payout (about $\$$3/hour) in our local region.
We obtained informed consent from all participants for reporting their scores in this research, and the Human Research Ethics Committee of the first author’s affiliation determined that the study was exempt from further human subjects review.
To our knowledge, well-known papers conducting user studies on the quality of adversarial images also do not involve ethics review~\cite{xiao2018spatially,bhattad2020Unrestricted,zhao2020adversarial_arxiv}.

Specifically, for each of the 10 original images shown in our GitHub repository, we generate C(24,2) + 24 = 300 pairs of adversarial images, involving all 24 attacks.
These 3000 pairs of adversarial images are given to 10 users for judgment.
For each pair, they are asked which of the two images are of higher visual quality and should choose from ``left'', ``right'', and ``similar''.
Statistics in Fig.~\ref{fig:human} further support our findings on stealthiness.
Specifically, Fig.~\ref{fig:human} Left confirms that generative attacks are the least stealthy, and Fig.~\ref{fig:human} Right confirms that the similarity between intra-category attacks is generally higher than that of inter-category attacks.

%% file: tex/Discussions.tex
\section{Discussion and Outlook}
\label{sec:dis}

\mypara{Forward compatibility of our evaluation}
Our attack systemization corresponds to the major components in the common attack pipeline, as shown in Fig.~\ref{fig:pipeline}.
Therefore, we believe that future attacks would also fall into those categories unless there appears a completely new design of adversarial attacks.
Our evaluation methodology can also be directly extended to other threat models. 
In the current stage, we have followed the most common threat model as described in Section~\ref{sec:thr}.
This threat model covers basic settings regarding models, datasets, constraints, attack goals, etc.
Therefore, specific practical considerations would not affect our scientific conclusions.   
In the future, exploring more realistic threat models would make more sense with the progress of transfer attacks.
For example, targeted attacks pose more realistic threats than non-targeted attacks but most of the current transfer attacks cannot achieve substantial targeted success~\cite{zhao2021success,naseer2021generating}.

\mypara{Further explanations of our findings}
We have drawn several new findings about transferable adversarial images through our systematic evaluation.
However, some of them still need further exploration to be fully understood.
For example, we find that for generative modeling attacks, using a moderate $\epsilon_{\textrm{train}}$ leads to consistently optimal performance for all $\epsilon_{\textrm{test}}$, although in normal machine learning, choosing a matched hyperparameter (here $\epsilon_{\textrm{test}}=\epsilon_{\textrm{train}}$) performs the best.
This uncommon phenomenon may be specifically related to generative perturbations and requires further study in the future. 
In addition, different attacks have dramatically different properties regarding perturbation and misclassification patterns.
For future work, it is important to figure out how these properties are related to the specific attack design. 
In particular, the reason why generative modeling attacks tend to produce image-agnostic perturbations falling into a dominant class should be explored.

%% file: tex/Conclusion.tex
\section{Conclusion}
\label{sec:conc}
In this paper, we have provided the first systematic evaluation of recent transferable attacks, involving 23 representative transferable attacks (in five categories) against 11 representative defenses (in four categories) on ImageNet.
In particular, we have established new evaluation guidelines that address the two common problems in existing work. 
Our evaluation focuses on both attack transferability and stealthiness, and our experimental results lead to new insights that complement or even challenge existing knowledge.
In particular, for stealthiness, we move beyond diverse imperceptibility measures and explore the potential of attack traceback based on image or misclassification features.
User study results on image quality further support our findings. 

Our work provides a thorough picture of the current progress of transfer attacks and demonstrate that existing problematic evaluations have indeed hindered the assessment of the actual progress in transferable adversarial images.
We hope that our analyses can guide future research towards good practices in transfer attacks.

%% file: tex/Appendix.tex
\appendix
\subsection{Comparions between Existing and Our Evaluations}
\label{app:com}
We compare the 
Detailed statistics about the transferability evaluations in existing work are summarized in Table~\ref{tab:comp}.
As can be seen, existing work considers 11 attacks at most and has not conducted intra-category analyses.
We make sure our threat model follows the most common settings in existing work regarding five aspects, as summarized in Table~\ref{tab:circle}.

\begin{table}[h]
\newcommand{\tabincell}[2]{\begin{tabular}{@{}#1@{}}#2\end{tabular}}
\caption{Comparions between existing and our evaluations.}

\renewcommand{\arraystretch}{1}
      \centering
        \begin{tabular}{l|ccc}
\toprule[1pt]
Studies&\tabincell{c}{\# of attacks\\(categories)}&\tabincell{c}{Fair\\comparisons}&\tabincell{c}{Intra-category\\analyses}\\

\midrule
MI~\cite{dong2018boosting}         &1 (1)&&\\
NI~\cite{lin2020nesterov}          &4 (2)&&\\
PI~\cite{wang2021boosting}         &5 (2)&\cmark&\\
DI~\cite{xie2019improving}         &2 (2)&\cmark&\\
TI~\cite{dong2019evading}          &3 (2)&&\\
SI~\cite{lin2020nesterov}          &4 (2)&&\\
VT~\cite{wang2021enhancing}        &6 (2)&&\\
Admix~\cite{wang2021admix}         &4 (1)&&\\
TAP~\cite{zhou2018transferable}    &2 (2)&\cmark&\\
AA~\cite{inkawhich2019feature}     &2 (2)&\cmark&\\
ILA~\cite{huang2019enhancing}      &4 (3)&&\\
FIA~\cite{wang2021feature}         &6 (3)&&\\
NAA~\cite{zhang2022improving}      &5 (2)&\cmark&\\
SGM~\cite{wu2020skip}              &4 (3)&&\\
LinBP~\cite{guo2020backpropagating}&5 (3)&\cmark&\\
RFA~\cite{springer2021little}      &1 (1)&&\\
IAA~\cite{zhu2021rethinking}       &7 (4)&\cmark&\\
DSM~\cite{yang2022boosting}        &4 (3)&&\\
GAP~\cite{poursaeed2018generative} &1 (1)& &\\
CDA~\cite{naseer2019cross}         &9 (4)&\cmark&\\
GAPF~\cite{kanth2021learning}      &3 (1)&\cmark&\\
BIA~\cite{zhang2022beyond}         &5 (3)& \cmark&\\
TTP~\cite{naseer2021generating}    &11 (5)& \cmark&\\
\midrule
Ours&\textbf{23} (\textbf{5})&\cmark&\cmark\\
\bottomrule[1pt]
\end{tabular}
\label{tab:comp}
\end{table}

\begin{table}[h]
\newcommand{\tabincell}[2]{\begin{tabular}{@{}#1@{}}#2\end{tabular}}
\caption{Threat models in existing work regarding five aspects. For each aspect (A/B), \fullcirc~ denotes only considering A, \emptycirc~ denotes only considering B, and \halfcirc~ denotes both.}

\renewcommand{\arraystretch}{1}
      \centering
      \resizebox{\columnwidth}{!}{
        \begin{tabular}{l|ccccc}
\toprule[1pt]
Attacks&\tabincell{c}{Scenario\\(public model/\\private system)}&\tabincell{c}{Attack goal\\ (untarget/\\target)}&\tabincell{c}{Dataset\\(ImageNet/\\others)}&\tabincell{c}{Model\\ (CNNs/\\others)}&\tabincell{c}{Source/target\\training data\\ (same/\\disjoint)}\\

\midrule
MI~\cite{dong2018boosting}          &\fullcirc&\fullcirc&\fullcirc&\fullcirc&\fullcirc\\
NI~\cite{lin2020nesterov}           &\fullcirc&\fullcirc&\fullcirc&\fullcirc&\fullcirc\\
PI~\cite{wang2021boosting}          &\fullcirc&\fullcirc&\fullcirc&\fullcirc&\fullcirc\\
DI~\cite{xie2019improving}          &\fullcirc&\fullcirc&\fullcirc&\fullcirc&\fullcirc\\
TI~\cite{dong2019evading}           &\fullcirc&\fullcirc&\fullcirc&\fullcirc&\fullcirc\\
SI~\cite{lin2020nesterov}           &\fullcirc&\fullcirc&\fullcirc&\fullcirc&\fullcirc\\ 
VT~\cite{wang2021enhancing}         &\fullcirc&\fullcirc&\fullcirc&\fullcirc&\fullcirc\\ 
Admix~\cite{wang2021admix}          &\fullcirc&\fullcirc&\fullcirc&\fullcirc&\fullcirc\\
TAP~\cite{zhou2018transferable}     &\fullcirc&\fullcirc&\fullcirc&\fullcirc&\fullcirc\\
AA~\cite{inkawhich2019feature}      &\fullcirc&\halfcirc&\fullcirc&\fullcirc&\fullcirc\\
ILA~\cite{huang2019enhancing}       &\fullcirc&\fullcirc&\fullcirc&\fullcirc&\fullcirc\\
FIA~\cite{wang2021feature}          &\fullcirc&\fullcirc&\fullcirc&\fullcirc&\fullcirc\\
NAA~\cite{zhang2022improving}       &\fullcirc&\fullcirc&\fullcirc&\fullcirc&\fullcirc\\
SGM~\cite{wu2020skip}               &\fullcirc&\fullcirc&\fullcirc&\fullcirc&\fullcirc\\
LinBP~\cite{guo2020backpropagating} &\fullcirc&\fullcirc&\fullcirc&\fullcirc&\fullcirc\\
RFA~\cite{springer2021little}       &\fullcirc&\halfcirc&\fullcirc&\halfcirc&\fullcirc\\
IAA~\cite{zhu2021rethinking}        &\halfcirc &\halfcirc&\fullcirc&\fullcirc&\fullcirc\\
DSM~\cite{yang2022boosting}         &\fullcirc&\fullcirc&\halfcirc&\fullcirc&\fullcirc\\
GAP~\cite{poursaeed2018generative}  &\fullcirc&\halfcirc&\halfcirc&\fullcirc&\fullcirc\\
CDA~\cite{naseer2019cross}          &\fullcirc&\fullcirc&\fullcirc&\fullcirc&\halfcirc\\
GAPF~\cite{kanth2021learning}       &\fullcirc&\fullcirc&\fullcirc&\fullcirc&\fullcirc\\
BIA~\cite{zhang2022beyond}          &\fullcirc&\fullcirc&\fullcirc&\fullcirc&\halfcirc\\
TTP~\cite{naseer2021generating}     &\fullcirc&\halfcirc&\fullcirc&\fullcirc&\halfcirc\\
\midrule
Ours     &\fullcirc&\fullcirc&\fullcirc&\halfcirc&\fullcirc\\

\bottomrule[1pt]
\end{tabular}
}
\label{tab:circle}
\end{table}

\subsection{Additional Experimental Results}
\label{app:add_res}

\begin{figure}[!t]
\centering
\includegraphics[width=\columnwidth]{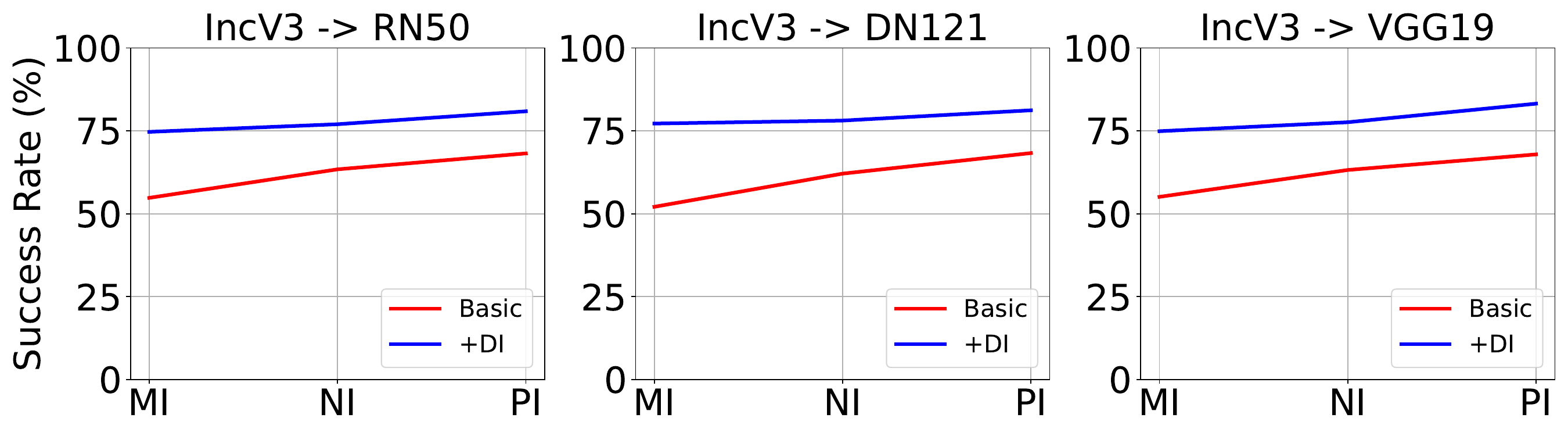}
\includegraphics[width=\columnwidth]{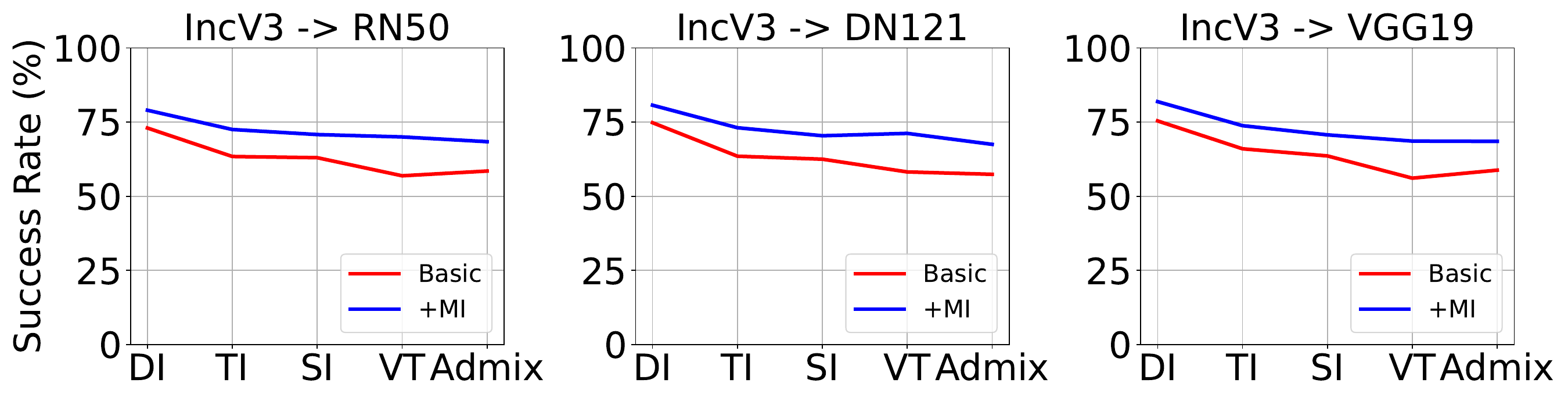}

\caption{\textcolor{blue}{Integrating one category of attacks with a well-known attack (i.e., DI or MI) from another category does not change the relative attack performance.}}
\label{fig:inte}
\end{figure}

\begin{figure}[h]
\centering
\includegraphics[width=\columnwidth]{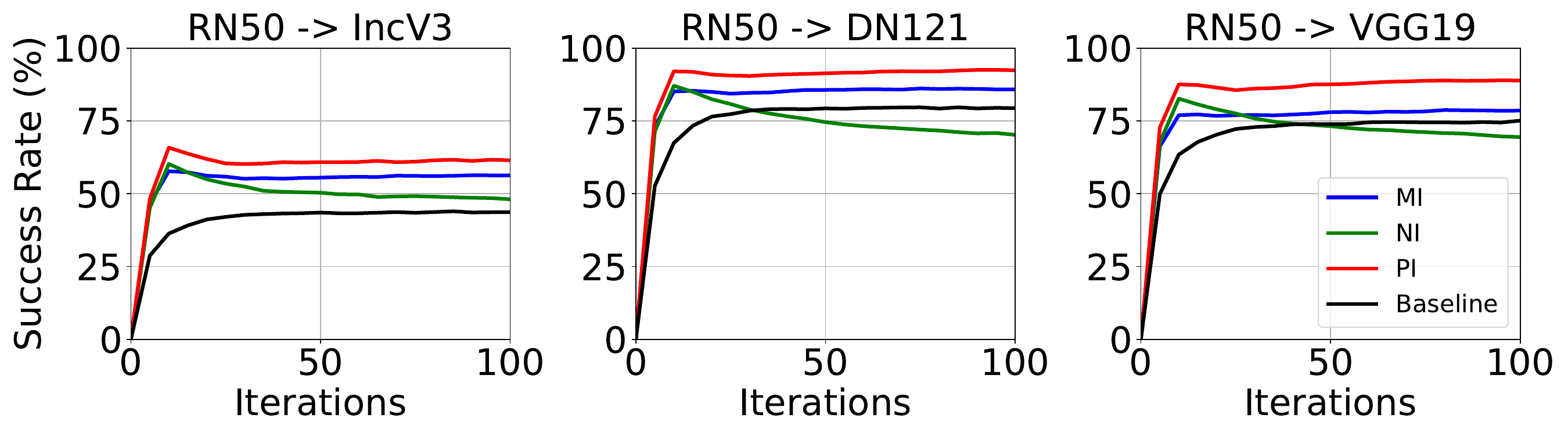}
\includegraphics[width=\columnwidth]{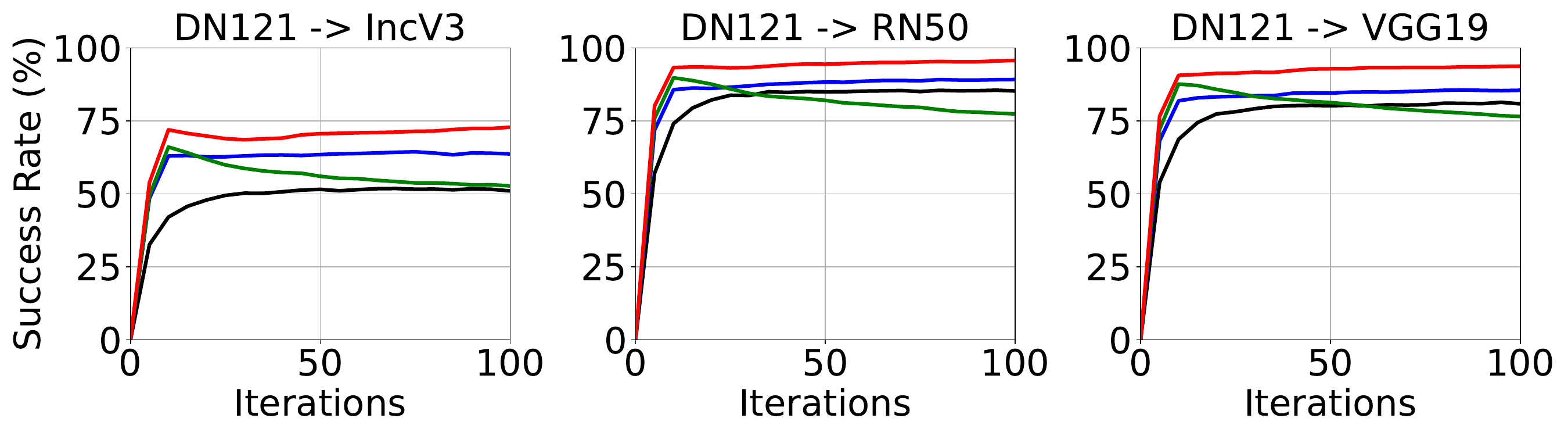}
\includegraphics[width=\columnwidth]{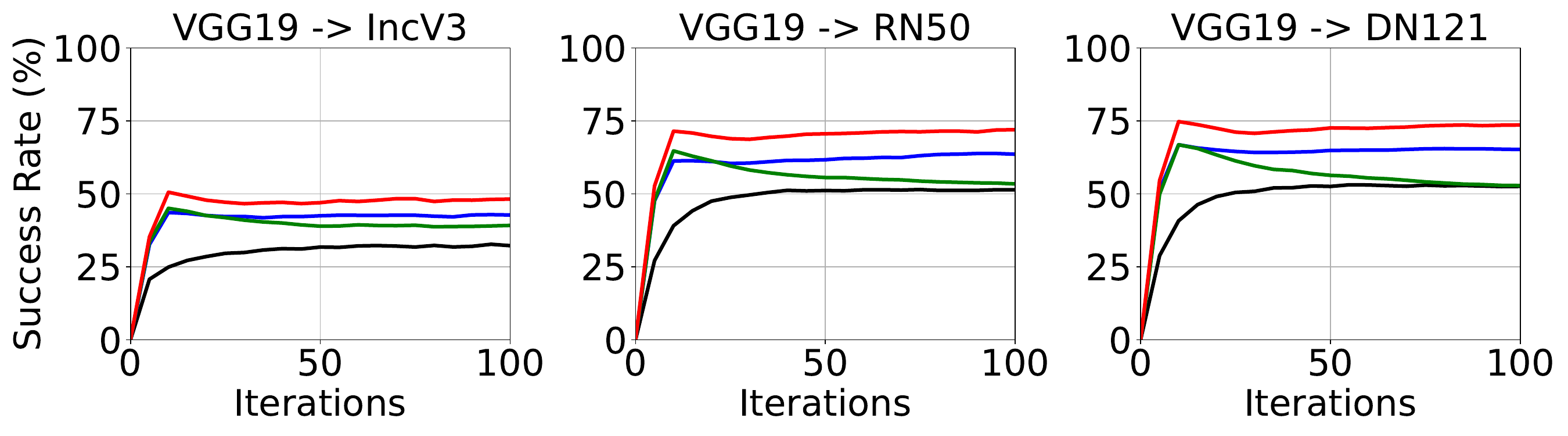}
\caption{Transferability vs. iteration curve on three different surrogates for gradient stabilization attacks. Our findings about transferability in Fig.~\ref{fig:gradient} still hold.}
\label{fig:gradient_app}
\end{figure}

\begin{figure}[h]
\centering
\includegraphics[width=\columnwidth]{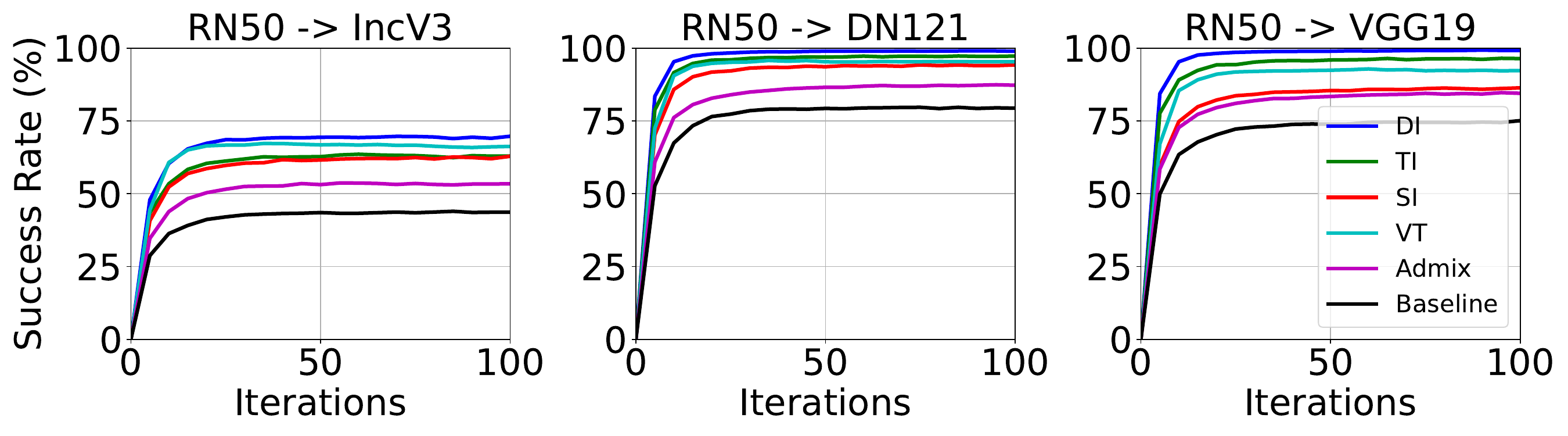}
\includegraphics[width=\columnwidth]{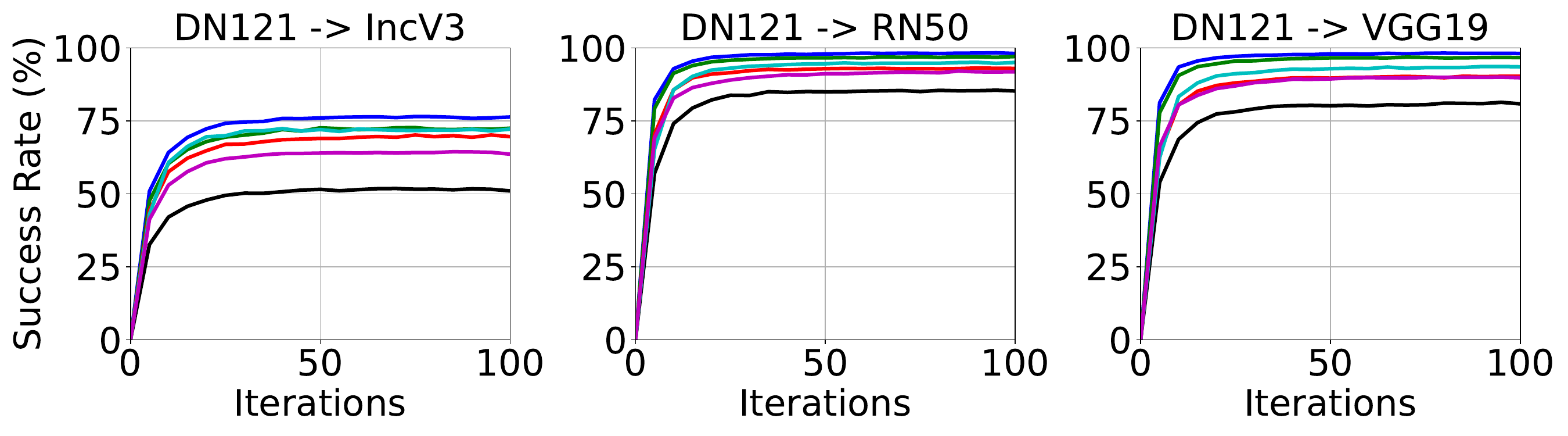}
\includegraphics[width=\columnwidth]{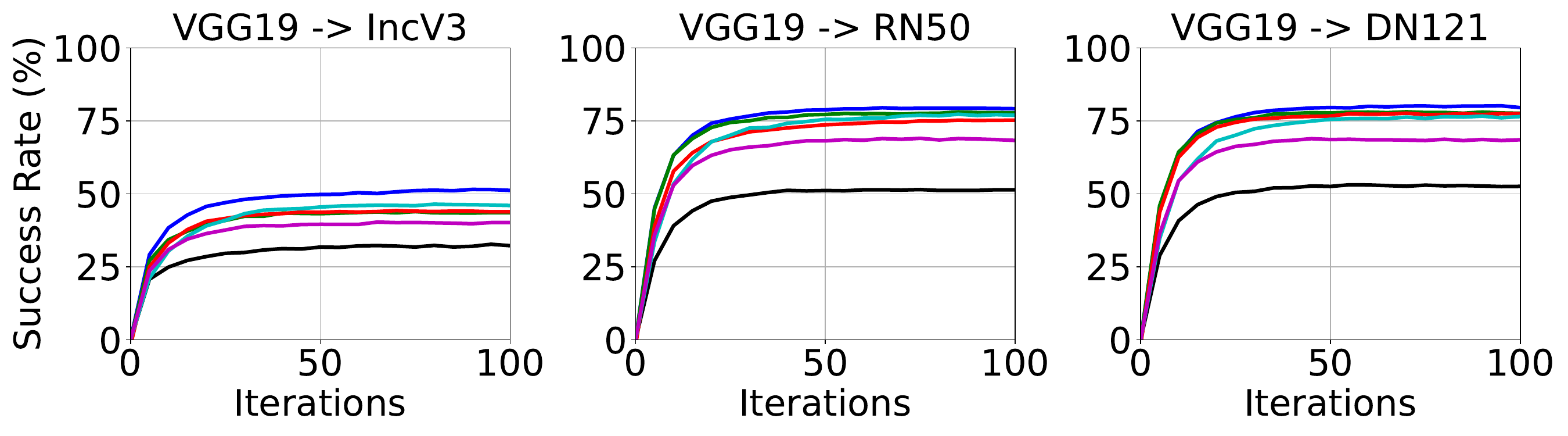}
\caption{Transferability vs. iteration curve on three different surrogates for input augmentation attacks. Our findings about transferability in Fig.~\ref{fig:copy_5} still hold.}
\label{fig:copy_5_app}
\end{figure}

In Fig.~\ref{fig:gradient_app}, Fig.~\ref{fig:copy_5_app}, Fig.~\ref{fig:inte}, Table~\ref{tab:trans_linf8}, and Table~\ref{tab:imper_linf8}, we present additional experimental results under the common settings in \label{tab:circle} to support our findings about transferability and stealthiness.

\begin{figure}[!t]
\centering
\includegraphics[width=1\columnwidth]{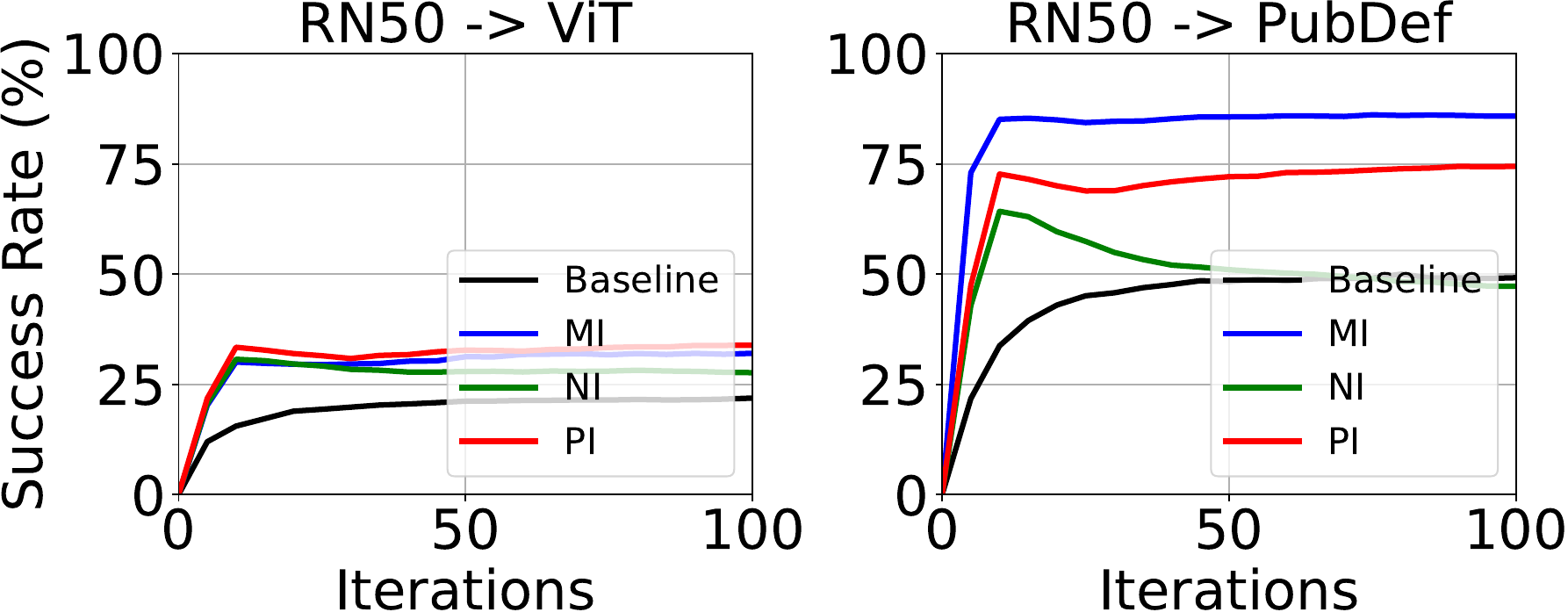}
\caption{\textcolor{blue}{Transferability vs. iteration curve on ViT and PubDef target models for gradient stabilization attacks.}}
\label{fig:GS_revision}
\end{figure}

\textcolor{blue}{Furthermore, we validate that our new insights drawn from the intra-category analyses still hold beyond the common settings in \label{tab:circle}.
Specifically, Fig.~\ref{fig:GS_revision} shows that for gradient stabilization attacks, ``using more iterations may even decrease the performance'' still holds on more advanced models, i.e., ViT and PubDef.
In addition, we look into a practical scenario with attack integration.
Following the common practice, we integrate the well-known attacks, DI or MI, and Fig.~\ref{fig:inte} demonstrates that the relative attack performance remains largely unchanged after integration.
This is also consistent with findings in existing work that their new method can achieve consistent superiority across different integrations.
However, it should be noted that integrating more attacks incurs more computation costs and yields less stealthy images.}

\begin{table}[!t]
\caption{Attack transferability (\%) under $L_{\infty}=8/255$. Our findings about transferability in Table~\ref{tab:transfer} still hold. Note that RFA$_{\infty}$ with $L_{\infty}=8/255$ is weak since its surrogate is an adversarially robust model also under $L_{\infty}=8/255$.}
\newcommand{\tabincell}[2]{\begin{tabular}{@{}#1@{}}#2\end{tabular}}

      \centering
        \begin{tabular}{l|cccc}
\toprule[1pt]
Attacks&DN121&VGG19&IncV3&ViT\\
\midrule
Clean Acc&100.0&100.0&100.0&94.7\\
PGD  &\gradient{53.9}&\gradient{50.7}&\gradient{28.0}&\gradient{12.7}\\
\hline
MI   &\gradient{67.2}&\gradient{59.9}&\gradient{39.4}&\gradient{17.3}\\
NI   &\gradient{67.8}&\gradient{63.3}&\gradient{40.0}&\gradient{17.6}\\
PI   &\gradient{76.2}&\gradient{70.7}&\gradient{44.2}&\gradient{19.0}\\
\hline
DI   &\underline{\gradient{91.6}}&\underline{\gradient{92.6}}&\gradient{46.4}&\gradient{20.7}\\
TI   &\gradient{85.6}&\gradient{84.8}&\gradient{41.5}&\gradient{19.3}\\
SI   &\gradient{74.4}&\gradient{64.0}&\gradient{40.4}&\gradient{16.3}\\
VT   &\gradient{76.7}&\gradient{72.7}&\gradient{41.1}&\gradient{18.0}\\
Admix&\gradient{71.7}&\gradient{70.1}&\gradient{36.2}&\gradient{15.7}\\
\hline
TAP&\gradient{56.0}&\gradient{56.3}&\gradient{34.0}&\gradient{15.1}\\
AA& \gradient{27.7}&\gradient{31.6}&\gradient{24.2}&\gradient{13.1}\\
ILA&\gradient{77.8}&\gradient{76.4}&\gradient{45.0}&\gradient{19.3}\\
FIA&\gradient{84.1}&\gradient{83.8}&\underline{\gradient{58.4}}&\underline{\gradient{26.2}}\\
NAA&\gradient{82.7}&\gradient{79.6}&\underline{\gradient{59.9}}&\underline{\gradient{28.4}}\\
\hline
SGM             &\gradient{69.8}&\gradient{68.1}&\gradient{35.1}&\gradient{17.4}\\
LinBP           &\gradient{82.6}&\gradient{83.3}&\gradient{44.2}&\gradient{18.9}\\
RFA$_2$         &\underline{\gradient{92.9}}&\gradient{86.8}&\textbf{\gradient{73.1}}&\textbf{\gradient{41.6}}\\
RFA$_{\infty}$  &\gradient{22.7}&\gradient{19.4}&\gradient{32.5}&\gradient{17.6}\\
IAA             &\gradient{91.2}&\underline{\gradient{90.9}}&\gradient{54.0}&\gradient{22.9}\\
DSM             &\gradient{81.2}&\gradient{75.3}&\gradient{36.1}&\gradient{15.8}\\
\hline
GAP &\gradient{39.4}&\gradient{50.8}&\gradient{32.0}&\gradient{14.7}\\
CDA &\gradient{45.4}&\gradient{69.4}&\gradient{34.4}&\gradient{15.9}\\
GAPF&\textbf{\gradient{93.5}}&\textbf{\gradient{96.6}}&\gradient{49.2}&\gradient{22.9}\\
BIA &\gradient{72.3}&\gradient{82.3}&\gradient{41.6}&\gradient{18.7}\\
TTP &\gradient{62.1}&\gradient{71.9}&\gradient{42.2}&\gradient{21.0}\\

\bottomrule[1pt]
\end{tabular}
\label{tab:trans_linf8}
\end{table}

\newcommand{\GPSNRE}[1]{
    \ifdimcomp{#1 pt}{<}{30.0 pt}{#1}{
        \ifdimcomp{#1 pt}{>}{35.0 pt}{#1}{
            \pgfmathparse{int(100.0*((#1-30.0)/(35.0-30.0)))}
            \xdef\tempa{\pgfmathresult}
            \cellcolor{gray!\tempa!lightgray} #1
    }}
}

\newcommand{\GSSIME}[1]{
    \ifdimcomp{#1 pt}{<}{0.75 pt}{#1}{
        \ifdimcomp{#1 pt}{>}{0.89 pt}{#1}{
            \pgfmathparse{int(100.0*((#1-0.75)/(0.89-0.75)))}
            \xdef\tempa{\pgfmathresult}
            \cellcolor{gray!\tempa!lightgray} #1
    }}
}

\newcommand{\GDEE}[1]{
    \ifdimcomp{#1 pt}{<}{0.34 pt}{#1}{
        \ifdimcomp{#1 pt}{>}{0.49 pt}{#1}{
            \pgfmathparse{int(100.0*((#1-0.34)/(0.49-0.34)))}
            \xdef\tempa{\pgfmathresult}
            \cellcolor{lightgray!\tempa!gray} #1
    }}
}

\newcommand{\GLPIPSE}[1]{
    \ifdimcomp{#1 pt}{<}{0.05 pt}{#1}{
        \ifdimcomp{#1 pt}{>}{0.17 pt}{#1}{
            \pgfmathparse{int(100.0*((#1-0.05)/(0.17-0.05)))}
            \xdef\tempa{\pgfmathresult}
            \cellcolor{lightgray!\tempa!gray} #1
    }}
}

\newcommand{\GFIDE}[1]{
    \ifdimcomp{#1 pt}{<}{14 pt}{#1}{
        \ifdimcomp{#1 pt}{>}{150 pt}{#1}{
            \pgfmathparse{int(100.0*((#1-14)/(150-14)))}
            \xdef\tempa{\pgfmathresult}
            \cellcolor{lightgray!\tempa!gray} #1
    }}
}

\begin{table}[!t]
\caption{Imperceptibility regarding five metrics under $L_{\infty}=8/255$. Our findings about stealthiness in Table~\ref{tab:imper} still hold.}
\newcommand{\tabincell}[2]{\begin{tabular}{@{}#1@{}}#2\end{tabular}}

\renewcommand{\arraystretch}{1}
      \centering
      \setlength{\tabcolsep}{1.5pt}
        \begin{tabular}{l|cccccc}
\toprule[1pt]
Attacks&PSNR$\uparrow$&SSIM$\uparrow$&$\Delta E$$\downarrow$&LPIPS$\downarrow$&FID$\downarrow$\\
\midrule

PGD&\textbf{\GPSNRE{33.651}}&\underline{\GSSIME{0.866}}&\textbf{\GDEE{0.343}}&\textbf{\GLPIPSE{0.058}}&\underline{\GFIDE{15.227}}\\

\hline
MI&\GPSNRE{32.13}&\GSSIME{0.816}&\GDEE{0.365}&\GLPIPSE{0.104}&\GFIDE{20.245}\\
NI&\GPSNRE{31.971}&\GSSIME{0.809}&\GDEE{0.372}&\GLPIPSE{0.112}&\GFIDE{22.398}\\
PI&\GPSNRE{31.918}&\GSSIME{0.809}&\GDEE{0.374}&\GLPIPSE{0.112}&\GFIDE{25.015}\\
\hline
DI   &\underline{\GPSNRE{33.415}}&\underline{\GSSIME{0.864}}&\underline{\GDEE{0.349}}&\underline{\GLPIPSE{0.062}}&\GFIDE{35.828}\\
TI   &\GPSNRE{33.285}&\underline{\GSSIME{0.864}}&\GDEE{0.354}&\GLPIPSE{0.063}&\GFIDE{29.333}\\
SI   &\GPSNRE{33.106}&\GSSIME{0.857}&\GDEE{0.350}&\GLPIPSE{0.068}&\GFIDE{24.483}\\
VT   &\GPSNRE{32.777}&\GSSIME{0.850}&\GDEE{0.370}&\GLPIPSE{0.074}&\GFIDE{25.213}\\
Admix&\GPSNRE{33.225}&\GSSIME{0.855}&\GDEE{0.350}&\GLPIPSE{0.070}&\GFIDE{22.596}\\
\hline
TAP&\GPSNRE{31.875}&\GSSIME{0.810}&\GDEE{0.365}&\GLPIPSE{0.101}&\GFIDE{44.118}\\
AA &\GPSNRE{31.949}&\GSSIME{0.842}&\GDEE{0.421}&\GLPIPSE{0.090}&\GFIDE{25.691}\\
ILA&\GPSNRE{31.511}&\GSSIME{0.812}&\GDEE{0.448}&\GLPIPSE{0.089}&\GFIDE{51.048}\\
FIA&\GPSNRE{31.886}&\GSSIME{0.833}&\GDEE{0.405}&\GLPIPSE{0.086}&\GFIDE{85.818}\\
NAA&\GPSNRE{31.962}&\GSSIME{0.837}&\GDEE{0.408}&\GLPIPSE{0.078}&\GFIDE{54.370}\\
\hline
SGM&\GPSNRE{33.020}&\GSSIME{0.855}&\underline{\GDEE{0.349}}&\GLPIPSE{0.066}&\underline{\GFIDE{19.759}}\\
LinBP&\GPSNRE{32.337}&\GSSIME{0.840}&\GDEE{0.376}&\GLPIPSE{0.073}&\GFIDE{44.614}\\
RFA$_{2}$&\GPSNRE{32.024}&\GSSIME{0.858}&\GDEE{0.391}&\GLPIPSE{0.086}&\GFIDE{40.841}\\
RFA$_{\infty}$&\GPSNRE{30.416}&\textbf{\GSSIME{0.884}}&\GDEE{0.455}&\GLPIPSE{0.074}&\textbf{\GFIDE{14.753}}\\
IAA&\GPSNRE{31.807}&\GSSIME{0.829}&\GDEE{0.417}&\GLPIPSE{0.081}&\GFIDE{109.173}\\
DSM&\underline{\GPSNRE{33.337}}&\GSSIME{0.862}&\underline{\GDEE{0.349}}&\underline{\GLPIPSE{0.062}}&\GFIDE{22.101}\\
\hline
GAP&\GPSNRE{30.171}&\GSSIME{0.769}&\GDEE{0.460}&\GLPIPSE{0.125}&\GFIDE{56.566}\\
CDA&\GPSNRE{30.455}&\GSSIME{0.796}&\GDEE{0.483}&\GLPIPSE{0.103}&\GFIDE{75.639}\\
GAPF&\GPSNRE{30.800}&\GSSIME{0.810}&\GDEE{0.473}&\GLPIPSE{0.095}&\GFIDE{146.506}\\
BIA&\GPSNRE{30.347}&\GSSIME{0.751}&\GDEE{0.462}&\GLPIPSE{0.161}&\GFIDE{123.255}\\
TTP&\GPSNRE{30.932}&\GSSIME{0.829}&\GDEE{0.430}&\GLPIPSE{0.107}&\GFIDE{69.490}\\

\bottomrule[1pt]
\end{tabular}
\label{tab:imper_linf8}
\end{table}

\begin{table}[!t]
\newcommand{\tabincell}[2]{\begin{tabular}{@{}#1@{}}#2\end{tabular}}
\caption{Hyperparameters of attacks and defenses.}

\renewcommand{\arraystretch}{1}
      \centering
        \begin{tabular}{l|c}
\toprule[1pt]
Attacks&Hyperparameter\\

\midrule
MI~\cite{dong2018boosting}&decay factor $\mu=1$\\
NI~\cite{lin2020nesterov} &decay factor $\mu=1$\\
PI~\cite{wang2021boosting} &decay factor $\mu=1$\\
\hline
DI~\cite{xie2019improving} &\tabincell{c}{resize\&pad range $R=[1,~1.1]$\\transformation probability $p=0.7$}\\
TI~\cite{dong2019evading} &translation range $R=[-2,2]$\\
SI~\cite{lin2020nesterov} &scale range $R=[0.1,1]$\\ 
VT~\cite{wang2021enhancing} &noise range $R=[-1.5\epsilon,1.5\epsilon]$\\ 
Admix~\cite{wang2021admix} &mixing factor $\eta=0.2$ \\
\hline
TAP~\cite{zhou2018transferable} & $\lambda = 0.005$, $\eta = 0.01$, $\alpha = 0.5$\\
AA~\cite{inkawhich2019feature}& $20$ random target images from 4 classes\\
ILA~\cite{huang2019enhancing} & ILA projection loss\\
FIA~\cite{wang2021feature}& $N=30$, $P_\mathrm{drop}=0.3$\\
NAA~\cite{zhang2022improving} & $N=30$, $\gamma=1.0$, linear transformation\\
\hline
SGM~\cite{wu2020skip} &decay parameter $\gamma=0.5$\\
LinBP~\cite{guo2020backpropagating} &first residual unit in the third meta block\\
RFA~\cite{springer2021little} & $L_{2}$ $\epsilon=0.1$ or $L_{\infty}$ $\epsilon=8$ PGD-AT\\
IAA~\cite{zhu2021rethinking} &$\beta=15$\\
DSM~\cite{yang2022boosting}&CutMix augmentation\\
\hline
GAP~\cite{poursaeed2018generative}&ResNet152 discriminator\\
CDA~\cite{naseer2019cross} &ResNet152 discriminator\\
GAPF~\cite{kanth2021learning} &ResNet152 discriminator\\
BIA~\cite{zhang2022beyond} &ResNet152 discriminator, RN module\\
TTP~\cite{naseer2021generating}&ResNet50 discriminator\\
\midrule

Defenses&Hyperparameters\\

\midrule
BDR~\cite{xu2017feature}&bit depth $D=2$\\
PD~\cite{prakash2018deflecting} &R-CAM: $k=5$, denoising: $\sigma=0.04$\\
R\&P~\cite{xie2017mitigating} &resize\&pad range $R=[1,~1.1]$\\
\hline
HGD~\cite{liao2018defense} &ResNet152-Wide\\
NRP~\cite{naseer2020self} &ResNet (1.2M parameters) \\
DiffPure~\cite{nie2022diffusion} & noise level $t=150$ WideResNet-50-2\\
\hline
AT$_{\infty}$~\cite{xie2019feature}&\tabincell{c}{$L_{\infty}$ $\epsilon=8$, PGD iters $n=30$, lr=$\alpha=1$}\\
FD$_{\infty}$~\cite{xie2019feature}&\tabincell{c}{$L_{\infty}$ $\epsilon=8$, PGD iters $n=30$, lr=$\alpha=1$}\\
AT$_{2}$~\cite{salman2020adversarially} &\tabincell{c}{$L_{2}$ $\epsilon=3.0$, PGD iters $n=7$, lr=$\alpha=0.5$}\\
\bottomrule[1pt]
\end{tabular}
\label{tab:hyper}
\end{table}

\subsection{Hyperparameters of Attacks and Defenses}
\label{app:hyper}
We follow the settings in the existing work for attack and defense hyperparameters, as summarized in Table~\ref{tab:hyper}.

\subsection{Descriptions of Attacks}
\label{app:att_def}

\noindent\textbf{Momentum Iterative (MI)}~\cite{dong2018boosting} integrates the momentum term into the iterative optimization of attacks, in order to stabilize the update directions and escape from poor local maxima. This momentum term accumulates a velocity vector in the gradient direction of the loss function across iterations.
   
   \noindent\textbf{Nesterov Iterative (NI)}~\cite{lin2020nesterov} is based on an improved momentum method that also leverages the look ahead property of Nesterov Accelerated Gradient (NAG) by making a jump in the direction of previously accumulated gradients before computing the current gradients. This makes the attack escape from poor local maxima more easily and faster.
   
   \noindent\textbf{Pre-gradient guided Iterative (PI)}~\cite{wang2021boosting} follows a similar idea as NI, but it makes a jump based on only the gradients from the last iteration instead of all previously accumulated gradients.
   
   \noindent\textbf{Diverse Inputs (DI)}~\cite{xie2019improving} applies random image resizing and padding to the input image before calculating the gradients in each iteration of the attack optimization. This approach aims to prevent attack overfitting (to the white-box, source model), inspired by the data augmentation techniques used for preventing model overfitting.
   
   \noindent\textbf{Translation Invariant (TI)}~\cite{dong2019evading} applies random image translations for input augmentation. It also introduces an approximate solution to improve the attack efficiency by directly computing locally smoothed gradients on the original image through the convolution operations rather than computing gradients multiple times for all potential translated images.
   
   \noindent\textbf{Scale Invariant (SI)}~\cite{lin2020nesterov} applies random image scaling for input augmentation. It scales pixels with a factor of $1/2^i$. In particular, in each iteration, it takes an average of gradients on multiple augmented images rather than using only one augmented image as in previous input augmentation attacks.
   
   \noindent\textbf{Variance Tuning (VT)}~\cite{wang2021enhancing} applies uniformly distributed additive noise to images for input augmentation and also calculates average gradients over multiple augmented images in each iteration. 
   
   \noindent\textbf{Adversarial mixup (Admix)}~\cite{wang2021admix} calculates the gradients on a composite image that is made up of the original image and another image randomly selected from an incorrect class. The original image label is still used in the loss function.
   
   \noindent\textbf{Transferable Adversarial Perturbations (TAP)}~\cite{zhou2018transferable} proposes to maximize the distance between original images and their adversarial examples in the intermediate feature space and also introduces a regularization term for reducing the variations of the perturbations and another regularization term with the cross-entropy loss.
   
   \noindent\textbf{Activation Attack (AA)}~\cite{inkawhich2019feature} drives the feature-space representation of the original image towards the representation of a target image that is selected from another class. Specifically, AA can achieve targeted misclassification by selecting a target image from that specific target class. 
   
   \noindent\textbf{Intermediate Level Attack (ILA)}~\cite{huang2019enhancing} optimizes the adversarial examples in two stages, with the cross-entropy loss used in the first stage to determine the initial perturbations, which will be further fine-tuned in the second stage towards larger feature distance while maintaining the initial perturbation directions.
   
   \noindent\textbf{Feature Importance-aware Attack (FIA)}~\cite{wang2021feature} proposes to only disrupts important features. Specifically, it measures the importance of features based on the aggregated gradients with respect to feature maps computed on a batch of transformed original images that are achieved by random image masking.
   
   \noindent\textbf{Neuron Attribution-based Attacks (NAA)}~\cite{zhang2022improving} relies on an advanced neuron attribution method to measure the feature importance more accurately. It also introduces an approximation approach to conducting neuron attribution with largely reduced computations.
   
   \noindent\textbf{Skip Gradient Method (SGM)}~\cite{wu2020skip} suggests using ResNet-like architectures as the source model during creating the adversarial examples. Specifically, it shows that backpropagating gradients through skip connections lead to higher transferability than through the residual modules.
   
   \noindent\textbf{Linear BackPropagation (LinBP)}~\cite{guo2020backpropagating} is proposed based on the new finding that the non-linearity of the commonly-used ReLU activation function substantially limits the transferability. To address this limitation, the ReLU is replaced by a linear function during only the backpropagation process.

   \noindent\textbf{Robust Feature-guided Attack (RFA)}~\cite{springer2021little} proposes to use an adversarially-trained model (with $L_{2}$ or $L_{\infty}$ bound) as the source model based on the assumption that modifying more robust features yields more generalizable (transferable) adversarial examples.
   
   \noindent\textbf{Intrinsic Adversarial Attack (IAA)}~\cite{zhu2021rethinking} finds that disturbing the intrinsic data distribution is the key to generating transferable adversarial examples. Based on this, it optimizes the hyperparameters of the Softplus and the weights of skip connections per layer towards aligned attack directions and data distribution. 
   
     \noindent\textbf{Dark Surrogate Model (DSM)}~\cite{yang2022boosting} is trained from scratch with additional ``dark'' knowledge, which is achieved by training with soft labels from a pre-trained teacher model and using data augmentation techniques, such as Cutout, Mixup, and CutMix. 
     
     \noindent\textbf{Generative Adversarial Perturbations (GAP)}~\cite{poursaeed2018generative} proposes a new attack approach that is based on generative modeling. Specifically, it uses the source classifier as the discriminator and trains a generator using the cross-entropy loss. Once trained, the generator can be used to generate an adversarial example for each input original image with only one forward pass.
     
   \noindent\textbf{Cross-Domain Attack (CDA)}~\cite{naseer2019cross} follows the GAP pipeline but uses a more advanced loss (i.e., relativistic cross entropy) to train the generator. This new loss explicitly enforces the probability gap between the clean and adversarial images, boosting the transferability, especially in cross-domain scenarios.
   
   \noindent\textbf{Transferable Targeted Perturbations (TTP)}~\cite{naseer2021generating} is focused on improving transferability of targeted attacks. It is based on learning target-specific generators, each of which is trained with the objective of matching the distribution of targeted perturbations with that of data from a specific target class. Specifically, input augmentation and smooth perturbation projection are used to further boost the performance.
   
   \noindent\textbf{Generative Adversarial Feature Perturbations (GAFP)}~\cite{kanth2021learning} follows the general pipeline of GAP but trains the generator using a loss that maximizes the feature map distance between adversarial and original images at mid-level CNN layers. 
   
   \noindent\textbf{Beyond ImageNet Attack (BIA)}~\cite{zhang2022beyond} also follows the general pipeline of GAP and specifically introduces a random normalization module to simulate different training data distributions and also a feature distance loss that is only based on important/generalizable features.

\subsection{Descriptions of Defenses}

   \noindent\textbf{Bit-Depth Reduction (BDR)}~\cite{xu2017feature} pre-processes input images by reducing the color depth of each pixel while maintaining the semantics. This operation can eliminate pixel-level adversarial perturbations from adversarial images but has little impact on model predictions of clean images.
   
   \noindent\textbf{Pixel Deflection (PD)}~\cite{prakash2018deflecting} pre-processes input images by randomly replacing some pixels with randomly selected pixels from their local neighborhood. It is specifically designed to happen more frequently to non-salient pixels, and a subsequent wavelet-based denoising operation is used to soften the corruption.
 
   \noindent\textbf{Resizing and Padding (R\&P)}~\cite{xie2017mitigating} pre-processes input images by random resizing, which resizes the input images to a random size, and then random padding, which pads zeros around the resized input images. 
   
   \noindent\textbf{High-level representation Guided Denoiser (HGD)}~\cite{liao2018defense} learns a purification network that can be used to purify/denoise the adversarial perturbations. Specifically, different from previous methods that focus on image-space denoising, HGD minimizes the difference between the clean image and the denoised image at intermediate feature layers. 
   
   \noindent\textbf{Neural Representation Purifier (NRP)}~\cite{naseer2020self} learns a purification network using a combined loss that calculates both the image- and feature-space differences. Specifically, the adversarial images used for training the purification network are generated by feature loss-based adversaries, which are shown to be more effective in handling unseen attacks.
   
   \noindent\textbf{Diffusion Purification (DiffPure)}~\cite{nie2022diffusion} uses a diffusion model as the purification network. It diffuses an input image by gradually adding noise in a forward diffusion process and then recovers the clean image by gradually denoising the image in a reverse generative process.
  The reverse process is shown to be also capable of removing adversarial perturbations.
  
\noindent\textbf{$L_{\infty}$-Adversarial Training (AT$_{\infty}$)}~\cite{xie2019feature} follows the basic pipeline of adversarial training, which is to train the robust model on adversarial images. Specifically, these adversarial images are generated using the PGD targeted attacks with the $L_{\infty}$ distance, and the model is trained distributedly on 128 Nvidia V100 GPUs.

\noindent\textbf{$L_{\infty}$-Adversarial Training with Feature Denoising (FD$_{\infty}$)}~\cite{xie2019feature} modifies the standard model architecture by introducing new building blocks that are designed for denoising feature maps based on non-local means or other filters. This modification can help suppress the potential disruptions caused by the adversarial perturbations, and the modified model is trained end-to-end.

   \noindent\textbf{$L_{2}$-Adversarial Training (AT$_{2}$)}~\cite{salman2020adversarially} trains the model on adversarial images that are generated using the PGD non-targeted attacks with the $L_{2}$ distance.

\subsection{Imperceptibility Metrics}
\label{app:metric}
  \noindent\textbf{Peak Signal-to-Noise Ratio (PSNR)} measures the ratio of the maximum possible power of a signal (image $\mathbf{x}$) to the noise (perturbations $\mathbf{\sigma}$) power.
  It is calculated by:
  \begin{equation}\label{eq:psnr}
  \textrm{PSNR}(\mathbf{x},\mathbf{\sigma})=10\cdot\log_{10}\frac{\max(\mathbf{x})^2}{\textrm{MSE}(\mathbf{x},\mathbf{\sigma})},
  \end{equation}
where the Mean Squared Error (MSE) measures the average squared difference between the two inputs.

  \noindent\textbf{Structural Similarity Index Measure (SSIM)}~\cite{wang2004image} is used for measuring the perceptual quality of digital images. In our case, it measures the structural similarity between the original image $\mathbf{x}$ and adversarial image $\mathbf{x}'$, which is considered to be a degraded version of $\mathbf{x}$.
  It is calculated by: 
    \begin{equation}\label{eq:ssim}
   \textrm{SSIM}(\mathbf{x},\mathbf{x}')=\frac{{ ( 2 \mu_{\mathbf{x}}\mu_{\mathbf{x}'}+ c_1 )+  (2 \sigma_{\mathbf{x}\mathbf{x}'}+c_2)}} 
{(\mu_{\mathbf{x}}^2+\mu_{\mathbf{x}'}^2+c_1)(\sigma_{\mathbf{x}}^2+\sigma_{\mathbf{x}'}^2+c_2)}.
  \end{equation}
Descriptions of $c_1$ and $c_1$ and more technical details can be found in~\cite{wang2004image}.

  \noindent\textbf{$\Delta$\textit{E}}~\cite{luo2001development} refers to the CIE $\Delta E$ standard formula used for measuring the perceptual color distance between two pixels. We follow~\cite{zhao2020towards} to use the latest variant, CIEDE2000, since it is shown to align very well with human perception.
It is calculated in the CIELCH space by:
\begin{equation}
\label{eq:deltae}
\begin{gathered}
\Delta E=\sqrt{(\frac{\Delta L'}{k_LS_L})^2+(\frac{\Delta C'}{k_CS_C})^2+(\frac{\Delta H'}{k_HS_H})^2+\Delta R},\\
\Delta R=R_T(\frac{\Delta C'}{k_CS_C})(\frac{\Delta H'}{k_HS_H}),
\end{gathered}
\end{equation}
where $\Delta L'$, $\Delta C'$, $\Delta H'$ denotes the distance between two pixels in their lightness, chroma, and hue channels, respectively. $\Delta R$ is an interactive term between chroma and hue differences.
Detailed definitions and explanations of the weighting functions ($S_L$, $S_C$, $S_H$ and $R_T$) and hyperparameters $k_L$, $k_C$ and $k_H$ can be found in~\cite{luo2001development}.
We use the averaged $\Delta E$ over all pixels.

  \noindent\textbf{Learned Perceptual Image Patch Similarity (LPIPS)}~\cite{zhang2018unreasonable} is developed for measuring the perceptual similarity between two images. It computes the cosine distance (in
the channel dimension) between features at each given convolutional layer $l$ and averages the results across spatial dimensions $H\times W$ and layers of a specific network $f$:
  \begin{equation}
\label{eq:lpips}
  \textrm{LPIPS}=\sum_{l}\frac{1}{H_{l}W_{l}}\sum _{h,w} \cos(f^{l}_{hw}(\mathbf{x}),f^{l}_{hw}(\mathbf{x}')),
  \end{equation}
where AlexNet is adopted as $f$ for computational efficiency.

\noindent\textbf{Frechet Inception Distance (FID)}~\cite{szegedy2016rethinking} is originally used to assess the quality of generated images and can also be used to assess the quality of adversarial images.
It compares the distribution of adversarial images to that of original images.
Specifically, the output features from the pool3 layer of an Inception-V3 for original and adversarial images are used to fit two multidimensional Gaussian distributions $\mathcal {N}(\mu,\Sigma )$ and $\mathcal{N}(\mu',\Sigma')$, and the FID score is calculated by:
  \begin{equation}
\label{eq:fid}
  \textrm{FID}(\mathcal{N}(\mu,\Sigma),\mathcal{N}(\mu' ,\Sigma' ))=\|\mu– \mu'\|_2^2 + \textrm{tr}(\Sigma + \Sigma'- 2\sqrt{\Sigma\Sigma'}),
  \end{equation}
where $\textrm{tr}$ denotes the trace of the matrix.

\subsection{Interpretability Measures}
\label{app:inter}
\input{tex/InterpretMetrics}

%% file: tex/InterpretMetrics.tex
For model interpretability, we adopt two metrics, Average Increase (AI) and Average Drop (AD), from~\cite{chattopadhay2018grad}.
Let $p^c_i$ and $o^c_i$ be the predicted probability for the ground-truth class $c$ given as input the $i$-th image $\mathbf{x}_i$ and its masked version, and let $n$ be the number of test images.
AI measures the percentage of images where the masked image yields a higher class probability than the original (higher is better):
\begin{equation}
	\AI~(\%) \defn \frac{1}{n} \sum_i^n \ind_{p^c_i < o^c_i} \cdot 100
\label{eq:ai}
\end{equation}

AD quantifies how much predictive power, measured as class probability, is lost when only the masked regions of the image are used (lower is better):
\begin{equation}
	\AD~(\%) \defn \frac{1}{n} \sum_{i=1}^n \frac{[p^c_i - o^c_i]_+}{p^c_i} \cdot 100.
\label{eq:ad}
\end{equation}
